\documentclass[aps,prb,twocolumn,superscriptaddress, floatfix]{revtex4-1} 

\usepackage[utf8]{inputenc} 

\usepackage{amsmath}
\usepackage{amssymb}
\usepackage{fullpage}
\usepackage[pdftex]{graphicx} 
\usepackage{bm}
\usepackage{natbib}
\usepackage{color}
\usepackage[squaren,Gray, cdot]{SIunits}        
\usepackage{stmaryrd}
\usepackage{hyperref}
\usepackage{cleveref}
\Crefformat{equation}{#2Eq.~(#1)#3}
\Crefformat{figure}{#2Fig.~#1#3}
\Crefrangeformat{equation}{Eqs.~(#3#1#4) to~(#5#2#6)}
\Crefmultiformat{equation}{Eqs.~(#2#1#3)}{ and~(#2#1#3)}{, (#2#1#3)}{ and~(#2#1#3)}
\Crefmultiformat{figure}{Figs.~#2#1#3}{ and~#2#1#3}{, #2#1#3}{ and~#2#1#3}

\newcommand{\eps}{\varepsilon}

\usepackage{url}

\usepackage[T1]{fontenc}

\begin{document}
\title{Elastically-mediated interactions between grain boundaries and precipitates in two-phase coherent solids}
\author{Ye-Chuan Xu}
\affiliation{Physics Department and Center for Interdisciplinary Research on Complex Systems, Northeastern University, Boston, Massachusetts 02115, USA} 
\affiliation{Department of Materials Physics, School of Physics and Optoelectronic Engineering, Nanjing University of Information Science \& Technology, Nanjing, China}
\author{Pierre-Antoine Geslin}
\affiliation{Physics Department and Center for Interdisciplinary Research on Complex Systems, Northeastern University, Boston, Massachusetts 02115, USA} 
\affiliation{Univ Lyon, Universit\'{e} Claude Bernard Lyon 1, CNRS, Institut Lumi\`{e}re Mati\`{e}re, F-69622, LYON, France
}
\author{Alain Karma}
	\email{a.karma@neu.edu}
\affiliation{Physics Department and Center for Interdisciplinary Research on Complex Systems, Northeastern University, Boston, Massachusetts 02115, USA} 
\date{\today} 

\begin{abstract}
We investigate analytically and numerically the interaction between grain boundaries and second phase precipitates in two-phase coherent solids in the presence of misfit strain. Our numerical study uses amplitude equations that describe the interaction of composition and stress   [R. Spatschek and A. Karma, Phys. Rev. B 81, 214201 (2010)] and free-energies corresponding to two-dimensional hexagonal and three-dimensional BCC crystal structures that exhibit isotropic and anisotropic elastic properties, respectively.  We consider two experimentally motivated geometries where (i) a lamellar precipitate nucleates along a planar grain boundary that is centered inside the precipitate, and (ii) a circular precipitate nucleates inside a grain at a finite distance to an initially planar grain boundary. For the first geometry, we find that the grain boundary becomes morphologically unstable due to the combination of long-range elastic interaction between the grain boundary and compositional domain boundaries, and shear-coupled grain boundary motion.
We characterize this instability analytically by extending the linear stability analysis carried out recently [P.-A. Geslin, Y.-C. Xu, and A. Karma, Phys. Rev. Lett. 114, 105501 (2015)] to the more general case of elastic anisotropy.  The analysis predicts that elastic anisotropy hinders but does not suppress the instability. Simulations also reveal that, in a well-developed non-linear regime, this instability can lead to the break-up of low-angle grain boundaries when the misfit strain exceeds a threshold that depends on the grain boundary misorientation. For the second geometry, simulations show that the elastic interaction between an initially planar grain boundary and an adjacent circular precipitate causes the precipitate to migrate to and anchor at the grain boundary.
\end{abstract}

\maketitle

\section{Introduction}
\label{sec:intro}

Phase separation into domain structures of distinct chemical compositions occurs in a wide range of technological materials.   
Nucleation and growth of second phase precipitates inside the matrix of a primary phase is commonly
used as a strengthening mechanism of structural materials \cite{Porter1992}. Domain structures also
commonly form by spinodal decomposition into two phases, which has been widely investigated in
various contexts
\cite{Cahn1961,Ramanarayan2003,Haataja2004,Haataja2005,Hu2004,Tang2010,Hoyt2011,Lu2012,
Tang2012,Tao2012,Wang2013}. Due to the dependence of the crystal lattice spacing on composition, domain formation typically generates a misfit strain that can be large in some cases, e.g. several percent in phase-separating lithium iron phosphate battery electrode materials \cite{Tang2010}. 

The effect of a coherency stress has been investigated theoretically in the context
of both single-crystalline and polycrystalline materials. In single-crystalline materials, Cahn demonstrated that coherency stress hinders spinodal decomposition, requiring a larger chemical driving force than in the absence of misfit to generate phase-separation inside a bulk material \cite{Cahn1961}. A recent extension of this analysis showed that stress relaxation near a free surface can lead to spinodal decomposition for smaller chemical driving forces than inside a bulk material, with compositional domain formation confined at the surface  \cite{Tang2012}. 
In polycrystalline materials, numerical simulations have been used to investigate the interaction
between compositional domain boundaries (DBs) and dislocations using continuum dislocation-based
models \cite{Haataja2004,Haataja2005,Hoyt2011} phase-field approaches \cite{Hu2004}.
More recently, the interaction between DBs and grain boundaries (GBs) has also been investigated using phase-field-crystal (PFC) simulations \cite{Tao2012,Wang2013}, and amplitude equations derived from the PFC framework \cite{Elder2010}. Those studies have shown that dislocations generically migrate to 
DBs to relax the coherency stress thereby strongly impacting microstructural evolution and domain coarsening behavior \cite{Haataja2005,Tao2012,Wang2013}. 

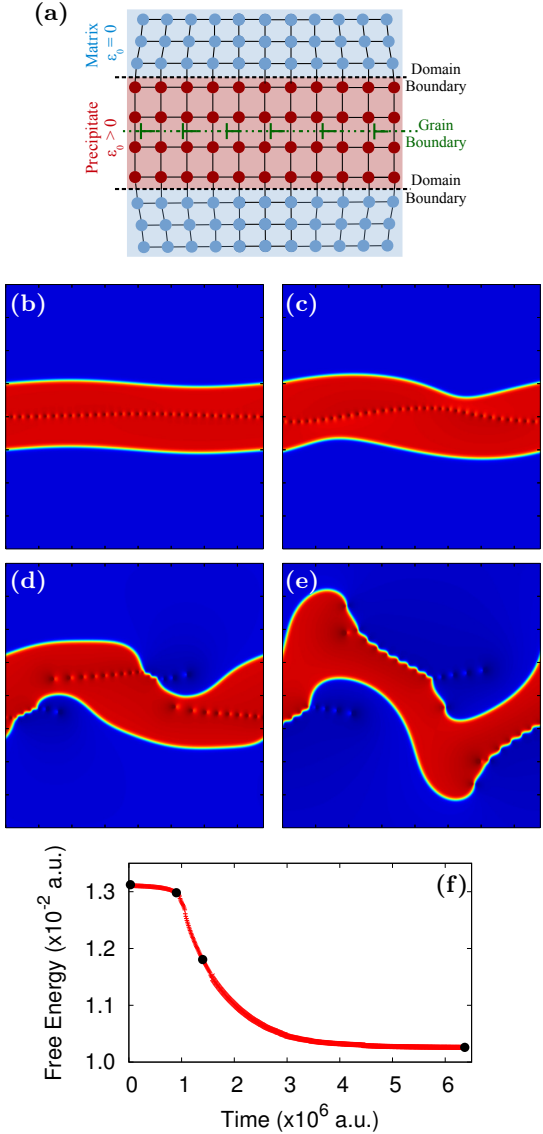
\begin{figure}
	\begin{center}
    	\includegraphics[width=0.65\linewidth]{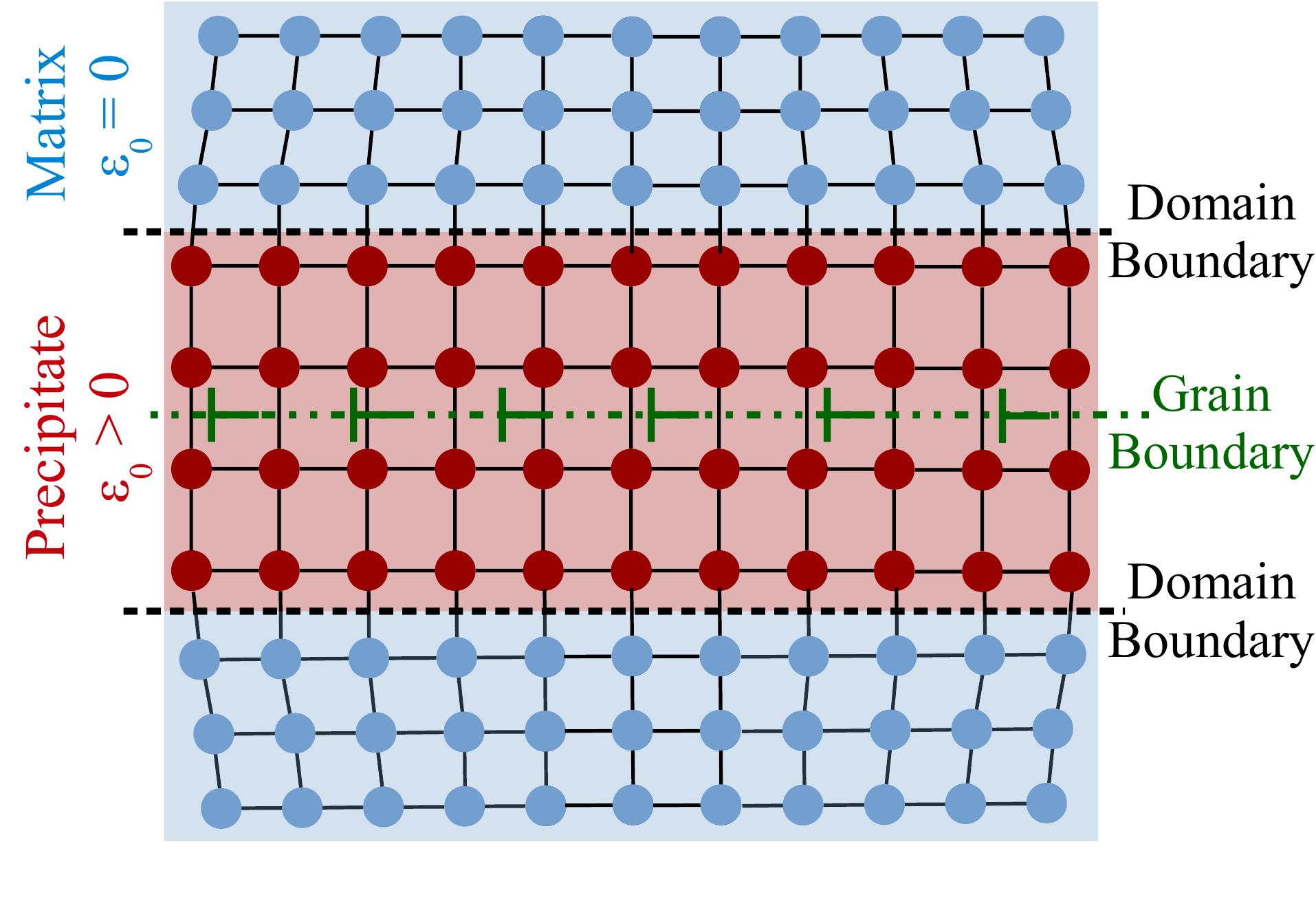}
		\includegraphics[width=0.45\linewidth]{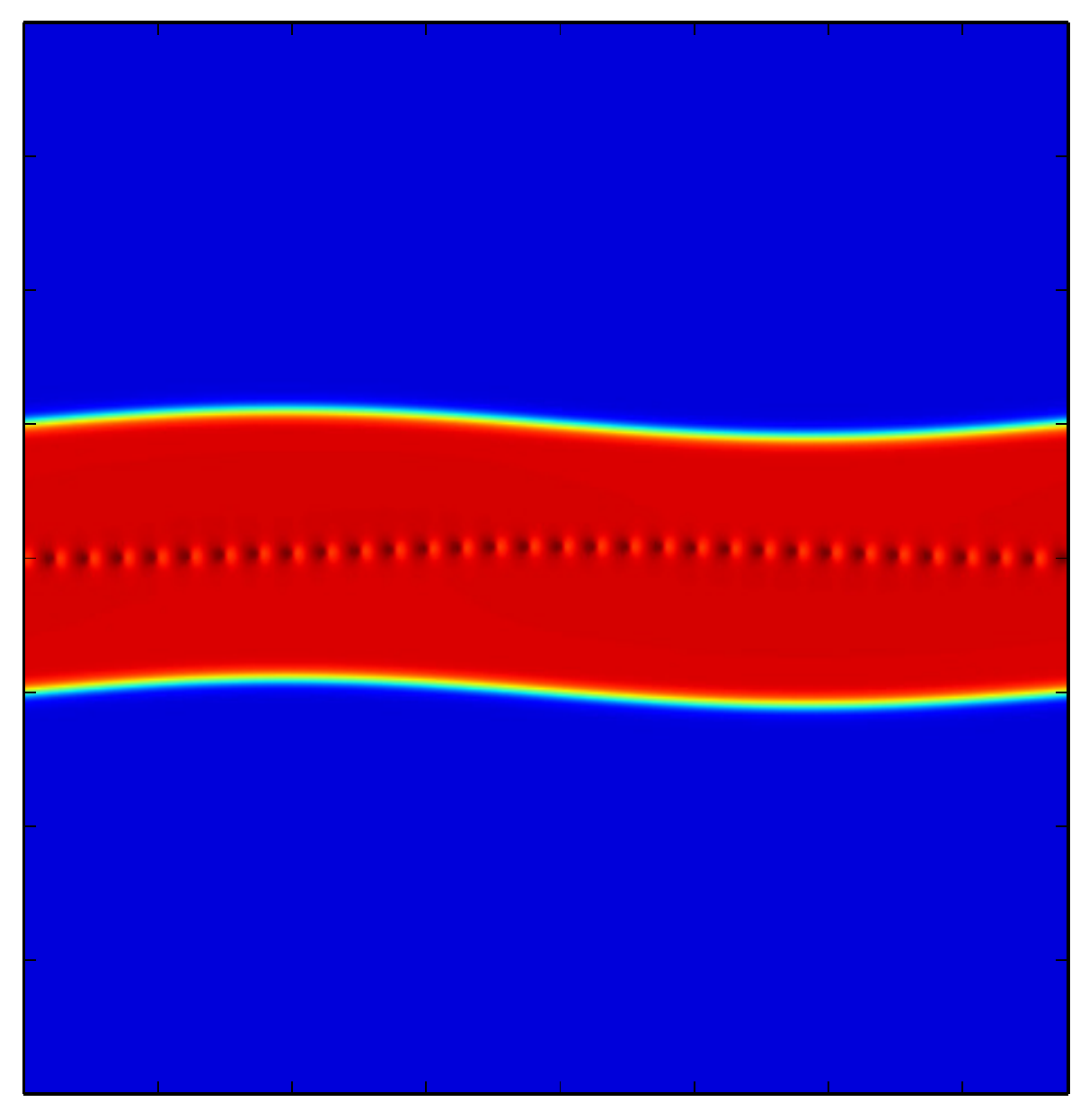}
		\includegraphics[width=0.45\linewidth]{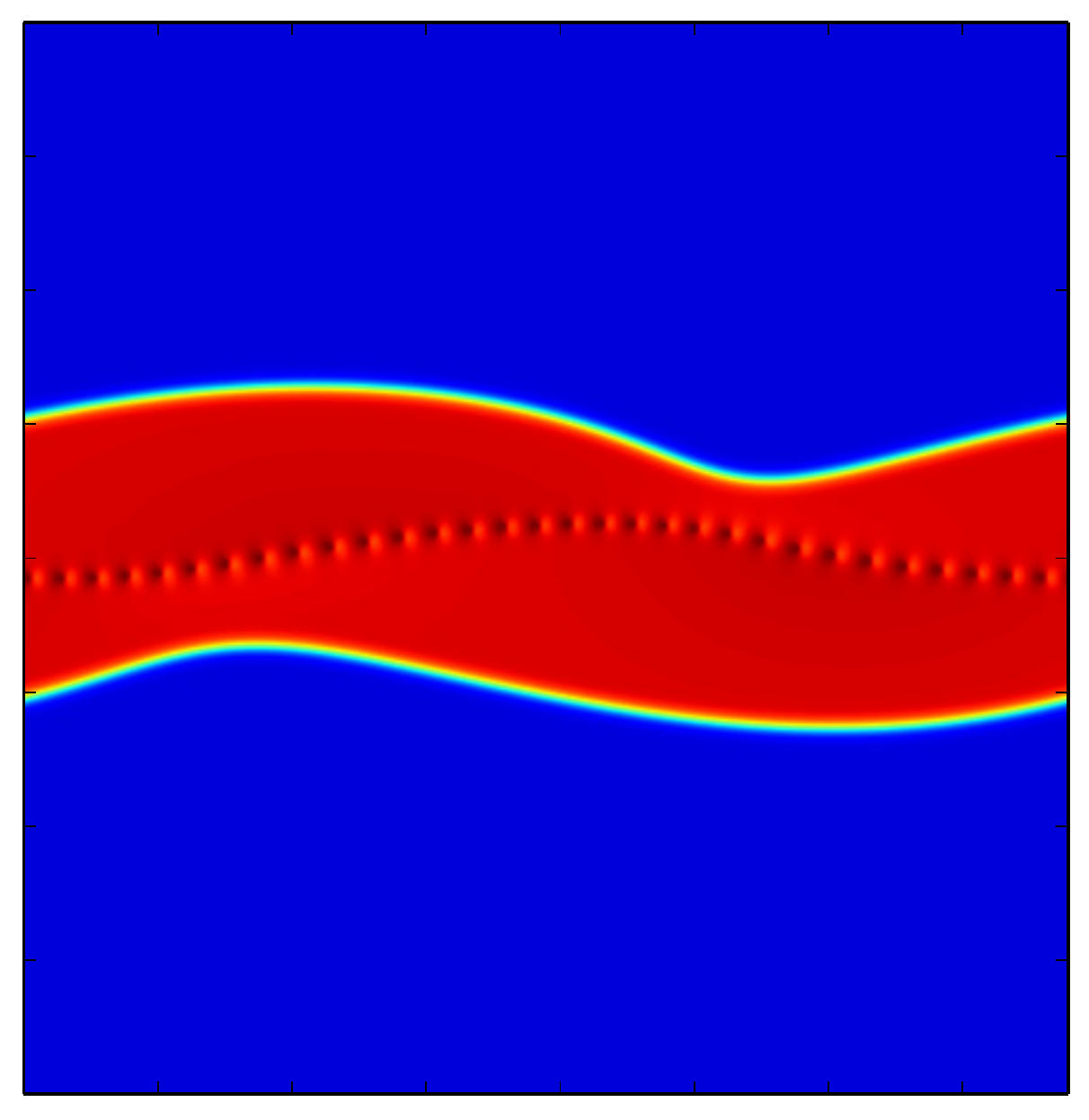}
		\includegraphics[width=0.45\linewidth]{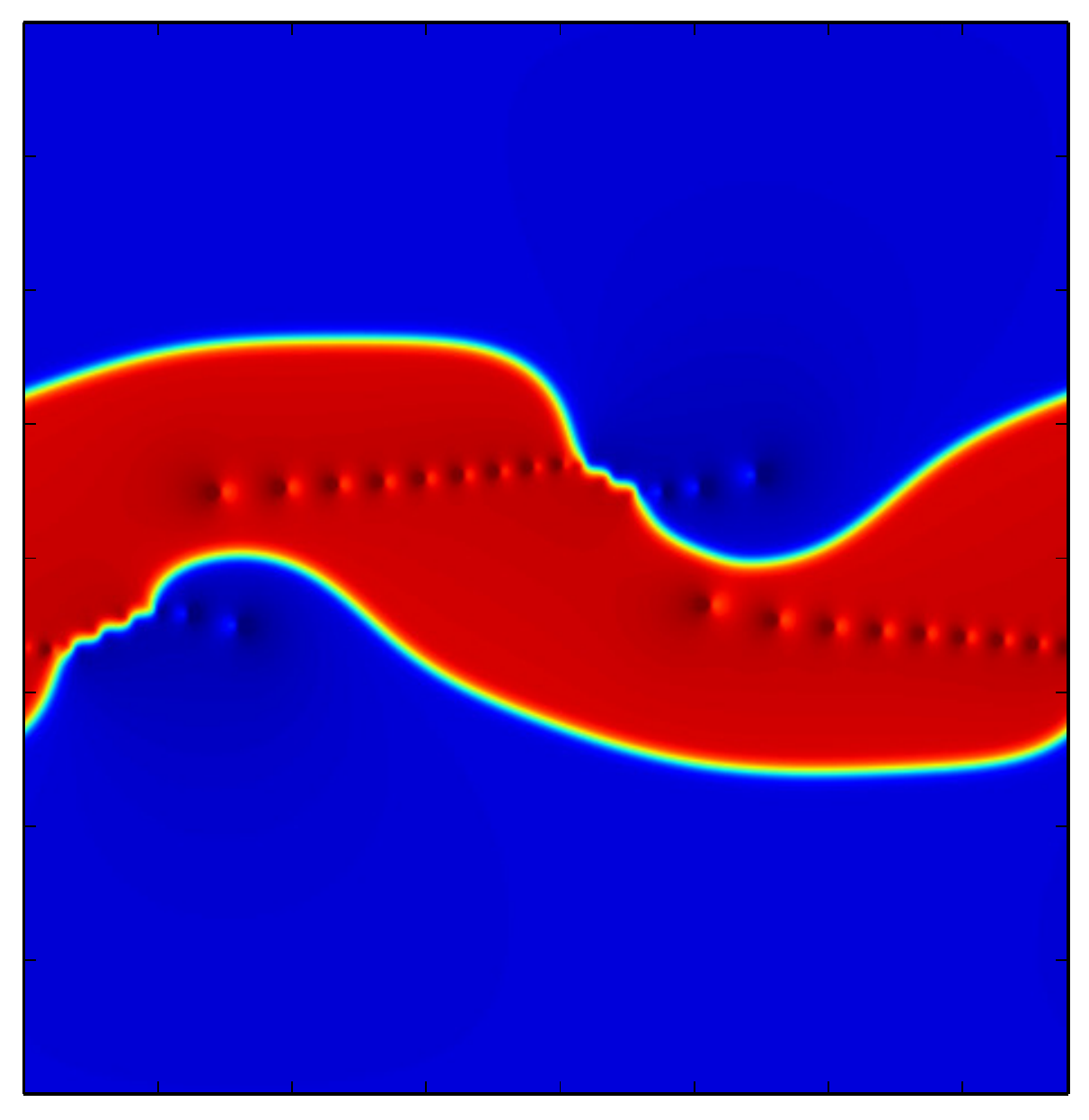}
		\includegraphics[width=0.45\linewidth]{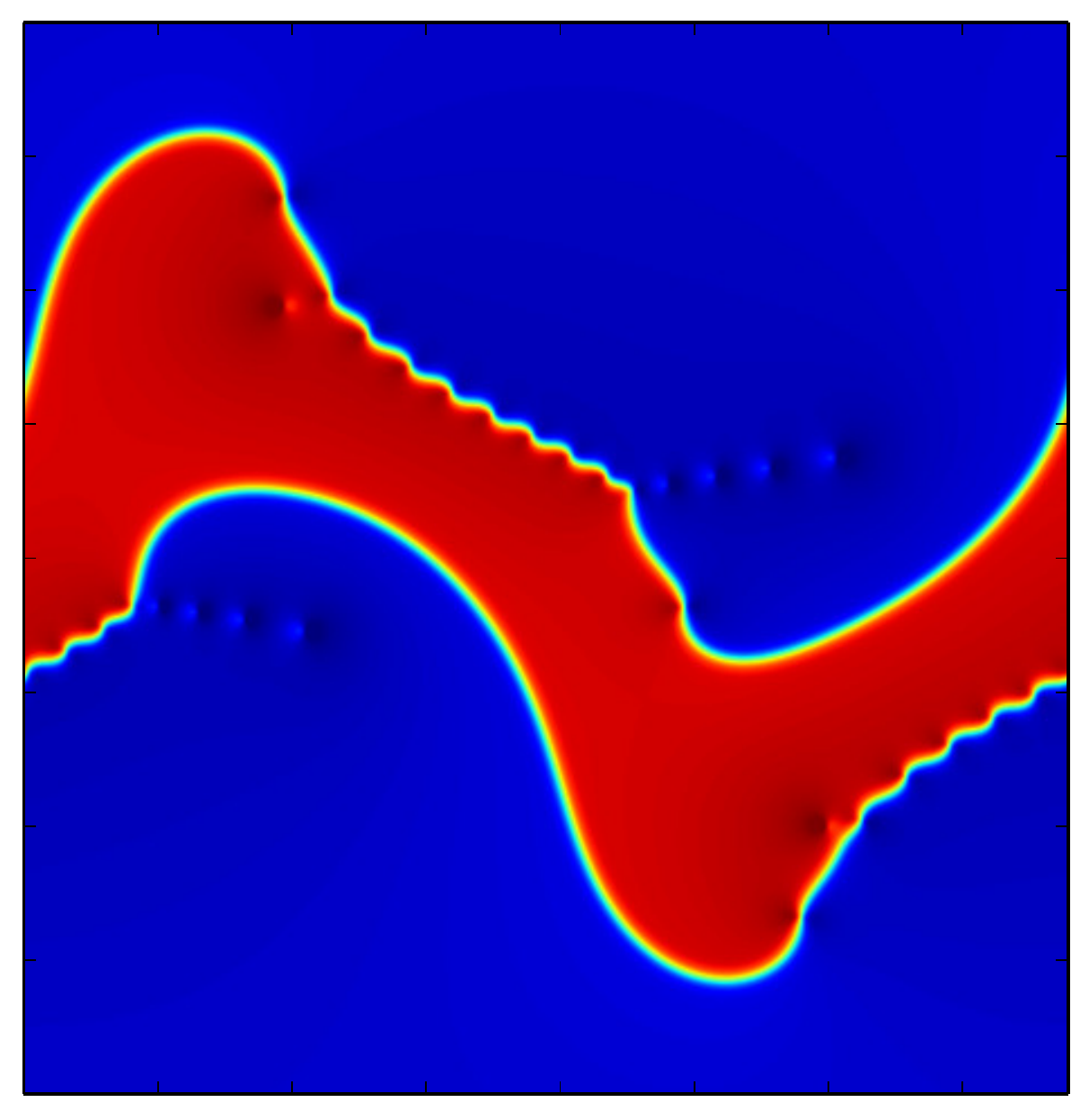}
		\includegraphics[width=0.8\linewidth] {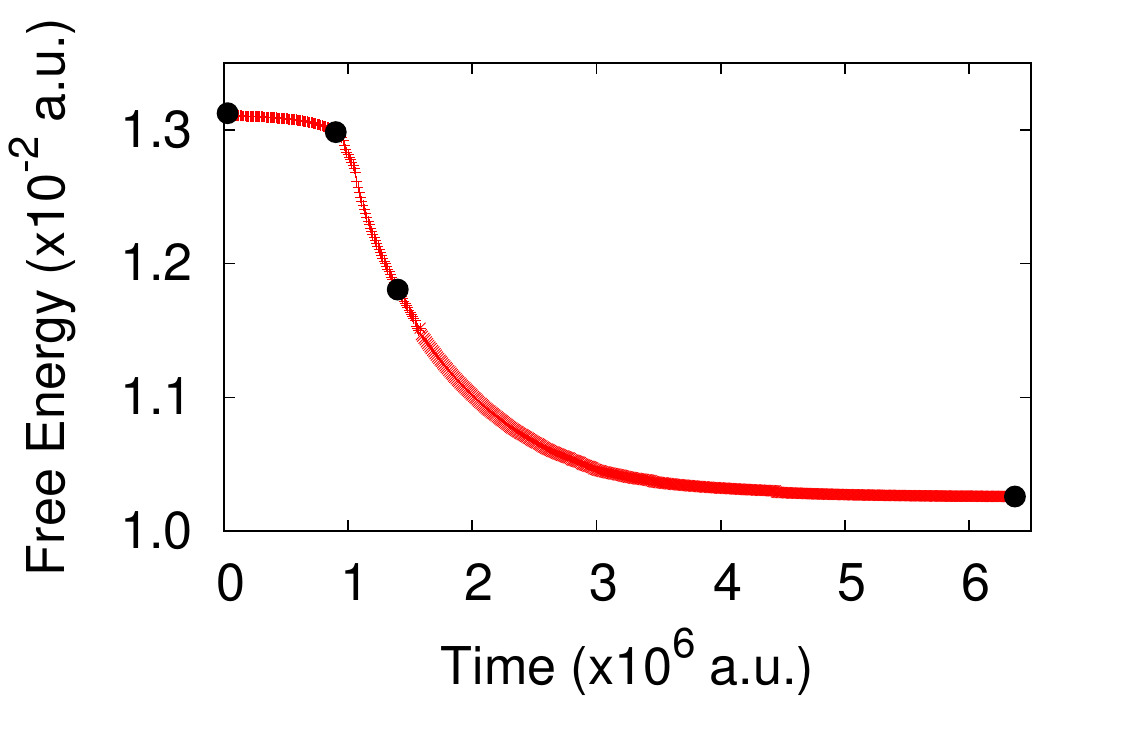}
		
		\begin{picture}(1,0)(0,0)
	     	\put(-90.,440.) {\mbox{\textcolor{black}{\textbf{(a)}}}}
			\put(-99.,330.) {\mbox{\textcolor{white}{\textbf{(b)}}}}
			\put(5.,330.) {\mbox{\textcolor{white}{\textbf{(c)}}}}
			\put(-99.,225.) {\mbox{\textcolor{white}{\textbf{(d)}}}}
			\put(5.,225.) {\mbox{\textcolor{white}{\textbf{(e)}}}}
			\put(62.,110.) {\mbox{\textcolor{black}{\textbf{(f)}}}}
		\end{picture}
	\end{center}
		\vskip -1.0cm
		\caption{(a) Scheme representing the configuration consisting of a misfitting lamellar precipitate centered on a low angle grain boundary.
		(b-e) Color plots of the composition field $c$ at times of (b) $0.03\times10^6$,
        (c) $0.9\times10^6$, (d) $1.4\times10^6$ and (e) $6.37\times10^6$ illustrating the destabilization of a 
        low-angle GB ($\theta=7.2\degree$) due to the presence of a misfitting precipitate (red domain) 
        computed with the amplitude equation model for hexagonal symmetry and with a misfit eigenstrain $\eps_0=0.043$. 
        Dislocations are visible because the composition profile is altered in their vicinity. 
        The system size is $249.6 a\times 361.6 a$ ($a$ being the lattice spacing) 
        and periodic boundary conditions are used in both $x$ and $y$ directions. 
        Only part of the system is shown in the $y$ direction. (f) Time evolution of the free energy, 
        showing the relaxation of the system towards an equilibrium state. Black dots locate the snapshots
        of panels (b-e).
		}
		\label{fig:nonlinear}
\end{figure}

Experimental observations also testify of strong interactions between GBs and precipitates. For example,
in Ni-Al superalloys, $\gamma'$ precipitates in the vicinity of GBs have been shown to be 
responsible for GB serration, leading to improved mechanical properties at high temperature \cite{Koul1983,Mitchell2009a}.
In addition, in steel and Ti-based alloys submitted to thermo-mechanical treatments, acicular Widmanst\"{a}tten precipitates, are observed to grow from the GBs in a direction normal 
to the GB plane \cite{DaCostaTeixeira2006,Cheng2010}. While the stationary growth kinetics of these 
structures have been recently clarified \cite{Cottura2014}, the initial stage of growth that involves
the nucleation of precipitates along the GBs is not understood. In both examples, elastic interactions
between GBs and precipitates might play a central role in the development of these microstructures. 
However, these interactions remain largely unexplored due to the complexity of the problem at hand 
that involves elastic interactions, grain boundary migration, and solute diffusion. 

In a recent study \cite{Geslin2015}, we provided further insight into the complex interaction between crystal defects and precipitates
by investigating the situation in which a planar GB is centered inside a misfitting lamellar precipitate (see Fig.~\ref{fig:nonlinear}.a). This
choice of geometry is physically motivated by the fact that dislocations act as preferred sites of
nucleation \cite{Cahn1957,Hu2001,leonard2005}. Hence GBs naturally seed the formation of lamellar
precipitates of this approximate geometry \cite{Ramanarayan2003,Zhao1998}. Using a nonlinear elastic
model \cite{Geslin2014a,Geslin2014b} and amplitude equations (AE) that describe the interaction
between composition and stress \cite{Spatschek2010}, we showed that this configuration is morphologically unstable. This instability is illustrated in Fig.~\ref{fig:nonlinear}.b-e that shows a sequence of GB and precipitate configurations obtained by AE simulations \cite{Geslin2015}. Furthermore, we carried out a linear stability analysis to predict the onset and wavelength of this instability. The starting point of this analysis is a free-boundary problem governing the coupled evolution of DBs and GBs, which corresponds to the sharp-interface limit of the AE model (i.e. the limit where the DBs and GB can be treated as sharp boundaries).
The physical mechanism of this instability can be qualitatively understood by considering a small
initial sinusoidal perturbation of DBs of wavelength $\Lambda=2\pi/k$. In the case of isotropic
elasticity, the elastic energy is unchanged by this perturbation because the Bitter-Crum theorem
\cite{Bitter1931,Fratzl1999} implies that this energy is independent of the shape of the precipitate
and only depends on its volume, which remains constant. In the absence of a GB inside the
precipitate, the DB is stable because the perturbation of its interface increases the total DB surface, therefore increasing the total energy of the system. In contrast, with a GB present, the elastic energy can be decreased by the relaxation of the shear stress, induced by the DB perturbation, along the GB plane via shear-coupled GB motion 
\cite{Cahn2004, Cahn2006, Ivanov2008, Gorkaya2009, Olmsted09b, Molodov2011, Trautt2012a, Karma2012, Rajabzadeh2014, Rupert2009, Sharon2011, Winning2010, Lim2012}.
Namely, the GB can move normal to its plane under an applied shear stress. This behavior referred as \emph{coupling} is characterized by the relation
\begin{equation}
	v_\parallel=\beta  v_n \label{eq:coupling}
\end{equation}  
between the velocity $v_n$ of the GB normal to the GB plane and the rate $v_\parallel$ of parallel grain 
translation. In the case of pure coupling, the coefficient $\beta$ is a geometrical factor depending 
only on GB bicrystallography with the coupling factor $\beta$ obtained from the geometrical relation between dislocation glide motion and crystal lattice translation \cite{Sutton1995, Cahn2004,Cahn2006}. Computations and experiments have shown that a wide range of both low- and high-angle GBs display shear-coupled motion \cite{Cahn2006, 
Ivanov2008, Gorkaya2009, Olmsted09b, Molodov2011, Trautt2012a, Karma2012, Rajabzadeh2014}.
Furthermore, GB coupling has been found to influence significantly
the coarsening behavior of polycrystalline materials in more complex multi-grain geometries where GBs form a complex network
\cite{Rupert2009, Sharon2011, Wu2012, Trautt2012a, Adland2013}. 

Our recent study \cite{Geslin2015} has highlighted the fundamental role of shear-coupled GB motion in the interaction between GBs and precipitates. However, this study only considered a limited range of misfit strain and GB misorientation and was limited to isotropic elasticity and a lamellar precipitate geometry. Materials forming second phase precipitates are often elastically anisotropic. This anisotropy is known to influence the shape of misfitting precipitates by inducing DBs to align along preferred crystallographic directions to minimize the elastic energy \cite{Khachaturyan2013}. Moreover, it also influences the elastic interaction between GBs and precipitates. In particular, the Bitter-Crum theorem \cite{Bitter1931,Fratzl1999} invoked above to explain the destabilization of the GB for isotropic elasticity no longer applies within elastic anisotropy. In this case, deformation of the precipitate shape increases the elastic energy and can hinder or even potentially suppress the GB morphological instability. Furthermore, in several important experimental situations, precipitates interacting with GBs have a circular or cuboid geometry if nucleation occurs away from the GB at multiple sites, e.g. $\gamma'$ precipitates in Ni-Al superalloys. It is unclear how in those situations, closed-shape precipitates interact with GBs and what role shear-coupled motion plays in this interaction. 
   
In this paper, we extend the study of Ref.~[\onlinecite{Geslin2015}] to investigate the interaction between GBs and precipitates of different shapes with and without elastic anisotropy. We first focus on the lamellar precipitate geometry and extend the linear stability analysis of Ref.~[\onlinecite{Geslin2015}] to anisotropic elastic behavior. This extension is conceptually straightforward even though the anisotropy makes the analysis more lengthy. The analysis predicts that elastic anisotropy hinders the instability, because of the energetic cost of deforming the lamellar precipitate, but does not suppress it. We test this prediction using the same AE approach as in Refs. [\onlinecite{Spatschek2010}] and [\onlinecite{Geslin2015}], albeit with a free-energy form that favors an elastically anisotropic 3D body-centered-cubic (BCC) structure. The simulation results are in good quantitative agreement with the predictions of the linear stability analysis. For a free-energy form that favors an elastically isotropic two-dimensional (2D) hexagonal structure,  we investigate the nonlinear development of the instability over a wider range of misfit strain and misorientation than in Ref.~[\onlinecite{Geslin2015}]. Simulations yield the novel insight that, in a well-developed non-linear regime, this instability can lead to the break-up of low-angle GBs when the misfit strain exceeds a threshold that depends on misorientation. Next, we investigate in 2D the interactions between a circular precipitate and a grain boundary. We find that a similar elastic interaction mediated by shear-coupled GB motion leads to the attraction of the precipitate to the GB. The GB and precipitate shape are simultaneously deformed in this process that can also lead to GB break-up for large enough misfit strain. 

Some properties of the amplitude equations (AE) approach relevant to the present study are worth pointing out. AE models can be generally derived from the phase-field-crystal (PFC) model \cite{Elder2002,Elder2004,Berry2006} via a multiscale expansion \cite{Goldenfeld2005,Athreya2006,Spatschek2010}. This expansion is formally valid in the limit where the correlation length of liquid density fluctuations (which sets of the width of the spatially diffuse solid-liquid interface at the melting point of a pure material) is much larger than the lattice spacing. AE models can also be derived in the spirit of Ginzburg-Landau expansions of a free-energy functional in terms of complex density wave amplitudes from symmetry considerations (see Ref. [\onlinecite{Wu2015}] and earlier references therein for a comparison of both approaches for solid-liquid interface properties). The latter approach provides more flexibility to formulate AE models with a minimal set of model parameters that can be related to material parameters. For this reason, it was used in Ref. [\onlinecite{Spatschek2010}] to derive the set of AEs that describes the interaction of composition, stress, and crystal defects. The parameters of this AE model, used here and in our previous study \cite{Geslin2015}, are uniquely fixed by the DB energy $\gamma$, the misfit strain $\eps_0$, linear elastic properties, and the dislocation core size that is proportional to the correlation length. Like PFC, the AE method describes dislocation glide, therefore reproducing salient features of GB shear-coupled motion for a wide range of GB bi-crystallography \cite{Trautt2012a}, and also dislocation climb. Since PFC and AE models do not track explicitly the vacancy concentration, the climb kinetics is modeled only heuristically. However, the incorporation of dislocation climb is important in that it provides an additional mechanism to relax the total free-energy as is apparent in Fig.~\ref{fig:nonlinear} where the final equilibrium configuration was attained by a combination of both dislocation glide and climb. Finally, as shown in Ref.~[\onlinecite{Spatschek2010}], the AE model can only describe GBs over a limited range of misorientation due to the choice of a fixed reference set of crystal axes to represent the crystal density waves. However, this limitation is not too stringent as the method is able to describe both low-angle GBs with separate dislocations and higher angle ones with overlapping dislocation cores. 

This paper is organized as follows. We start by introducing in \cref{sec:AE} the AE model for both hexagonal (isotropic elasticity) and BCC ordering (anisotropic elasticity).
The following \cref{sec:LSA} is dedicated to generalizing the linear stability analysis 
to the case of anisotropic elasticity. In particular, we show that the introduction of 
elastic anisotropy inhibits the instability by reducing the growth rate and decreasing the range of unstable wavelengths. 
Next, in \cref{sec:GBB}, we investigate more closely the nonlinear regime of instability for isotropic elasticity, showing that
a sufficiently large misfit strain can lead to GB break-up. 
Finally, in \cref{sec:CPG}, we investigate the interactions between circular precipitates
and a GB, showing that similar elastic interaction leads to the attraction of the precipitate
to the GB and can also lead to GB break-up.

\section{Amplitude-Equation model}
\label{sec:AE}

\subsection{Free-energies}

In the present study, we used the AE approach developed by Spatschek and Karma \cite{Spatschek2010}, which provides a general methodology for modeling the interaction of composition and stress \cite{Spatschek2010}. 
In this AE framework, the atomic density field is expanded as a sum of crystal density waves  
\begin{equation}
n(\vec r,t)=n_0+\delta n_s\sum_{n=1}^{N/2} \left(A_n e^{i\vec k_n\cdot \vec r}+A_n^* e^{-i\vec k_n\cdot \vec r}\right),
\end{equation}
where $\pm \vec{k}_n$ ($1\le n\le N/2$) correspond to the $N$ principal reciprocal lattice vectors (RLVs) 
of equal magnitude $|\vec k_n|=q_0=2\pi/ a$, where $a$ is the lattice spacing,
$n_0$ is a reference average value of this field, and $\delta n_s$ is a scale factor that 
can be adjusted to match arbitrary values of solid density wave amplitudes. 
The amplitudes have a constant value $|A_n|=A_s$ in a perfect crystal, and decrease to low values in the atomically disordered core region
of dislocations, which is similar to the liquid phase where the amplitudes vanish.

The total free-energy of the system is given by the functional:
\begin{equation}
F = \int dV  f_c+ \int dV f_{el},
\end{equation}
where the chemical and elastic parts of the free-energy density are defined respectively by
\begin{equation}
f_c=\frac{K}{2}|\nabla c|^2 
  +f_{dw}(c) 
\end{equation}
and
\begin{align}
 f_{el} =    F_0 \bigg[ & \xi_d^2\sum_{n=1}^{N}|(\Box_n +i\eps_0 q_0 c)A_n|^2 \nonumber\\
                        &+ f_{b}(\{A_n\},\{A_n^*\}) \bigg],	\label{eq:aefed}
\end{align}
where the ``box operator'' is defined by $\Box_n=\hat{k}_n\cdot{\nabla}-\frac{i}{2q_0}{\nabla}^2$. This elastic free energy density 
represents the energy cost of an arbitrary perturbation 
of the atomic density field associated with linear elastic deformations 
and crystal defects (nonlinear deformations) such as dislocations or grain boundaries.

The free-energy cost of defects is captured by the box operator that is introduced to insure that the elastic part 
of the free-energy is rotationally invariant \cite{Spatschek2010}. In addition, the operator $i\eps_0 q_0 c$, 
accounts for the influence of the 
compositional field on the lattice spacing through the misfit strain $\eps_{0}$, where
we assume a linear relationship between strain and concentration (Vegard's law). 
The parameter $\xi_d$ is a dimensionless coefficient that is proportional to the width of the solid-liquid interface at the melting point and
also sets the scale of the dislocation core.

As in our previous study \cite{Geslin2015}, we use a version of the AE model where the bulk chemical free-energy density has a standard double-well Cahn-Hilliard-like contribution \cite{Cahn1959} $f_{dw}(c)$ that favors phase separation into two solid phases of distinct chemical compositions \cite{Geslin2015}.

The bulk chemical free-energy density has the double-well form
\begin{equation}
	f_{dw}(c) = g (c-c_0^-)^2(c-c_0^+)^2,
   \label{eq:fch}
\end{equation}
where the minima ($c_0^\pm$) represent the equilibrium concentrations in the composition domains in the absence of stress and the expressions
\begin{align}
  g=&\frac{12\gamma}{w_i(c_0^+-c_0^-)^4}\label{gdef}\\
  K=&\frac{3w_i\gamma}{2(c_0^+-c_0^-)^2},\label{Kdef}
\end{align}
relate the parameters $g$ and $K$ to the width $w_i$ and excess free-energy $\gamma$ of the spatially diffuse boundary between those domains.

The RLVs and the bulk part of the elastic energy density, $f_b(\{A_n\}, \{A_n^*\})$, can be chosen to stabilize different crystalline structures \cite{Wu2010,Greenwood2011,Toth2010}. 
In this study, we consider the 2D hexagonal lattice described by $N=6$ RLVs $\vec k_n=\pm q_0\hat k_n$ where
\begin{equation}
	\hat{k}_1=\left( -\frac{\sqrt{3}}{2}, -\frac{1}{2} \right),\enskip 
	\hat{k}_2=\left(0, 1\right),\enskip 
	\hat{k}_3=\left( \frac{\sqrt{3}}{2},-\frac{1}{2} \right) \nonumber
\end{equation}
and
\begin{widetext}
\begin{equation}
	f_b^{HEX}(\{A_n\},\{A_n^*\})=\frac{1}{6} \sum_{n=1}^{3} A_n A_n^* + \frac{1}{2} (A_1 A_2 A_3 + A_1^* A_2^* A_3^*) + \frac{1}{15} \left( \sum_{n=1}^{3} A_n A_n^* \right)^2 - \frac{1}{30} \sum_{n=1}^{3} |A_n|^4,
\end{equation}
which reproduce isotropic elasticity for small deformations \cite{Spatschek2010}. To investigate the effect of anisotropic elasticity, we also consider BCC ordering described by $N=12$ RLVs $\vec k_n=\pm q_0\hat k_n$ where:
\[ \hat{k}_1=\frac{1}{\sqrt{2}}(1,1,0),\quad 
\hat{k}_2=\frac{1}{\sqrt{2}}(1,0,1), \quad
\hat{k}_3=\frac{1}{\sqrt{2}}(0,1,1),\quad
\hat{k}_4=\frac{1}{\sqrt{2}}(1,-1,0),\quad 
\hat{k}_5=\frac{1}{\sqrt{2}}(1,0,-1), \quad
\hat{k}_6=\frac{1}{\sqrt{2}}(0,1,-1), \]
and the function \cite{Spatschek2010} 
\begin{align}
    f_b^{BCC}(\{A_n\},\{A_n^*\}) =& \frac{1}{12}\sum_{n=1}^6A_nA_n^* + 
    							    + \frac{1}{90}\Bigg\{\bigg(\sum_{n=1}^6A_nA_n^*\bigg)^2 
							  	    - \frac{1}{2}\sum_{n=1}^6|A_n|^4 
								    + 2A_{1}^*A_{2}  A_{4}^*A_{5}
								    + 2A_{1}  A_{2}  A_{4}^*A_{5}^* \nonumber \\
    							  & + 2A_{1}^*A_{3}  A_{4}  A_{6}
							  	    + 2A_{1}  A_{3}^*A_{4}^*A_{6}^*
								    + 2A_{2}  A_{3}^*A_{5}^*A_{6}  
								    + 2A_{2}^*A_{3}  A_{5}  A_{6}^*\Bigg\}\nonumber \\
    							  & -\frac{1}{8}\Big\{
							  		  A_{2}  A_{3}^*A_{4}^*
							        + A_{2}  A_{3}^*A_{4}
							        + A_{1}  A_{3}^*A_{5}^*
							        + A_{1}^*A_{3}  A_{5}	\nonumber \\
							      & + A_{1}  A_{2}^*A_{6}^*   
								    + A_{1}^*A_{2}  A_{6}
								    + A_{4}^*A_{5}  A_{6}^*  
								    + A_{4}  A_{5}^*A_{6}  \Big\}.
    \label{eq:fbBCC}
  \end{align}
\end{widetext}

\begin{figure}
  \centering
  \includegraphics[width=.45\linewidth]{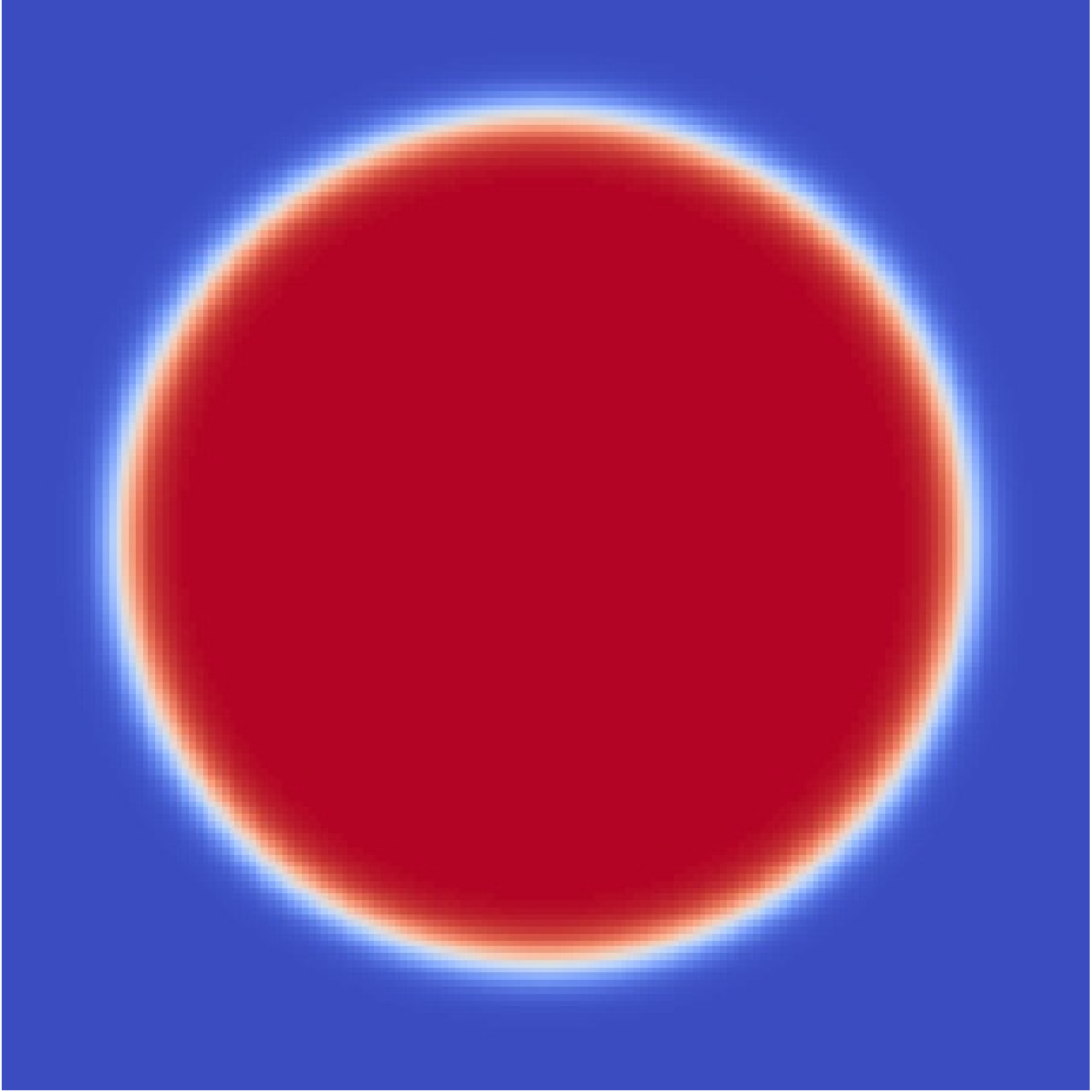}
  \hspace{0.1cm}
  \includegraphics[width=.45\linewidth]{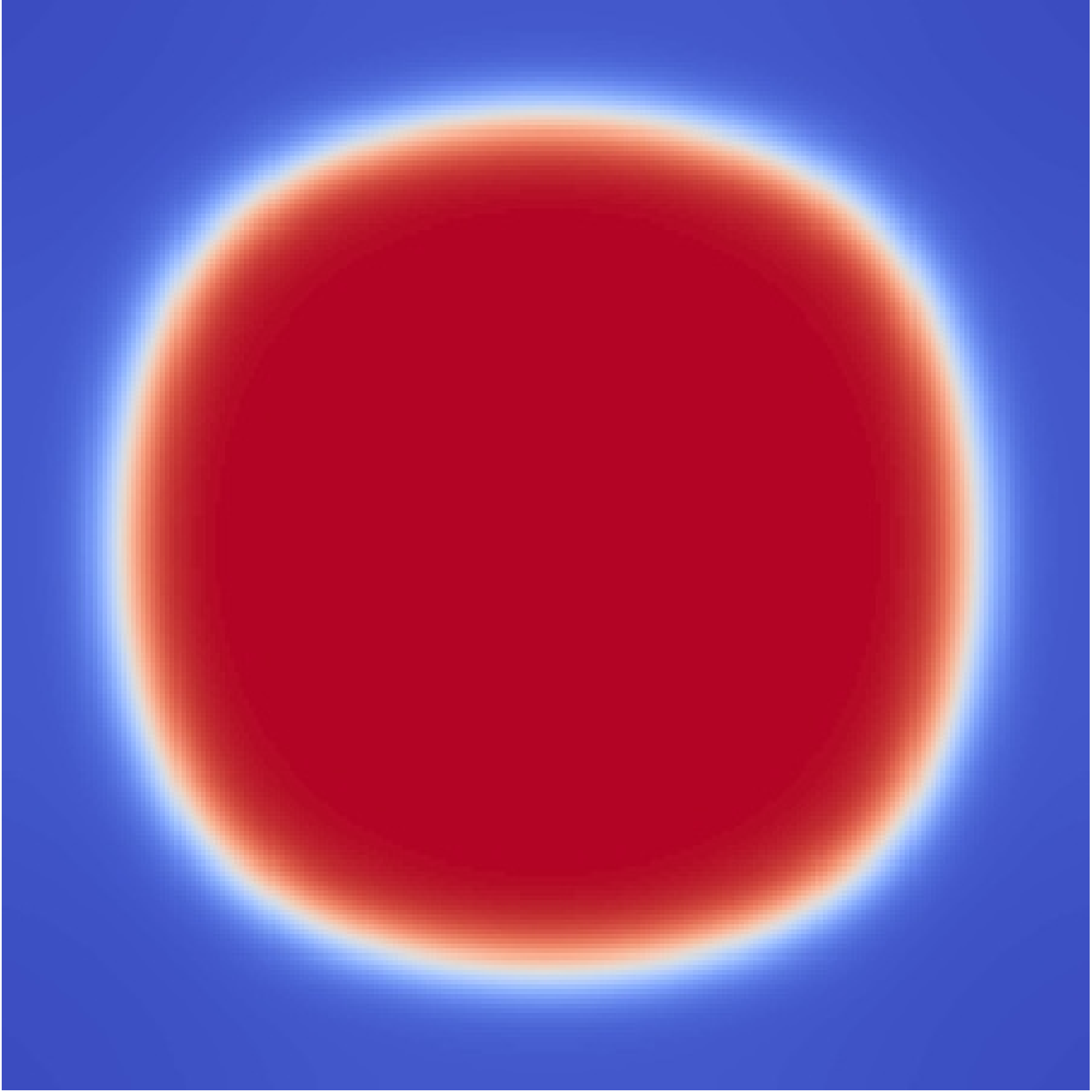}

   \begin{picture}(1,0)(0,0)
     \put(-97.,140.) {\mbox{\textcolor{white}{\textbf{(a)}}}}
     \put(8.,140.) {\mbox{\textcolor{white}{\textbf{(b)}}}}
   \end{picture}

  \caption{An initially circular precipitate of radius $R=40.5a$ and eigenstrain $\eps_0=0.043$
      (a) relaxes to a square shape with rounded corners (b) due to the anisotropic elasticity in the BCC AE model.}
  \label{fig:bcc-sketch}
\end{figure}
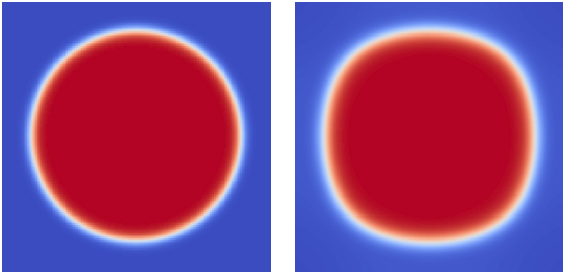

The effect of the anisotropic elasticity of the BCC structure on the precipitate morphology is
illustrated in \Cref{fig:bcc-sketch}. An initially circular precipitate
of radius $R=40.5a$ and eigenstrain $\eps_0=0.043$ (\Cref{fig:bcc-sketch}.a) evolves into 
a square with rounded corners (\Cref{fig:bcc-sketch}.b). 
Even though this morphology increases the surface energy, it is the equilibrium state of the system
because of the drop of elastic energy due to anisotropic elasticity effects \cite{Khachaturyan2013}.  

\subsection{Determination of model parameters}
\label{param}

The free-energies of the AE model depend on eight parameters $c_0^\pm$, $g$, $K$, $q_0$, $\xi_d$, $\eps_0$, and $F_0$. Their value can be generally determined uniquely in terms of material parameters as follows. The phase diagram determines $c_0^\pm$, the lattice spacing $a$ determines $q_0=2\pi/a$, the compositional domain width $w_i$ and the interface free-energy $\gamma$ determine $g$ and $K$ via Eqs. (\ref{gdef}) and (\ref{Kdef}). The misfit strain $\eps_0$ is a known material parameter and the microscopic   
length $\sim \xi_d$ can be in principle estimated by matching the dislocation core size to experimental measurement or the results of   atomistic simulations; for simplicity here we choose $\xi_dq_0=1$. In addition, 
$F_0$ can be related to elastic constants of the material using relations derived in Ref.~[\onlinecite{Spatschek2010}].
In the case of the elastically isotropic hexagonal model, the elastic constants are
$C_{11}=C_{22}=\lambda+2G=\frac{9}{4}F_0A_s^2\xi_d^2q_0^2$ and  
$C_{12}=C_{44}=\lambda=\frac{3}{4}F_0A_s^2\xi_d^2q_0^2$, yielding a Poisson ratio
$\nu=\lambda/[2(\lambda+G)]=0.25$ ($\lambda$ and $G$ denote the Lam\'{e} coefficients). For the elastically anisotropic BCC model, $C_{11}=C_{22}=C_{33}=2F_0A_s^2\xi_d^2q_0^2$ and 
$C_{12}=C_{23}=C_{44}=F_0A_s^2\xi_d^2q_0^2$. In these definition, the coefficient $A_s$ denotes the amplitude of solid density waves in a perfect crystal and can be expressed as
\begin{equation}
A_s=\frac{3+\sqrt{1 - Z \eps_0^2 c_0^2\xi_d^2q_0^2}}{4},  \label{As}
\end{equation}
where $Z=48$ ($Z=96$) for the hexagonal (BCC) lattice. Let us notice that for small values of $\eps_0$, $A_s$ depends weakly on composition $c_0$ via a shift of the lattice constant induced by the misfit strain.

In this study, simulations are performed for a generic set of material parameters  
similar to the one used to model phase separation in Li-ion
battery materials \cite{Tang2012}. In particular, we choose $c_0^-=0.05$, $c_0^+=0.95$, $\gamma=\unit{0.2}{J/m^2}$, and $w_i=\unit{2}{nm}$, yielding $g=\unit{1.8\times 10^9}{J/m^3}$ and $K=\unit{7.4 \times 10^{-10}}{J/m}$ using Eqs.~(\ref{gdef}) and (\ref{Kdef}). In addition, we take $a=\unit{0.5}{nm}$, $\xi_dq_0=1$, and $G=C_{44}=\unit{39}{GPa}$, yielding $F_0=4G/(3A_s^2\xi_d^2q_0^2)=\unit{5.2\times 10^{10}}{J/m^3}$ for the hexagonal lattice and $F_0=C_{44}/(A_s^2\xi_d^2q_0^2)=\unit{3.9\times 10^{10}}{J/m^3}$ for the BCC lattice where $A_s \approx 1$ is used in those relations to compute $F_0$. This is equivalent to neglecting the dependence of $A_s$ on $c_0$ in Eq.~(\ref{As}), which is negligible for small misfit strain. In the following, the simulations used to test the predictions of the linear stability analysis are carried out with $\eps_0=0.043$ for both the hexagonal and BCC models. Additional simulations are carried out for various values of $\eps_0$ to explore the influence of the misfit strain on the equilibrium state of the microstructure. Mathematically, the AE model is only valid for small misorientations between grains. However, it has been shown \cite{Spatschek2010} that the predictions of the GB energy from AE remain valid over roughly half the complete range allowed by the full crystal symmetry. Therefore, the limitations of the AE model on grain rotations does not influence significantly the results obtained for misorientations below $30\degree$ investigated in this article.

\subsection{Dynamical equations}

The concentration field $c$ is assumed to follow a conserved dynamics
\begin{equation}
   \frac{\partial c}{\partial t} = M\nabla^2 \frac{\delta F}{\delta c}. 
   \label{eq:aemecd} 
\end{equation}
where the mobility $M=Df_{dw}''(c_0^\pm)$ is chosen such that Fickian diffusion is recovered for vanishing stresses and composition close to the equilibrium values $c_0^{\pm}$. We note here that for finite misfit, the equilibrium concentrations in the low ($c^-$) and high  ($c^+$) concentration domains are slightly shifted from their equilibrium values $c_0^\pm$ as described further in the paper, but this shift has a negligible effect on the effective value of the mobility.

On the other hand, The amplitudes $A_n$ are evolved using a formulation of non-conserved dynamics introduced previously in the context of the PFC model \cite{Stefanovic2006,Stefanovic2009} to relax the elastic field rapidly over the entire system by the damped propagation of density waves: 
\begin{equation}
	\label{eq:aemead}
	\rho\frac{\partial^2 A_n}{\partial t^2} + \frac{\partial A_n}{\partial t} = -M_A \frac{\delta F}{\delta A_n^*},\quad (n=1\dots N)
\end{equation}
where the parameters $M_A$ and $\rho$, which control the wave damping rate and propagation velocity are chosen such that the amplitudes and hence the elastic field relax quickly on the diffusive time scale of the concentration field evolution.  

To see how to choose those parameters, and for the purpose of numerical implementation, it is useful to rewrite Eqs. (\ref{eq:aemead}) and (\ref{eq:aemecd}) in dimensionless form by measuring lengths in unit of $1/q_0$ and time 
in unit of $1/(M g q_0^{d})$ where the space dimension is $d=2$ ($d=3$) for the hexagonal (BCC) lattice.
After rescaling space and time, Eqs.~(\ref{eq:aemead}-\ref{eq:aemecd}) become for the hexagonal lattice:
\begin{widetext}
\begin{align}
	c_w^{-2}\frac{\partial^2 A_n}{\partial t^2} + \beta_w\frac{\partial A_n}{\partial t} 
		&= \alpha_d^2\left[\Box_n^2A_n + 2i\eps_0 c\Box_n A_n + i\eps_0 A_n\Box_n c +\eps_0\nabla A_n\cdot\nabla c -\eps_0^2 c^2A_n\right] \label{eq:aemea} \\ 
		&- \frac{1}{6}A_n- \frac{1}{2}\prod_{j\neq n}^{3}A_j^{*} -\frac{2}{15}A_n\sum_{j=1}^{3}A_jA_j^{*} +\frac{1}{15}A_n|A_n|^2, \nonumber  \\
  	\frac{\partial c}{\partial t} 
		&= \nabla^2 \Bigg\{-\alpha_c^2\nabla^2c + 2(c-c_0^-)(c-c_0^+)(2c-c_0^--c_0^+) \label{eq:aemec} \\
  		& + 2F_0'\alpha_d^2 \bigg( \eps_0 \sum_{j=1}^{3} \operatorname{Im} \left\{A_j^{\ast}\hat{k}_{j} \cdot \nabla A_j \right\}
      -\frac{1}{2}\eps_0 \sum_{j=1}^{3} \operatorname{Re} \left\{A_j^{\ast}\nabla^2 A_j \right\}
  + \eps_0^2 c\sum_{j=1}^{3}A_j A_j^{*}\bigg) \Bigg\}, \nonumber 
\end{align}
\end{widetext}
where we have defined the dimensionless parameters $\alpha_c=q_0\sqrt{K/g}=q_0w_i(c_0^+-c_0^-)/(2\sqrt{2})$, 
$\alpha_d=q_0\xi_d$, $F'_0=F_0/g$, $c_w^{-2}=\rho(M g q_0^d)^2/(M_AF_0)$, and $\beta_w=M g q_0^d/(M_AF_0)$. 
For the choice of parameters given in section~\ref{param}, $\alpha_d=1$, $\alpha_c=8$, and $F_0'=29.1$ ($F_0'=21.7$) 
for the hexagonal (BCC) lattice. Furthermore, in rescaled units, $c_w$ and $\beta_w$ determine 
the wave propagation velocity and damping rate, respectively. 
Since the diffusion constant is of order unity in those units, choosing $c_w=1$ and $\beta_w=0.05$ insures that 
the mechanical degrees of freedom relax faster than the concentration field. 

For the BCC lattice, the dimensionless dynamical equations analogous to Eqs.~(\ref{eq:aemea}) and (\ref{eq:aemec}) 
are quite lengthy and are detailed in \cref{sec:AE-BCC}.

\subsection{Numerical implementation}

We use a pseudo-spectral method to solve the dynamical Eqs.~(\ref{eq:aemea}) and (\ref{eq:aemec}). 
Following the same steps as in Ref.~[\onlinecite{Adland2013a}],
the evolution equations of the amplitude equations in Fourier space read

\begin{equation}
  c_w^{-2}\partial_{tt}\tilde{A}_n^k + \beta_w \partial_t\tilde{A}_n^k = L_A^k\tilde{A}_n^k+\tilde{f}_A^k(\{A_n\},c)
  \label{eq:aemeak}
\end{equation}
where the linear operator $L_A^k=\alpha_d^2(\tilde\Box_n^k)^2-1/6$ is the Fourier 
transform of $L_A=\alpha_d^2\Box_n^2-1/6$, and 
$\tilde{f}_A^k$ is the Fourier transform of the non-linear term of $f_A$ which 
contains all the remaining terms in the right hand side of \Cref{eq:aemea}. We use
the algorithm described in appendix A.2 of Ref.~[\onlinecite{Adland2013a}] to 
solve efficiently \Cref{eq:aemeak}.

The evolution equation for the concentration field becomes in Fourier space
\begin{equation}
  \partial_t\tilde{c}^k = L_c^k\tilde{c}^k+\tilde{f}_c^k(\{A_n\},c),
  \label{eq:aemeck}
\end{equation}
where the Fourier transform of the linear operator $L_c=-\alpha_c^2\nabla^4$ is
$L_c^k=-\alpha_c^2k^4$, and $\tilde{f}_c^k$ is the Fourier
transform of the non-linear term $f_c$ containing all the remaining terms in the right
hand side of \Cref{eq:aemec}. The algorithm described in appendix A1 of Ref.
[\onlinecite{Adland2013a}] is used to solve \Cref{eq:aemeck}.

Periodic boundary conditions are used in both directions. 
Thus, two grain boundaries are introduced in the simulation cell, located respectively 
at the center and the edge of the simulation box. The domain size in the 
direction $y$ (normal to the GBs) is chosen sufficiently large to consider that the 
influence of the second grain boundary is negligible. The simulations  
to obtain the growth rate of the instability (see \Cref{fig:omegak}) are performed 
using a fine grid spacing $\Delta x \approx 1$ and a time step $\Delta t=0.05$ to obtain fully converged numerical results for an accurate quantitative comparison with analytical predictions. The simulations presented in 
\Cref{fig:nonlinear,fig:nobreak,fig:hvbreak,fig:sumbreak,fig:circlelow,fig:circlehigh} 
are performed with coarser discretization parameters $\Delta x\approx$ 2 and $\Delta t=0.2$ to
follow the fully nonlinear development of the instability on much longer time scales 
while retaining a reasonable level of convergence.

\section{Linear stability analysis}
\label{sec:LSA}

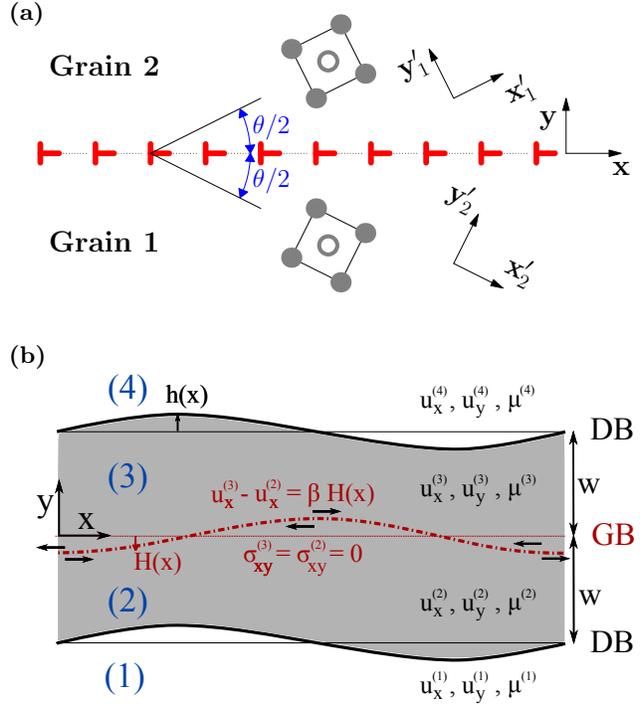
\begin{figure}[h]
	\begin{center}
		\includegraphics[width=1.0\linewidth]{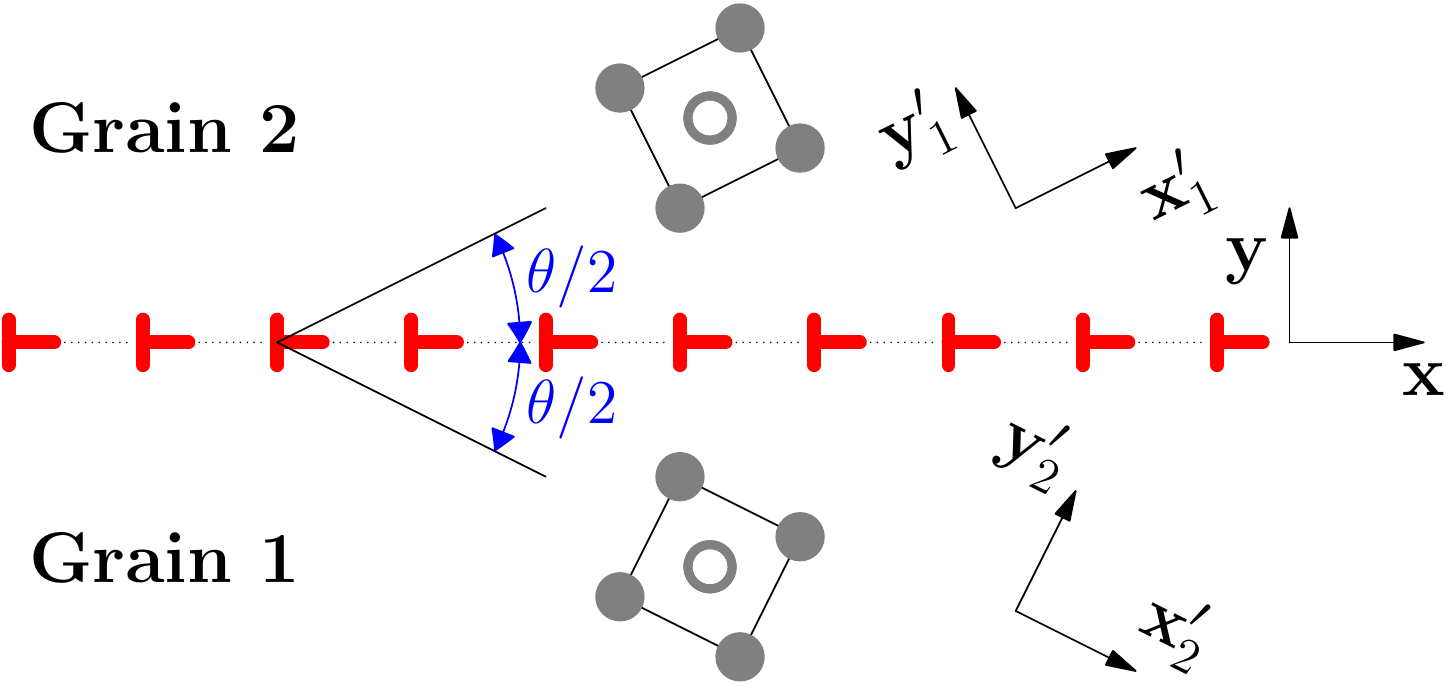}\\
		\vspace{1.0cm}
		\includegraphics[width=1.0\linewidth]{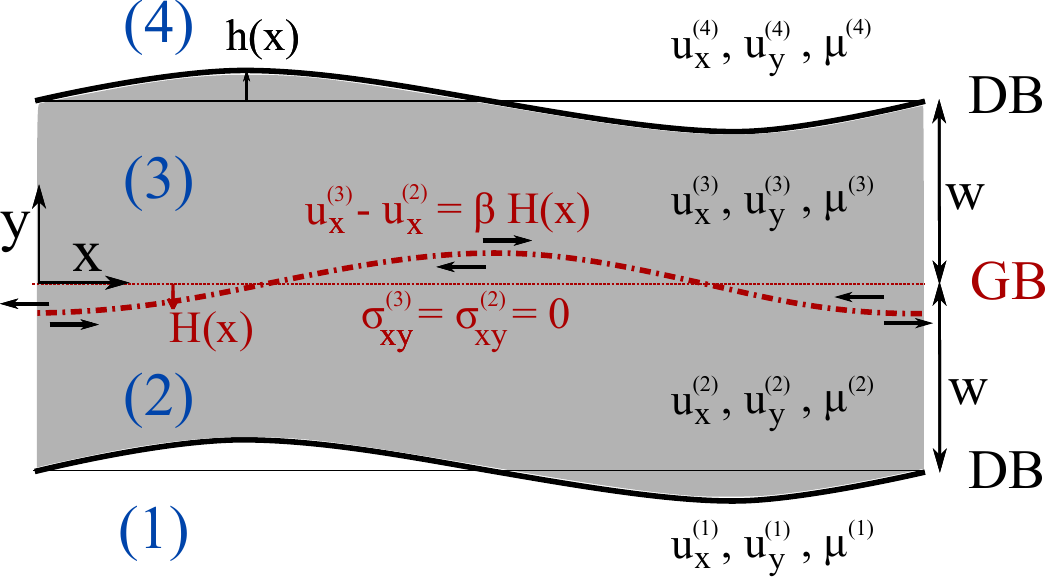}
   \end{center}
   \begin{picture}(1,0)(0,0)
     \put(-123.,280.) {\mbox{\textcolor{black}{\textbf{(a)}}}}
     \put(-123.,150.) {\mbox{\textcolor{black}{\textbf{(b)}}}}
   \end{picture}

	\caption{Schematic representation of lamellar precipitate geometry. (a) A low-angle symmetrical tilt GB of angle $\theta$ with the reference frame $(x,y)$
		and the frames $(x_1',y_1')$ and $(x_2',y_2')$ associated with both grains. 
		(b) a GB centered on a slightly perturbed lamellar precipitate. We 
        distinguish four regions (numbered 1 to 4). The displacement fields and chemical potentials 
        in the different regions are respectively denoted $u_x^{(n)}$, $u_y^{(n)}$ and $\mu^{(n)}$ ($n=1...4$).
        The horizontal arrows show the directions of grain translation 
        resulting from GB coupled motion via Eq.~(\ref{eq:coupling}), which relaxes the 
        shear stress induced by the perturbation of the surface of the precipitate.}
	\label{fig:schema}
\end{figure}

We now analyze the morphological stability of a lamellar precipitate centered on a GB such that the GB is sandwiched between two parallel DBs as depicted schematically in Fig.~\ref{fig:schema}. This geometry arises naturally when it is energetically favorable for the second phase precipitate to nucleate along the GB. We denote by $w$ the half-width of the lamellar precipitate. Its value depends on the composition and growth history of the second phase after nucleation. In the case of isotropic elasticity, the linear stability is detailed in the supplemental 
material of Ref.~[\onlinecite{Geslin2015}]. In this section, we will follow similar steps to extend this calculation to
the more complicated case of anisotropic elasticity for a cubic crystal symmetry. 

We take advantage of the fact that the GB shape and the elastic field (i.e. the displacive degrees of freedom) adapt instantaneously to a change of DB shape that occurs on a slow diffusive time scale. In other words, the elastic fields and GB evolutions are slaved to the DB evolution. This allows us to split the analysis into two main steps. 

In a first step, carried out in subsections A, B, and C, we compute the equilibrium GB shape and stress field resulting from imposing a wavy perturbation of the DBs. We first write down the anisotropic elastostatic equations in subsection A. We then solve those equations in the four separate domains depicted in Fig.~\ref{fig:schema}.b by imposing appropriate boundary conditions on the 
displacement and stress fields at the different interfaces (GB and DBs) separating those domains. We then compute the solutions for unperturbed planar interfaces in subsection B, and for an imposed DB perturbation of the form $h(x)\sim \sin (kx)$ in subsection C. In particular, it will be shown that under the geometrical coupling relation given by Eq.~(\ref{eq:coupling}), the GB relaxes to a stationary shape $H(x)\sim \cos kx$ with vanishing shear stress on the GB. 

The second main step of the analysis carried out in subsection D consists of computing the growth rate of the instability. For this we write down the equivalent free-boundary problem governing the evolution of the DBs in the limit where the DB width is much smaller than the wavelength of the perturbation, which allows to treat the DB as a sharp interface. This free-boundary problem consists of the diffusion equation for concentration coupled to two boundary conditions that must be self-consistently satisfied at the DBs: a Stefan-like mass conservation condition that relates the normal interface (DB) velocity to the normal gradient of chemical potential, and a local equilibrium condition that determines how the value of the chemical potential at the interface is shifted by stresses and interface curvature (as in the standard Gibbs-Thomson condition).

\subsection{Elastostatic equations}
\label{ES_eqs}
We consider a straight symmetric tilt grain boundary of angle $\theta$ obtained from a rotation of the two grains
of angles $\pm \theta/2$ around the $z$-axis as depicted in Fig.~\ref{fig:schema}.a. For low angle GBs, this tilt grain boundary can be seen as a wall of edge dislocations.

We consider that the reference frame $(x,y)$ coincides with the cubic axes of the crystal structure. In this frame, 
the elastic constants are $\bar{C}_{11}=\bar{C}_{22}=\bar{C}_{33}$, 
$\bar{C}_{12}=\bar{C}_{13}=\bar{C}_{23}$ and $\bar{C}_{44}=\bar{C}_{55}=\bar{C}_{66}$.
The system is invariant along the $z$ direction such that we can consider plain strain conditions. 
We define $S$ as the anisotropy coefficient by:
\begin{equation}
  S=1-\frac{\bar{C}_{11}-\bar{C}_{12}}{2\bar{C}_{44}}.
  \label{eq:eqS}
\end{equation}
We note that for isotropic elasticity, we have $S=0$. In the frames $(x'_1,y'_1)$ and $(x'_2,y'_2)$
associated to the grains rotated by an angle $\pm \theta/2$, the values of the elastic constants are given by:

\begin{align}
  C_{11}(\psi) =& \bar{C}_{11} + S \bar{C}_{44}\sin^2 (2\psi),\\
  C_{12}(\psi) =& \bar{C}_{12} - S \bar{C}_{44}\sin^2 (2\psi),\\
  C_{44}(\psi) =& \bar{C}_{44} - S \bar{C}_{44}\sin^2 (2\psi),\\
  C_{14}(\psi) =& \frac{S\bar{C}_{44}}{2} \sin (4 \psi)
\end{align}
where $\psi=\pm \theta/2$ is the rotation angle between the crystal axis of the 
grains 1 and 2 and the reference frame. 

To keep the elastostatic equations analytically solvable, we consider the limit of small $\theta$ where the elastic constants
are the same in both grains and in the reference frame and we note them $C_{11}$, $C_{12}$ and $C_{44}$.

We consider that the concentration is homogeneous in the different domains and is denoted by
$c^{(n)}$ (where the superscript $(n)$ denotes different domains, $n = 1 \dots 4$).
The stresses can therefore be simply expressed in terms of the displacements in the different domains:
 \begin{align}
   \sigma^{(n)}_{xx} =& C_{11}(\partial_x u^{(n)}_x-\eps_0c^{(n)})+C_{12}(\partial_y u^{(n)}_y-\eps_0c^{(n)}) \label{eq:stsxx} \\
   \sigma^{(n)}_{yy} =& C_{12}(\partial_x u^{(n)}_x-\eps_0c^{(n)})+C_{11}(\partial_y u^{(n)}_y-\eps_0c^{(n)}) \label{eq:stsyy} \\
   \sigma^{(n)}_{xy} =& C_{44}(\partial_x u^{(n)}_y+\partial_y u^{(n)}_x) \label{eq:stsxy}
 \end{align}
where the coordinate $x$ and $y$ refer to the reference basis $(x,y)$. Substituting these equations 
for stresses into the elastic equilibrium $\nabla \cdot \sigma=0$, we obtain the following 
elastostatic equations in terms of the displacements fields:
\begin{align}
 	C_{11} \frac{\partial^2u^{(n)}_x}{\partial x^2}         + C_{44}\frac{\partial^2u^{(n)}_x}{\partial y^2}
  + (C_{12} + C_{44})\frac{\partial^2u^{(n)}_y}{\partial x\partial y} =& 0 \label{eq:ese1} \\
  (C_{12}+C_{44})\frac{\partial^2u^{(n)}_x}{\partial x\partial y}+C_{44}\frac{\partial^2u^{(n)}_y}{\partial x^2}
  + C_{11}\frac{\partial^2u^{(n)}_y}{\partial y^2} =& 0 \label{eq:ese2}
\end{align}

\subsection{Non-perturbed problem}

We first consider the non-perturbed problem where the DBs and the GB
are perfectly straight ($h(x)=H(x)=0$) and solve for the equilibrium displacement field $\bar{u}$ and composition field.
In this case, the problem is invariant along the $x$ direction and $\bar{u}_x=0$ in all the domains. For the component $\bar{u}_y$, Eqs.~(\ref{eq:ese1}-\ref{eq:ese2}) admit the following solution in the different domains
\begin{align}
  \bar{u}_y^{(1)} =& \alpha^-(y+w)-\alpha^+w   \nonumber\\
  \bar{u}_y^{(2)} =& \bar{u}_y^{(3)}=\alpha^+y \label{eq:nonpd}\\
  \bar{u}_y^{(4)} =& \alpha^-(y-w)+\alpha^+w   \nonumber
\end{align}
where $\alpha^+=(C_{11}+C_{12})\eps_0c^+/C_{11}=\eps_0 c^+\left(1+\frac{1}{\zeta} \right)$ and 
$\alpha^-=(C_{11}+C_{12})\eps_0c^-/C_{11}=\eps_0 c^-\left(1+\frac{1}{\zeta} \right)$ with
$\zeta=C_{11}/C_{12}$.

Because of the stresses arising from the precipitate eigenstrain, the equilibrium concentrations $c^+$ and $c^-$ 
inside and outside the precipitate differ slightly from $c_0^+$ and $c_0^-$. 
This deviation can be computed by minimizing the total free energy
with respect to $\Delta c=c_0^+-c^+=c^--c_0^-$. The free energy is minimum for
\begin{equation}
  \Delta c=\frac{1}{2}\left[(c_0^+-c_0^-)-\sqrt{(c_0^+-c_0^-)^2-\frac{(C_{11}^2-C_{12}^2)\eps_0^2}{gC_{11}}}\right].
\end{equation}

\subsection{Perturbed problem}

\subsubsection{Solutions of the elastostatic equations}

We consider now that the DBs position are perturbed by a periodic function $h(x)=h_0\sin(kx)$ 
whose amplitude $h_0$ is assumed to be small compared to its wave length $2\pi/k$ and the width of the precipitate $w$.
The total displacement $u^{(n)}(x,y)$ can be decomposed into a non-perturbed part 
$\bar{u}^{(n)}(x,y)$ derived in \Cref{eq:nonpd} and a perturbed part $\tilde{u}^{(n)}(x,y)$ 
arising from the perturbation: 
\begin{align}
  \label{eq:totdisx}
  u_x^{(n)}(x,y) =& \bar{u}_x^{(n)}(x,y)+\tilde{u}_x^{(n)}(x,y)\\
  u_y^{(n)}(x,y) =& \bar{u}_y^{(n)}(x,y)+\tilde{u}_y^{(n)}(x,y)
  \label{eq:totdisy}
\end{align}
Following the supplemental material of Ref.~[\onlinecite{Karma2012}], we consider that 
the perturbed displacements take the following form in the different domains:

\begin{widetext}
\begin{itemize}
  \item In domain (1):
    \begin{align}
      \label{eq:ux1}\tilde{u}_x^{(1)} =& \operatorname{Re}\left\{e^{ikx}(A_1e^{-ikp_1(y+w)}+B_1e^{-ikp_2(y+w)})\right\}\\
      \label{eq:uy1}\tilde{u}_y^{(1)} =& \operatorname{Re}\left\{e^{ikx}(M_1A_1e^{-ikp_1(y+w)}+M_2B_1e^{-ikp_2(y+w)})\right\}
    \end{align}
  \item In domain (2):
    \begin{align}
      \label{eq:ux2}\tilde{u}_x^{(2)} =&
      \operatorname{Re}\left\{e^{ikx}(A_2e^{ikp_1(y+w)}+B_2e^{ikp_2(y+w)})\right\}+\operatorname{Re}\left\{e^{ikx}(C_2e^{-ikp_1y}+D_2e^{-ikp_2y})\right\}\\\nonumber
      \label{eq:uy2}\tilde{u}_y^{(2)} =&
      -\operatorname{Re}\left\{e^{ikx}(M_1A_2e^{ikp_1(y+w)}+M_2B_2e^{ikp_2(y+w)})\right\}+\operatorname{Re}\left\{[e^{ikx}(M_1C_2e^{-ikp_1y}+M_2D_2e^{-ikp_2y})\right\}\\
    \end{align}
  \item In domain (3):
    \begin{align}
      \label{eq:ux3}\tilde{u}_x^{(3)} =&
      \operatorname{Re}\left\{e^{ikx}(A_3e^{-ikp_1(y-w)}+B_3e^{-ikp_2(y-w)})\right\}+\operatorname{Re}\left\{e^{ikx}(C_3e^{ikp_1y}+D_3e^{ikp_2y})\right\}\\\nonumber
      \label{eq:uy3}\tilde{u}_y^{(3)} =&
      \operatorname{Re}\left\{e^{ikx}(M_1A_3e^{-ikp_1(y-w)}+M_2B_3e^{-ikp_2(y-w)})\right\}-\operatorname{Re}\left\{e^{ikx}(M_1C_3e^{ikp_1y}+M_2D_3e^{ikp_2y})\right\}\\
    \end{align}
  \item In domain (4):
    \begin{align}
      \label{eq:ux4}\tilde{u}_x^{(4)} =& \operatorname{Re}\left\{e^{ikx}(A_4e^{ikp_1(y-w)}+B_4e^{ikp_2(y-w)})\right\}\\
      \label{eq:uy4}\tilde{u}_y^{(4)} =& -\operatorname{Re}\left\{e^{ikx}(M_1A_4e^{ikp_1(y-w)}+M_2B_4e^{ikp_2(y-w)})\right\},
    \end{align}
\end{itemize}
\end{widetext}
where $A_i$, $B_i$, $C_i$, $D_i$, $M_i$ and $p_i$ are constants left to be determined. 
One can show (see \onlinecite{Karma2012}) that these equations are solutions of the elastostatic equations \Cref{eq:ese1,eq:ese2} only if the coefficients $M_1$ and $M_2$ are written as:
\begin{align}
  M_1 =& \frac{(C_{12}+C_{44})p_1}{C_{44}+C_{11}p_1^2}\\
  M_2 =& \frac{(C_{12}+C_{44})p_2}{C_{44}+C_{11}p_2^2},
\end{align}
where $p_1$ and $p_2$ are solutions of the equation
\begin{equation}
  p^4+\frac{C_{11}^2-C_{12}^2-2C_{12}C_{44}}{C_{11}C_{44}}p^2+1=0.
  \label{eq:eqp}
\end{equation}
This polynomial admits two complex roots of the form
\begin{align}
   p_1=e^{i\xi}, \qquad p_2=-e^{-i\xi}
\end{align}
with
\begin{equation}
   \xi=\frac{1}{2}\arccos\frac{C_{12}^2-C_{11}^2+2C_{12}C_{44}}{2C_{11}C_{44}}\label{eq:xi}.
\end{equation}
In the limiting case of isotropic elasticity, $p_1=p_2=i$ and \Crefrange{eq:ux1}{eq:uy4} reduce to the displacements function used in Refs.~[\onlinecite{Srolovitz1989}],[\onlinecite{Karma2012}] and [\onlinecite{Geslin2015}]. 

The other coefficients $A_i$, $B_i$, $C_i$, $D_i$ can be determined from the boundary 
conditions at the DBs and GB as detailed in the following sections.

\subsubsection{Boundary conditions on the DBs}
\label{sec:DB_BC}

We first examine the boundary conditions at the DB located at $y=w+h(x)$ separating domains (3) and (4).
Because the DB is coherent, the total displacement $u(x,y)$ must be continuous across the boundary.
We first consider the continuity of the $x$-component: $u_x^{(3)}(x,w+h(x)) = u_x^{(4)}(x,w+h(x))$.
Using Taylor expansions around $y=w$ and keeping only the lowest order terms in $h_0 k$ yields a continuity equation 
on the perturbed part of the displacements:
\begin{equation}
  \label{eq:bcd1} \tilde{u}_x^{(4)}(x,w) = \tilde{u}_x^{(3)}(x,w).
\end{equation}

For the same boundary, a similar procedure applied to the component $u_y$ yields
\begin{equation}
  \label{eq:bcd2} \tilde{u}_y^{(4)}(x,w) = \tilde{u}_y^{(3)}(x,w) + \alpha h(x),
\end{equation}
where $\alpha=\alpha^+-\alpha^-=\eps_0\left(1+\frac{1}{\zeta}\right)(c^+-c^-)$.

Also, the stress vector across the DBs of normal $\bold{n}$ defined as $\bold{T} = [T_x,T_y] = \boldsymbol{\sigma} \cdot \bold{n}$ must be continuous\footnote{The components of normal $\bold{n}$ to the perturbed interface are $n_x=-\frac{\partial_x h}{\sqrt{1+(\partial_x h)^2}}$ and $n_y=\frac{1}{\sqrt{1+(\partial_x h)^2}}$}.

Substituting \Cref{eq:totdisx,eq:totdisy} into \Cref{eq:stsxx,eq:stsyy,eq:stsxy} and using \Cref{eq:nonpd}, 
we get the stress expressed in terms of the perturbed displacements in different domains:
\begin{align}
  \sigma_{xx}^{(n)} =& C_{11}\partial_x\tilde{u}^{(n)}_x+C_{12}\partial_y\tilde{u}^{(n)}_y-\frac{C_{11}^2-C_{12}^2}{C_{11}}\eps_0c^{(n)} \nonumber\\
  \label{eq:stsp}\sigma_{yy}^{(n)} =& C_{11}\partial_y\tilde{u}^{(n)}_y+C_{12}\partial_x\tilde{u}^{(n)}_x\\\nonumber
  \sigma_{xy}^{(n)} =& C_{44}[\partial_y\tilde{u}^{(n)}_x+\partial_x\tilde{u}^{(n)}_y]
\end{align}
where $c^{(2)}=c^{(3)}=c^+$ and $c^{(1)}=c^{(4)}=c^-$. 

Substituting \Cref{eq:stsp} for domains (3) and (4) into the continuity of the stress vector 
and keeping only the lowest order terms after performing Taylor expansions in $h_0k\ll 1$, 
we obtain two additional boundary conditions on the perturbed displacement components $\tilde{u}_x$
and $\tilde{u}_y$. Finally, the boundary conditions on the DB located at $y=w+h(x)$ can be summarized as follow:
\begin{widetext}
\begin{align}
  \label{eq:bc34-1}
  \tilde{u}_x^{(4)}(x,w)-\tilde{u}_x^{(3)}(x,w) =& 0\\
  \label{eq:bc34-2}
  \tilde{u}_y^{(4)}(x,w)-\tilde{u}_y^{(3)}(x,w) =& \alpha h(x)\\
  \label{eq:bc34-3}
  \left[\partial_y\tilde{u}_x^{(4)}+\partial_x\tilde{u}_y^{(4)}\right]_{y=w}-\left[\partial_y\tilde{u}_x^{(3)}+\partial_x\tilde{u}_y^{(3)}\right]_{y=w}=& \frac{C_{11}-C_{12}}{C_{44}}\alpha h^\prime(x)\\
  \label{eq:bc34-4}
  \left[\partial_x\tilde{u}_x^{(4)}+\zeta\partial_y\tilde{u}_y^{(4)}\right]_{y=w}-\left[\partial_x\tilde{u}_x^{(3)}+\zeta\partial_y\tilde{u}_y^{(3)}\right]_{y=w}=& 0
\end{align}

We derive similar boundary conditions for the DB between domains (1) and (2) located at $y=-w+h(x)$:
\begin{align}
  \label{eq:bc12-1}
  \hspace{-0.3cm}\tilde{u}_x^{(1)}(x,-w)-\tilde{u}_x^{(2)}(x,-w) =& 0\\
  \label{eq:bc12-2}
  \hspace{-0.3cm}\tilde{u}_y^{(1)}(x,-w)-\tilde{u}_y^{(2)}(x,-w) =& \alpha h(x)\\
  \label{eq:bc12-3}
  \hspace{-0.3cm}\left[\partial_y\tilde{u}_x^{(1)}+\partial_x\tilde{u}_y^{(1)}\right]_{y=-w}-\left[\partial_y\tilde{u}_x^{(2)}+\partial_x\tilde{u}_y^{(2)}\right]_{y=-w}=& \frac{C_{11}-C_{12}}{C_{44}}\alpha h^\prime(x)\\
  \label{eq:bc12-4}
  \hspace{-0.3cm}\left[\partial_x\tilde{u}_x^{(1)}+\zeta\partial_y\tilde{u}_y^{(1)}\right]_{y=-w}-\left[\partial_x\tilde{u}_x^{(2)}+\zeta\partial_y\tilde{u}_y^{(2)}\right]_{y=-w}=& 0
\end{align}
\end{widetext}

\subsubsection{Boundary conditions on the GB}
\label{sec:GB_BC}

The perturbation of DBs produces shear stresses on the GB which is considered to relax entirely 
the shear stresses by coupling. We note $H(x)$ 
the perturbation of the GB position whose amplitude is assumed to be of the order of $h_0$. 
Boundary conditions accounting for the GB coupling behavior can then be written assuming that the 
GB behaves like a sharp interface located at $H(x)$.

As explained in \Cref{sec:intro}, the coupling behavior of the GB can be translated into 
the well-known geometrical relation of \Cref{eq:coupling} 
between the normal GB velocity $v_n$ and the velocity of parallel grain translation $v_{\parallel}$ 
\cite{Cahn2004,Cahn2006}. A simple time integration of this equation leads to
a relationship between the GB perturbation, $H(x)$, and the jump of the total displacement  
$u_x$ across the GB. After performing Taylor expansions around $y=H(x)$ and keeping the dominant term, we obtain
\begin{equation}
  \label{eq:bcgb-2}
	\tilde{u}_x^{(3)}(x,0) - \tilde{u}_x^{(2)}(x,0) = \beta H(x).
\end{equation}

Substituting the displacements $\tilde{u}_x^{(2)}$ and $\tilde{u}_x^{(3)}$ 
described in \Cref{eq:ux2,eq:ux3} into this equation, we deduce that the 
function $H(x)$ takes the form $H(x)=H_0 \cos(kx)$, where $H_0$ is a constant unknown at the 
moment. Therefore, the GB perturbation $H(x)$ is out of phase compared to the DB perturbation $h(x)$,
as depicted in \Cref{fig:nonlinear}.

The coupling behavior of the GB does not influence the component $u_y$ of the displacement field,
which remains continuous across the boundary. The procedure explained in \cref{sec:DB_BC} can 
be applied straightforwardly to the component $u_y$, yielding:
\begin{equation}
  \label{eq:bcgb-1}
	\tilde{u}_y^{(3)}(x,0) - \tilde{u}_y^{(2)}(x,0) = 0.
\end{equation}

Just like in the case of DBs, the components of the stress vector $\bold{T}$ is continuous across the GB. 
The continuity of the component $T_x$ leads to the following equation:

\begin{equation}
  \label{eq:bcgb-3}
  \left[\partial_x\tilde{u}_x^{(3)}+\zeta\partial_y\tilde{u}_y^{(3)}\right]_{y=0}-\left[\partial_x\tilde{u}_x^{(2)}+\zeta\partial_y\tilde{u}_y^{(2)}\right]_{y=0} =  0
\end{equation}

In addition to the continuity of the component $T_y$, we assume that the GB relaxes completely 
the shear stresses through coupling. In other words, the GB adapts its shape to the shear stress environment produced 
by the perturbation on the DBs such that the shear stresses vanish at $y=H(x)$. This leads to the following
relation on the perturbed displacements:

\begin{align}
  \label{eq:bcgb-4}
	\left[ \partial_y \tilde{u}_x^{(2)} + \partial_x \tilde{u}_y^{(2)} \right]_{y=0} =& 0\\
  \label{eq:bcgb-5}
   \left[ \partial_y \tilde{u}_x^{(3)} + \partial_x \tilde{u}_y^{(3)} \right]_{y=0} =& 0
\end{align}

Finally, we obtained five boundary conditions (\Crefrange{eq:bcgb-1}{eq:bcgb-5}) 
that have to be fulfilled on the GB by the displacement field.

\subsubsection{Solution of the elastostatic equations}

Substituting the expression of the displacements \Crefrange{eq:ux1}{eq:uy4} into
the boundary conditions (\ref{eq:bc34-1})-(\ref{eq:bcgb-5}) yields
13 linear equations. The 13 unknowns of the problem ($A_i$, $B_i$, 
$C_i$, $D_i$, $H_0$) are then determined uniquely by solving the linear system of equations.
In particular, we obtain an expression of the GB amplitude $H_0$:
\begin{widetext}
\begin{align}
  H_0 =& \left(2 i h_0 (M_1 p_1 - M_2 p_2) \alpha (1 + \zeta) \left[(e^{i k p_1 w} M_2 (-M_1 + p_1) - e^{i k p_2 w} M_1 (-M_2 + p_2)) (-1 + \zeta) C_{12}\right.\right.\nonumber\\
  \label{eq:eqH0}
  &+ \left.\left.(e^{i k p_1 w} - e^{i k p_2 w}) (-M_1 + p_1) (-M_2 + p_2) C_{44}\right]\right)\\
  & /\left((M_2 p_1 - M_1 p_2) \left[-p_1 + p_2 + M_1 (1 + M_2 (p_1 - p_2) \zeta - p_1 p_2 \zeta) + M_2 (-1 + p_1 p_2 \zeta)\right] C_{44} \beta\right)\nonumber
\end{align}
\end{widetext}
The full expression of the other unknowns $A_i$, $B_i$ $C_i$ and $D_i$ are quite lengthy and are detailed in \cref{sec:RLE}.
To highlight the influence of the anisotropic elasticity on the instability, 
we express the elastic constants $C_{11}$, $C_{12}$ and $C_{44}$ as a function of an equivalent shear modulus $G=C_{44}$, Poisson's 
ratio $\nu=C_{12}/(C_{11}+C_{12})$ and the anisotropic factor $S=1-\frac{C_{11}-C_{12}}{2 C_{44}}$:
\begin{align}
  C_{44} = & G \label{eq:C12G} \\ 
  C_{12} = & \frac{2(1-S)G\nu}{1-2\nu}\\
  C_{11} = & \frac{2(1-S)G(1-\nu)}{1-2\nu}
\label{eq:C11G}
\end{align}
Finally, we expand \Cref{eq:xi} in the limit of small $S$: $\xi=\left(\pi-\sqrt{2S/(1-\nu)}\right)/2$.
After substituting these expressions into \Cref{eq:eqH0} and performing a Taylor expansion 
for small $S$, we obtain
\begin{widetext}
\begin{equation}
  H_0 = -\frac{4\eps_0(c^+-c^-)h_0e^{-kw}}{\beta}+\frac{\eps_0(c^+-c^-)h_0e^{-kw}(1-3kw+k^2w^2)S}{\beta(1-\nu)}+O(S^{3/2})
  \label{eq:eqH0te}
\end{equation}
\end{widetext}

\begin{figure}[htb]
  \begin{center}
    \includegraphics[width=1.0\linewidth]{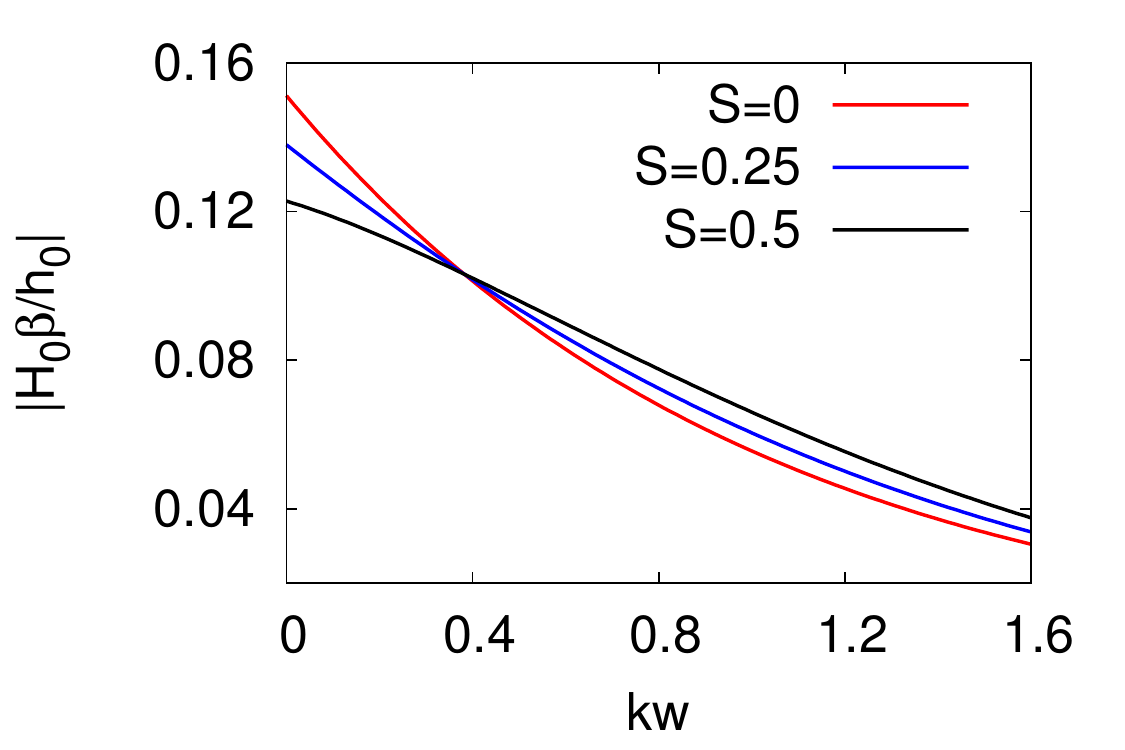}
  \end{center}
  \caption{Ratio $|H_0\beta/h_0|$ versus the dimensionless wave-vector $kw$ obtained from \Cref{eq:eqH0te} 
           for different values of the anisotropic coefficient $S$, showing the influence of the elastic anisotropy on 
           the relative amplitudes of the GB and DB perturbations.}
  \label{fig:Hbeta-kw}
\end{figure}

In the isotropic limit ($S=0$), we recover exactly our previously derived result  
(Eq.~(13) in Ref.~[\onlinecite{Geslin2015}]).
The role of anisotropic elasticity enters as a correction 
proportional to $\eps_0 h_0/\beta$ in the limit of small $S$.
We note that the term $(1-3kw+k^2w^2)$ entering this corrective term 
can be positive or negative depending on value the $kw$. This is illustrated in \Cref{fig:Hbeta-kw} 
where $|H_0\beta/h_0|$ is plotted as a function of $kw$ for different values of $S$.


\subsection{Linear stability analysis}
\label{sec:SIA}

In the previous section, we solved the elastostatic equations when the DB position is modified by a periodic
perturbation. 
In this section, we formulate a Stefan-like free boundary problem that governs the diffusion-controlled motion of the DBs  in the sharp-interface limit where the DB width is much smaller than the perturbation wavelength. Furthermore, we perform a linear stability analysis of the evolution equations for the DBs to obtain the growth rate of the morphological instability driven by the elastic interaction between the DBs and GB. This analysis makes use of the results of the previous section for the stresses on the perturbed DBs. 

Far from the DBs, the concentration is close to its equilibrium value $\bar{c}^{\pm}$
such that \Cref{eq:aemecd} reduces to the diffusion equation. Moreover, in this limit, the
chemical potential defined as $\mu=\delta F/\delta c$ is proportional to the solute concentration. 
Therefore, the same diffusion equation holds for $\mu$:

\begin{equation}
  \frac{\partial\mu^{(n)}}{\partial t} = D\nabla^2\mu^{(n)},\quad n=1\dots4,
  \label{eq:mume}
\end{equation}
where $D=Mf_{dw}^{\prime\prime}(c^{\pm})$ is the diffusion coefficient.

Similarly to what has been done for the displacement field in the previous section, the chemical potential 
is decomposed as a sum $\mu=\bar{\mu}+\tilde{\mu}$ where $\bar{\mu}$ is the equilibrium chemical 
potential for the non-perturbed configuration and $\tilde{\mu}$ is a small variation due to DB perturbations.

We first consider the non-perturbed DB located between domains (1) and (2), at $y=-w$. 
The composition field $c(x,y)$ does not depend on $x$ and adopts an equilibrium profile along $y$ denoted by $\bar{c}(y)$, reaching the values $\bar{c}^-$ and $\bar{c}^+$ in domains (1) and (2) respectively. 
At equilibrium, the chemical potential is constant across the DB interface and is given by 
\begin{equation}
  \bar{\mu}=f_{dw}^\prime(\bar{c}(y))-\eps_0(\bar{\sigma}_{xx}(y)+\bar{\sigma}_{yy}(y))-K\partial_y^2\bar{c}(y)
  \label{eq:unpmu}
\end{equation}
where $\bar{\sigma}_{xx}(y)$ and $\bar{\sigma}_{yy}(y)$ are the stress profiles along the $y$ direction.
The second term emerges from the derivative of the elastic energy density \Cref{eq:aefed} lineralized for small deformations $f_{el}^{lin} = \frac{1}{2}\sigma_{ij}(\eps_{ij} - \delta_{ij}\eps_0 c)$.

We then consider a perturbation $h(x)$ of the DBs and elastic displacements. 
Using the linearity of elasticity, the total stress fields can be written as 
$\sigma_{ij} = \bar{\sigma}_{ij}+\tilde{\sigma}_{ij}$, where $\sigma_{ij}$ 
is the stress field induced by the perturbation.
Considering that the perturbation $h(x)$ is a slowly varying function of $x$, 
we can assume that, in the vicinity of the DB, the concentration field takes the form
$c(x,y) \approx \bar{c}(y-h(x))$. Substituting these expressions for the stress and composition fields
 into the definition of the chemical potential and keeping only the dominant terms, we obtain
\begin{equation}
  \mu=\bar{\mu}-\eps_0(\tilde{\sigma}_{xx}+\tilde{\sigma}_{yy})+K\partial_y\bar{c}\kappa
  \label{eq:totmu}
\end{equation}
where $\kappa$ is the domain interface curvature. The chemical potential $\mu$ and the stress fields
$\tilde{\sigma}_{xx}$ and $\tilde{\sigma}_{yy}$ vary on a  much larger length-scale than the interface width and can be 
assumed to be constant across the DB. We then multiply \Cref{eq:totmu} by $\partial_y\bar{c}$
and integrate over the interval $[-w-\delta, -w+\delta]$ where $\delta$ is an arbitrary intermediate length, 
larger than the interface width but much smaller than the characteristic scale on which the stresses 
and chemical potential vary. We finally obtain the chemical potential acting on the DBs:
\begin{equation}
\mu_{\text{\tiny DB}} = \bar{\mu} - \eps_0[\tilde{\sigma}_{xx}+\tilde{\sigma}_{yy}]_{\text{\tiny DB}} + \frac{\gamma\kappa}{\bar{c}^+-\bar{c}^-}
  \label{eq:DBmu}
\end{equation}
where $\gamma=K\int_{-w-\delta}^{-w+\delta}(\partial_y\bar{c})^2dy$ is the 
interfacial energy and $[\tilde{\sigma}_{xx}+\tilde{\sigma}_{yy}]_{\text{\tiny DB}}$ is the sum 
of the stresses at the DB.

In the case of a periodic perturbation $h(x) = h_0 \sin(kx)$, the stresses at the DB 
are obtained by substituting \Crefrange{eq:ux1}{eq:uy4} into 
\Cref{eq:stsp} and using the expression of $A_i$, $B_i$, $C_i$ and $D_i$ listed
in \Cref{sec:RLE}. Similarly to the expression of the GB perturbation, 
the stresses can be expressed as a Taylor expansion, treating the anisotropic coefficient 
$S$ as a small parameter:
\begin{widetext}
\begin{align}
   \label{eq:stsxxyy1} [\tilde{\sigma}_{xx}+\tilde{\sigma}_{yy}]_{DB} =& \operatorname{Re}\{-k(C_{11}+C_{12})[(A_1+B_1)-(M_1p_1A_1+M_2p_2B_1)]_{y=-w}\}\\\nonumber
   =& \operatorname{Re}\{-k(C_{11}+C_{12})[(A_4+B_4)-(M_1p_1A_4+M_2p_2B_4)]_{y=w}\}\\\nonumber
  \label{eq:stsxxyy2}=& \left\{-\frac{4\eps_0(\bar{c}^+-\bar{c}^-)k G e^{-2kw}}{1-\nu}+\frac{\eps_0(\bar{c}^+-\bar{c}^-) k G S[1+(4(1-\nu)-4kw+2k^2w^2)e^{-2kw}]}{(1-\nu)^2}+O(S^{3/2})\right\}h\\
\end{align}
\end{widetext}

In the limit of isotropic elasticity ($S=0$), we recover the stresses obtained in Ref.~[\onlinecite{Geslin2015}].

We now perform a linear stability analysis by considering that the amplitude of the perturbation $h_0(t)$ evolves exponentially in time: $h_0(t) = h_i \exp(\omega_k t)$, with $\omega_k$ the growth rate of the instability and $h_i$ the initial amplitude of the perturbation.

For simplicity, we define the function\footnote{We note that $\Gamma(k)$ does not depend on time because the stresses $\sigma_{xx}$ and $\sigma_{yy}$ depend linearly on $h(x,t)$}
\begin{equation}
  \Gamma(k)=-\frac{\eps_0[\tilde{\sigma}_{xx}+\tilde{\sigma}_{yy}]_{\text{\tiny DB}}}{h(x,t)} + \frac{\gamma k^2}{\bar{c}^+-\bar{c}^-},
\end{equation}
such that the chemical potential on the DB between domains (1) and (2) is simply given by
\begin{equation}
  \mu_{\text{\tiny DB}}^{(1,2)}(t) = \bar{\mu} + \Gamma(k)h(x,t)
  \label{eq:DB12mu}
\end{equation}

Similarly, the chemical potential on the DB between domains (3) and (4) is:
\begin{equation}
  \mu_{\text{\tiny DB}}^{(3,4)}(t) = \bar{\mu} - \Gamma(k)h(x,t).
  \label{eq:DB34mu}
\end{equation}

\Cref{eq:DB12mu,eq:DB34mu} serve as boundary conditions for the solution of \Cref{eq:mume}. 
In addition, two additional boundary conditions are obtained by considering that the chemical potential reaches $\bar{\mu}$ far from the DBs (i.e. for $y \to \pm \infty$). The solution of  
\Cref{eq:mume} satisfying these boundary conditions is of the form
\begin{align}
  \label{eq:mu1}
  \mu^{(1)} =& \bar{\mu} + \Gamma(k)e^{q(y+w)}h(x,t) \\
  \label{eq:mu23}
  \mu^{(n)} =& \bar{\mu} - \Gamma(k)\frac{\sinh(qy)}{\sinh(qw)}h(x,t), \quad n=2,3\\
  \label{eq:mu4}
  \mu^{(4)} =& \bar{\mu} - \Gamma(k)e^{-q(y-w)}h(x,t)
\end{align}
where $q=\sqrt{k^2+\omega_k/D}$.

Next, the normal velocity of the DBs is given by the mass conservation (Stefan-like) condition
\begin{equation}
  v_{\text{\tiny DB}} =  -\frac{M}{\bar{c}^+-\bar{c}^-}\llbracket\partial_y\mu\rrbracket_{\text{\tiny DB}},
  \label{eq:DBv}
\end{equation}
where the double brackets denotes the jump of the normal gradient of chemical potential $\partial_y\mu$ across the DB, neglecting higher order nonlinear terms originating from the change of normal direction induced by the perturbation of the DB (i.e. $\sqrt{1+(\partial_xh)^2}\approx 1$). Using the fact that $v_{\text{\tiny DB}}=\partial_t h=\omega_kh(x,t)$ and
\Crefrange{eq:mu1}{eq:mu4} to evaluate the right-hand-side of \Cref{eq:DBv}, we obtain an implicit transcendental equation for $\omega_k$:
\begin{equation}
  \omega_k=\frac{M\Gamma(k)}{c^+-c^-}\sqrt{\frac{\omega_k}{D}+k^2}\left(1+\coth\left(w\sqrt{\frac{\omega_k}{D}+k^2}\right)\right).
  \label{eq:omegak_full}
\end{equation}

We can consider the quasistatic limit where the concentration field that evolves on a time-scale $1/Dk^2$ 
reaches quickly an equilibrium profile compared to the time-scale of the evolution of the DB $1/\omega_k$. Our simulations are performed within this quasistatic limit. With $\omega_k \ll Dk^2$, $q \simeq k$ and \Cref{eq:omegak_full} reduces to a straightforward expression for $\omega_k$:
\begin{equation}
  \omega_k=\frac{M\Gamma(k)k}{c^+-c^-}\left(1+\coth(kw)\right)
  \label{eq:omegak_approx}
\end{equation}

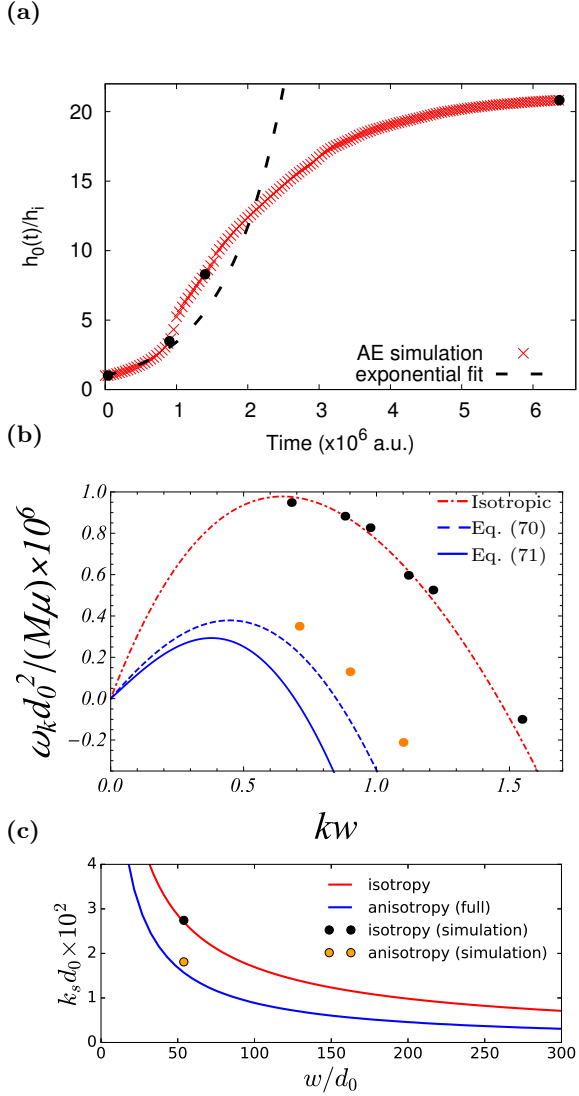
\begin{figure}
  \begin{center}
    \includegraphics[width=1.0\linewidth]{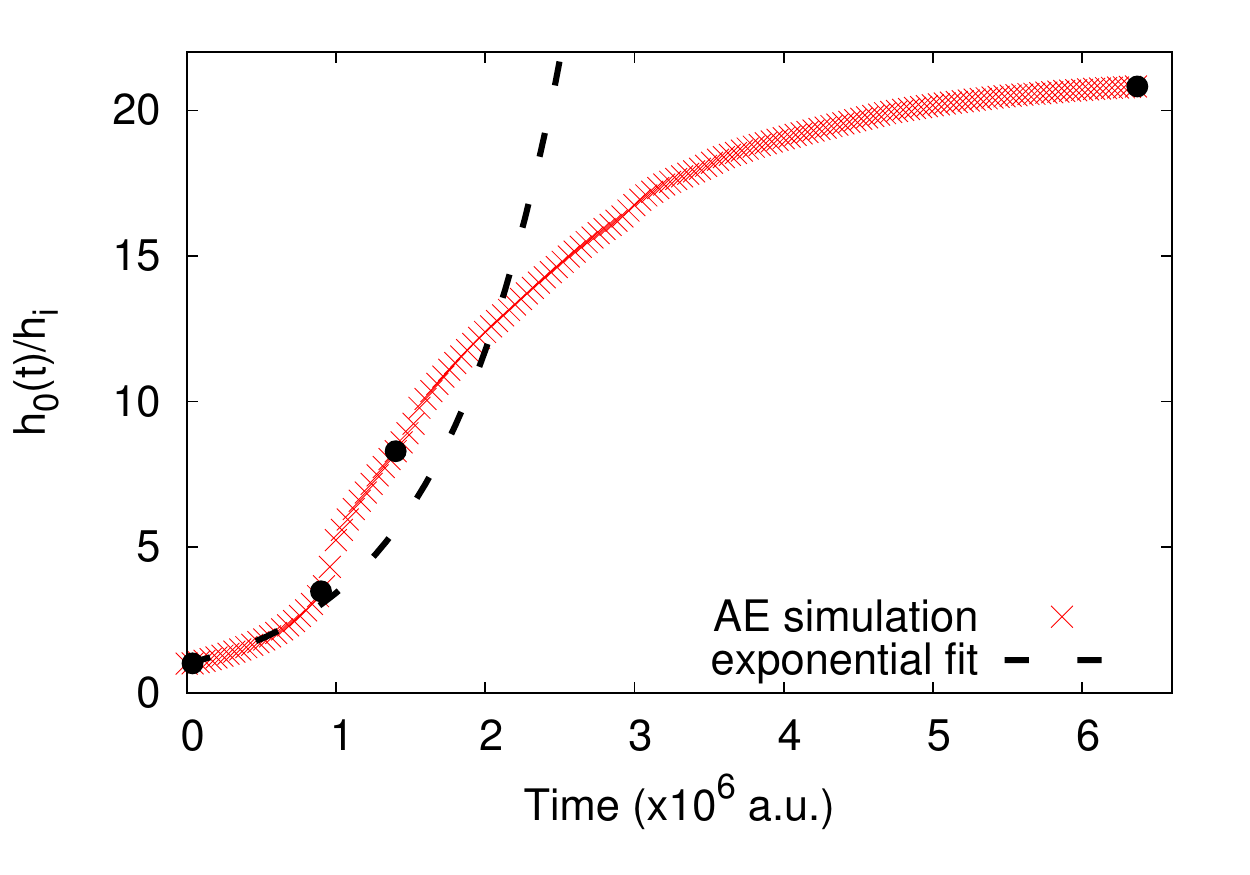}\\
    \vspace{-0.5cm}
    \includegraphics[width=1.0\linewidth]{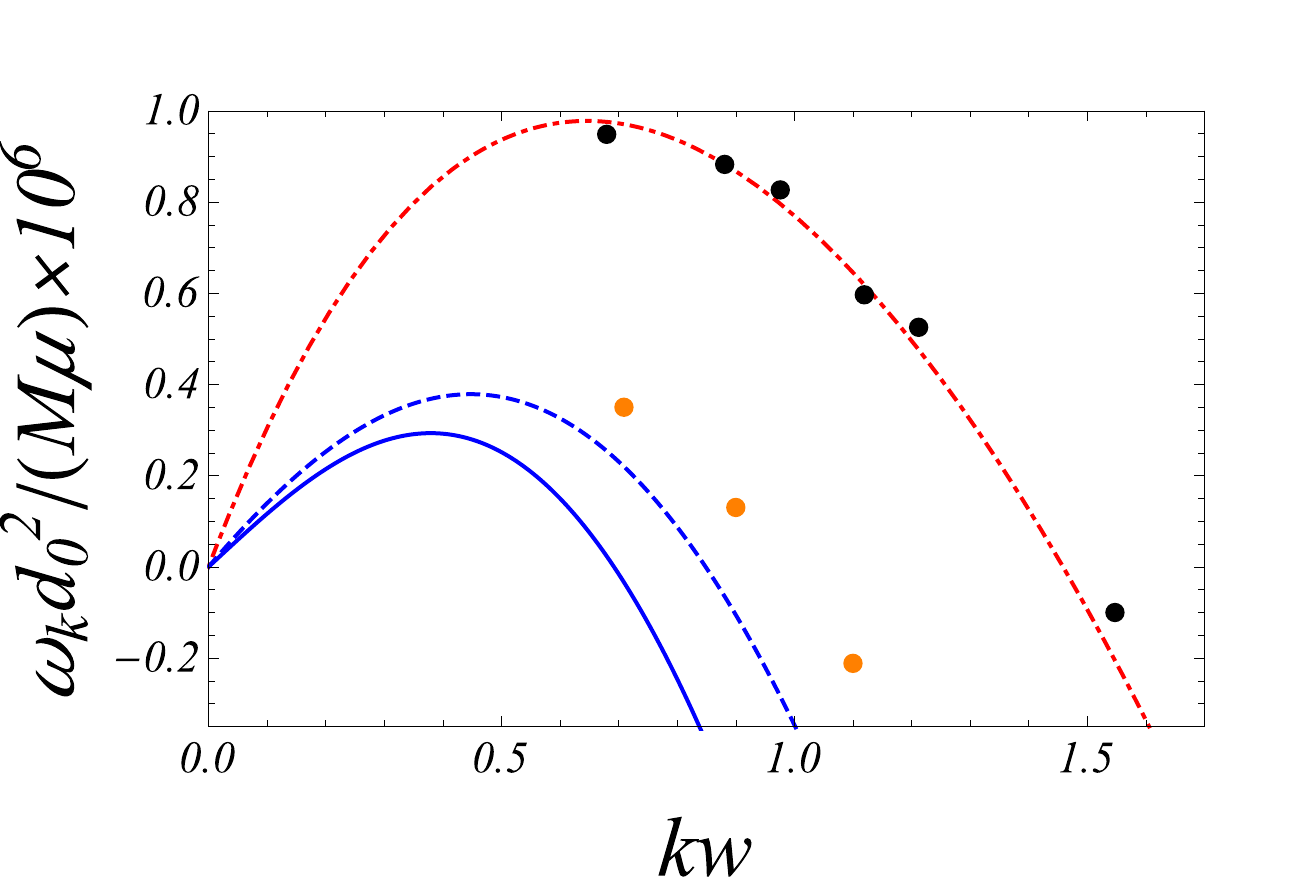}
    \includegraphics[width=0.9\linewidth]{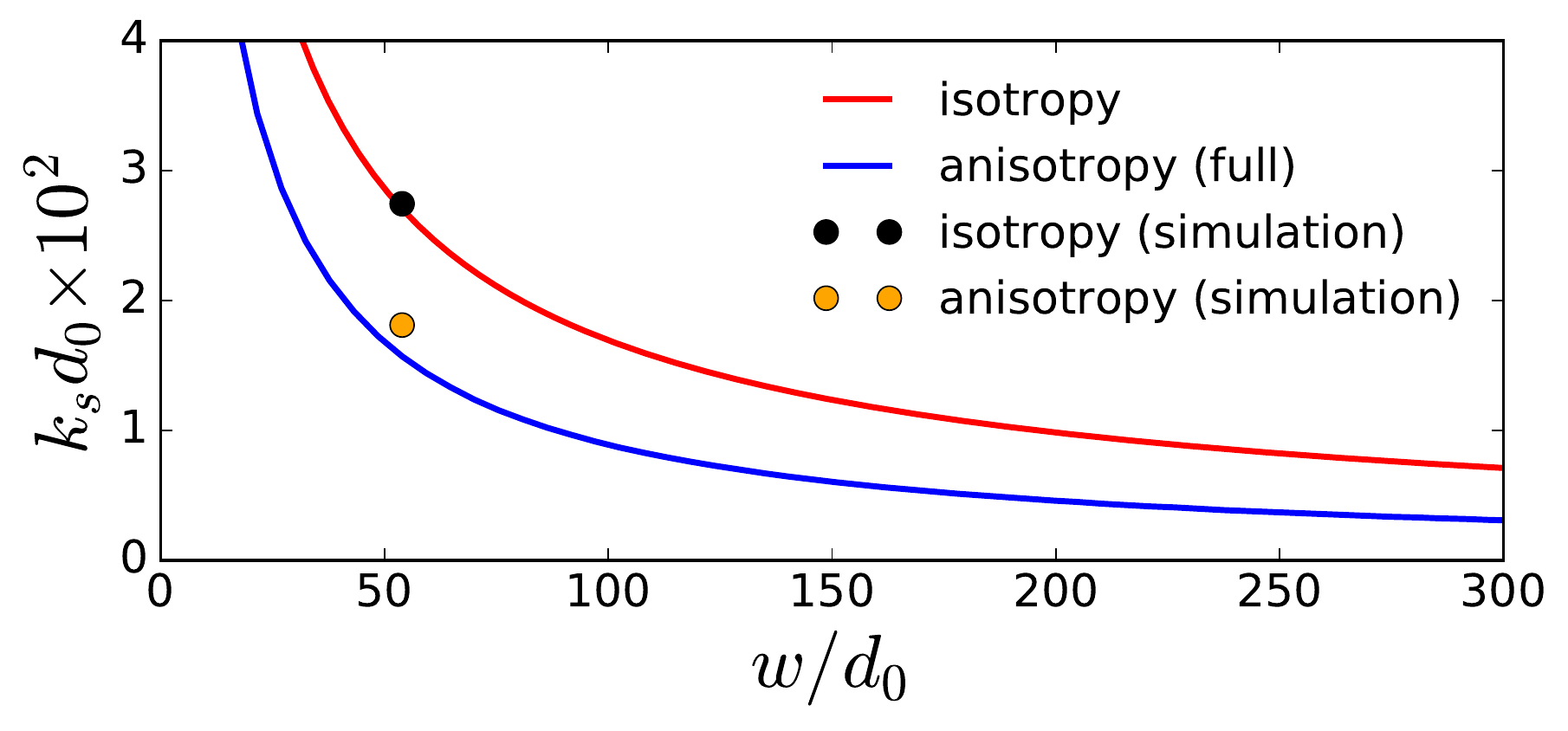}
  \end{center}

  \begin{picture}(1,0)(0,0)
    \put(77.,245.) {\mbox{\textcolor{black}{\scriptsize {Isotropic}}}}
    \put(77.,235.) {\mbox{\textcolor{black}{\scriptsize {\Cref{eq:stsxxyy1}}}}}
    \put(77.,225.) {\mbox{\textcolor{black}{\scriptsize {\Cref{eq:stsxxyy2}}}}}
    \put(-99.,430.) {\mbox{\textcolor{black}{\textbf{(a)}}}}
    \put(-99.,270.) {\mbox{\textcolor{black}{\textbf{(b)}}}}
    \put(-99.,120.) {\mbox{\textcolor{black}{\textbf{(c)}}}}
  \end{picture}
  \setlength{\unitlength}{1mm}
  \begin{picture}(10,1)(0,0)
    \linethickness{0.3mm}
    \color{red}
    \put(19.,87.0) {\line(1,0){1.4}}
    \put(21.0,87.0) {\circle*{0.1}}
    \put(21.6,87.0) {\line(1,0){1.4}}
    \color{blue}
    \put(19.,83.5) {\line(1,0){1.6}}
    \put(21.8,83.5) {\line(1,0){1.6}}
    \put(19.,80.0) {\line(1,0){4}}
  \end{picture}

  \vspace*{-0.7cm}
  \caption{(a) Amplitude of perturbed DB v.s. time for the simulation shown in
    \Cref{fig:nonlinear}. Black dots locate the snapshots in \Cref{fig:nonlinear}. The 
    dashed line represents the exponential fit performed to obtain the 
    growth rate of the instability. (b) Dimensionless growth rate $\omega_k$ as a function of the normalized wavevector $kw$ 
    ($w$ is the half-width of the compositional domain) in the isotropic case (red) and anisotropic case
    (blue) with $S=0.5$. In both cases, $w=32a$ and $\eps_0=0.043$ while the
    GB misorientation is $7.2\degree$ and $11.5\degree$ in the isotropic and anisotropic case, respectively. 
    For the anisotropic case, both the exact solution (derived using \Cref{eq:stsxxyy1}) and the approximate
    solution (derived using \Cref{eq:stsxxyy2}) are represented respectively with a dashed and continuous line.
    (c) Marginal wavevector $k_s$ as a function of the domain half-width $w$ 
    for both isotropic (red) and anisotropic (blue) elasticity. The theoretical prediction in the anisotropic case
    is obtained from the exact solution (derived using \Cref{eq:stsxxyy1}).}
  \label{fig:omegak}
\end{figure}

We note that for typical material values of $\nu<1/2$, the second term of \Cref{eq:stsxxyy2} has the same sign
as the anisotropic coefficient $S$. Therefore, if $S>0$ ($S<0$), the elastic anisotropy inhibits (promotes) the 
development of the instability compared to the isotropic case. 

To check the validity of this analysis, we performed AE simulations with both the isotropic hexagonal and anisotropic BCC models.
In both cases, we choose $C_{44}=\unit{39}{GPa}$. In the BCC AE model, we have necessarily $C_{11}=2C_{12}=2C_{44}$, fixing $S=0.5$. 
The misfit eigenstrain is $\eps_0=0.043$. In addition, 
the GB misorientation is $7.2\degree$ and $11.5\degree$ in the isotropic and anisotropic simulations, respectively.

To obtain the growth rate numerically, we perform simulations where the DBs
are initially gently deformed from their planar configuration with a small amplitude sinusoidal perturbation.
As demonstrated by the linear stability analysis, the stresses induced by this perturbation lead
to an increase of the perturbation amplitude $h_0(t)$. \Cref{fig:omegak}.a displays the amplitude of the DB perturbation
as a function of time for the simulation presented in \Cref{fig:nonlinear}. The black dots along the curve locate the snapshots shown in 
\Cref{fig:nonlinear}. We can distinguish two regimes. First, the perturbation amplitude grows exponentially with time as 
predicted by the linear stability analysis. The growth rate of the simulation is obtained by performing 
an exponential fit on this part of the curve. Second, at longer times, nonlinearities play 
a significant role and are responsible for
the deviation of the simulation results from the exponential fit. As depicted in \Cref{fig:nonlinear}.c, the DBs collide
with the GB, leading to a highly non-linear regime where GB breaks-up and the position of the individual dislocations are
relaxed by both glide and climb (see \Cref{fig:nonlinear}.d).

The growth rates are obtained for different simulations performed with various wavevector $k$ 
and for a precipitate width $w=32\,a$. 
We note that large wave-lengths (i.e. $kw<0.7$) are not investigated computationally due to the large simulation box sizes
necessary to explore this part of the dispersion diagram.
For both the isotropic and anisotropic AE models, the results are compared to analytical predictions in \Cref{fig:omegak}.b. 
For the sake of consistency with Ref.~[\onlinecite{Geslin2015}], the growth rate $\omega_k$ is normalized 
by the characteristic time $d_0^2/G M$ where $d_0$ is 
defined for an isotropic material by $d_0=\gamma(1-\nu)/[8G\eps_0^2(c^+-c^-)]$. 
As already discussed in Ref.~[\onlinecite{Geslin2015}], the simulation results in the isotropic case ($S=0$) 
agree well with the analytical prediction. 

As discussed previously, \Cref{fig:omegak}.b clearly shows that for our choice of parameters ($S=0.5$), 
the anisotropic elasticity reduces significantly the growth rate and shifts the unstable range (where $\omega_k>0$) 
to larger wavelengths, therefore inhibiting the instability. This can be understood with the following 
qualitative argument: in the isotropic case, the Bitter-Crum theorem \cite{Bitter1931,Fratzl1999}
insures that the elastic energy of a precipitate does not depend on its shape. Therefore, the perturbation
of the DB interface leads automatically to a decrease of the elastic energy due to the relaxation of the shear stresses 
at the GBs. If this energy drop compensates the increase of energy attributed to the lengthening of the perturbed DBs, the system is unstable. This reasoning does not hold in the anisotropic case where the Bitter-Crum theorem does not apply. In our case, the lamellar precipitate is oriented along an elastically soft direction. Any perturbation of such a well-oriented precipitate increases the elastic energy. Therefore, the destabilization of the system occurs only if the stress relaxation
at the GB compensates this additional amount of energy. We note that a lamellar precipitate oriented along an elastically hard direction (e.g. with a $45 \degree$ angle with the $x$-axis) is intrinsically unstable \cite{Khachaturyan2013}.

The simulations performed with the BCC AE model show a good agreement with the growth rate predicted by the linear stability analysis. The small discrepancy between the numerical and analytical results is attributed to the homogeneous elasticity approximation.
Indeed, to perform the linear stability analysis, we considered that the elastic constants are the same in both grains, regardless of the 
rotations introduced by the GB. Also, numerical limitations such as limited system sizes might also
contribute to this small discrepancy.

The marginally stable wavevetor $k_s$ defined as the positive root of $\omega_k=0$ can be deduced 
for both numerical and theoretical results and is plotted as a function of the normalized composition 
domain half-width $w$ in the isotropic and anisotropic case in \Cref{fig:omegak}.c. This plot shows 
again that the elastic anisotropy shifts the domain of instability to longer wavelengths, thus 
inhibiting the morphological instability. Even though we only presented numerical results for one value of 
$w$, the dependence of the results on $w$ can be deduced from the predictions of the linear stability analysis. For isotropic elasticity, this analysis predicts that, in the physically relevant limit $w\gg d_0$ where the precipitate width is much larger than the microscopic capillary length scale $d_0$, the marginally stable wavector $k_s\approx \frac{1}{2w}\ln(w/d_0)$ and the fastest growing wavector $k_0\approx C/w$ where $C=0.797...$ is a numerical constant  \cite{Geslin2015}.  
The same scalings holds for the anisotropic case but with the constant $C$ depending generally on the magnitude $S$ of the anisotropy. As can be seen in Fig. \ref{fig:omegak}.b, the fastest growing wavector is smaller in the anisotropic case than the isotropic case, consistent with the fact that anisotropy has a stabilizing effect when the lamellar precipitate is oriented along an elastically soft direction. However, in both the isotropic and anisotropic cases, the most unstable wavelength $2\pi/k_0$ is proportional to $w$ so that the instability will generally develop on the scale of the precipitate width.

\section{Grain boundary break-up}
\label{sec:GBB}

\begin{figure}[h]
  \begin{center}
    \includegraphics[width=0.45\linewidth]{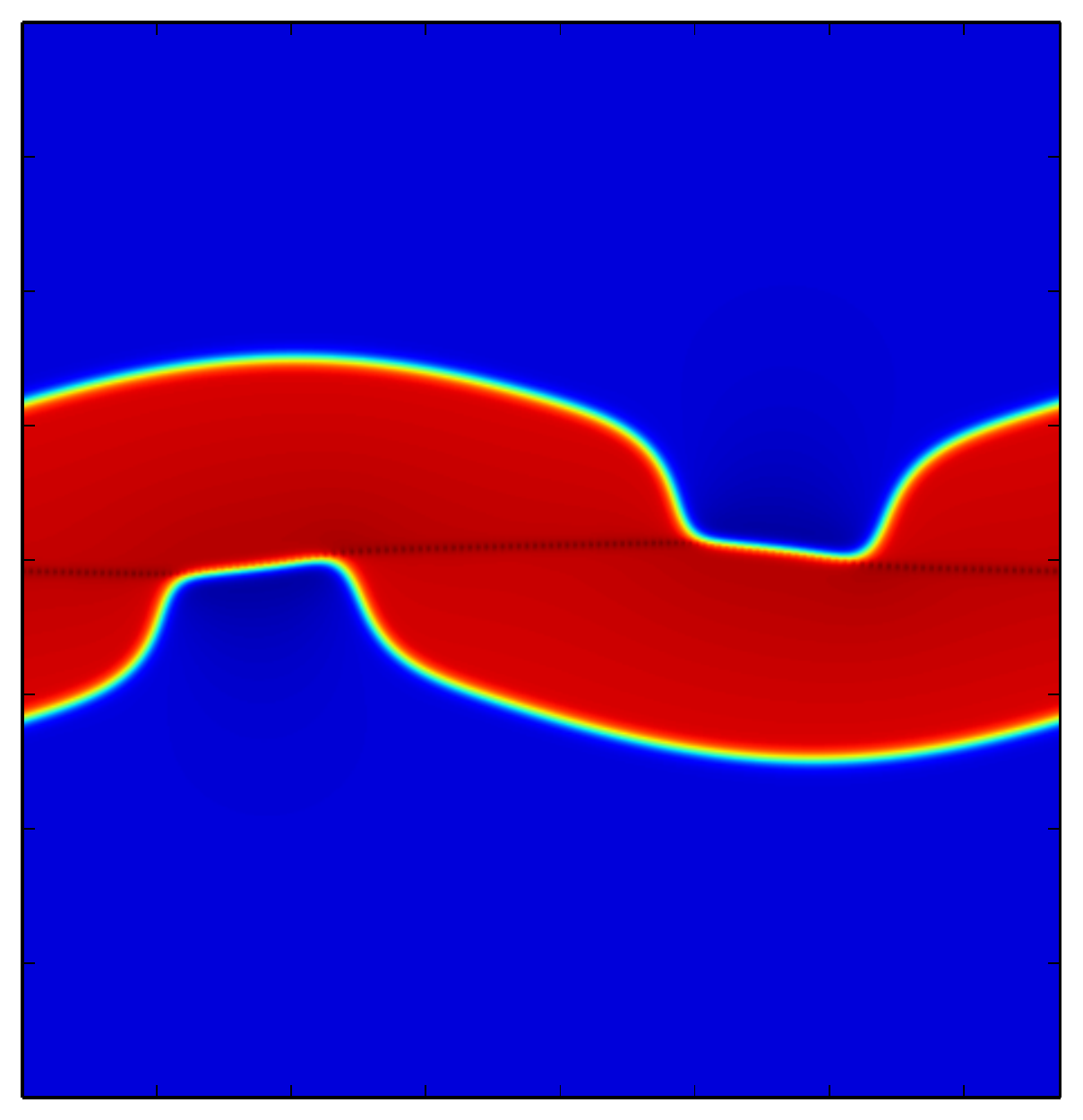}
    \includegraphics[width=0.45\linewidth]{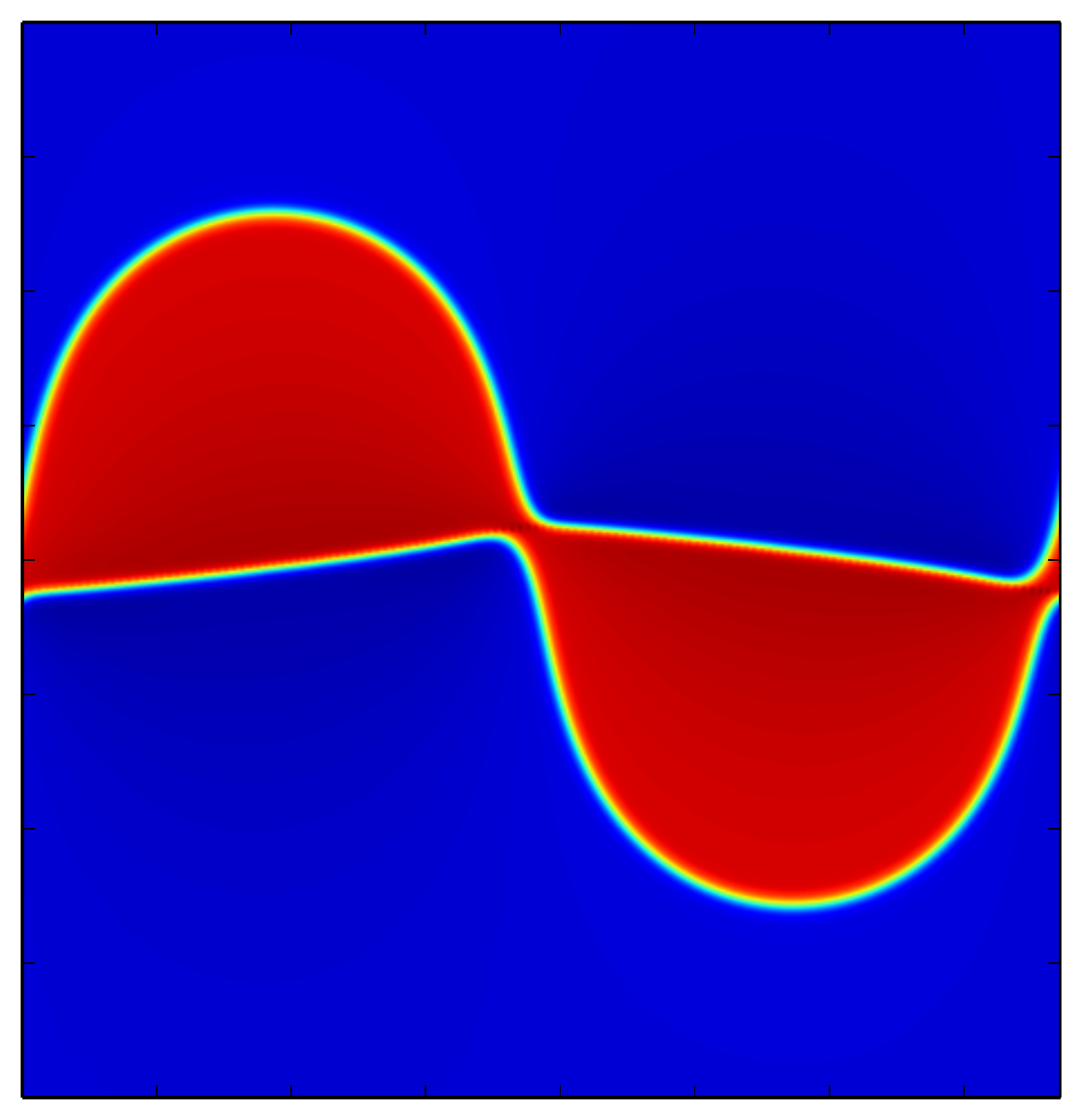}
		
		\begin{picture}(1,0)(0,0)
		  \put(-100.,19.) {\mbox{\textcolor{white}{\textbf{(a)}}}}
          \put(5.,19.) {\mbox{\textcolor{white}{\textbf{(b)}}}}
		\end{picture}
  \end{center}

   \caption{Snapshots of the concentration field during a simulation performed with
    a GB misorientation angle $\theta=30.4\degree$ centered on a lamellar precipitate of
    eigenstrain $\eps_0=0.043$.
    The system size is $7.8w \times 10.8w$ where $2w$ is the initial width of the precipitate 
    (the vertical length of each frame is smaller than the box dimension).
    The snapshots correspond to dimensionless times (a) $1.2\times10^6$ and
    (b) $5.12\times10^6$. See online supplementental material\cite{SupplMat} for the movie of this simulation (file movie1.avi).}
  \label{fig:nobreak}
\end{figure}
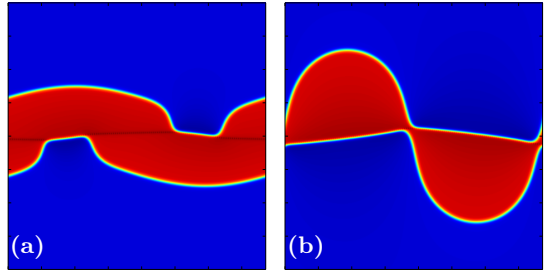

\begin{figure}[h]
  \begin{center}
    \includegraphics[width=0.45\linewidth]{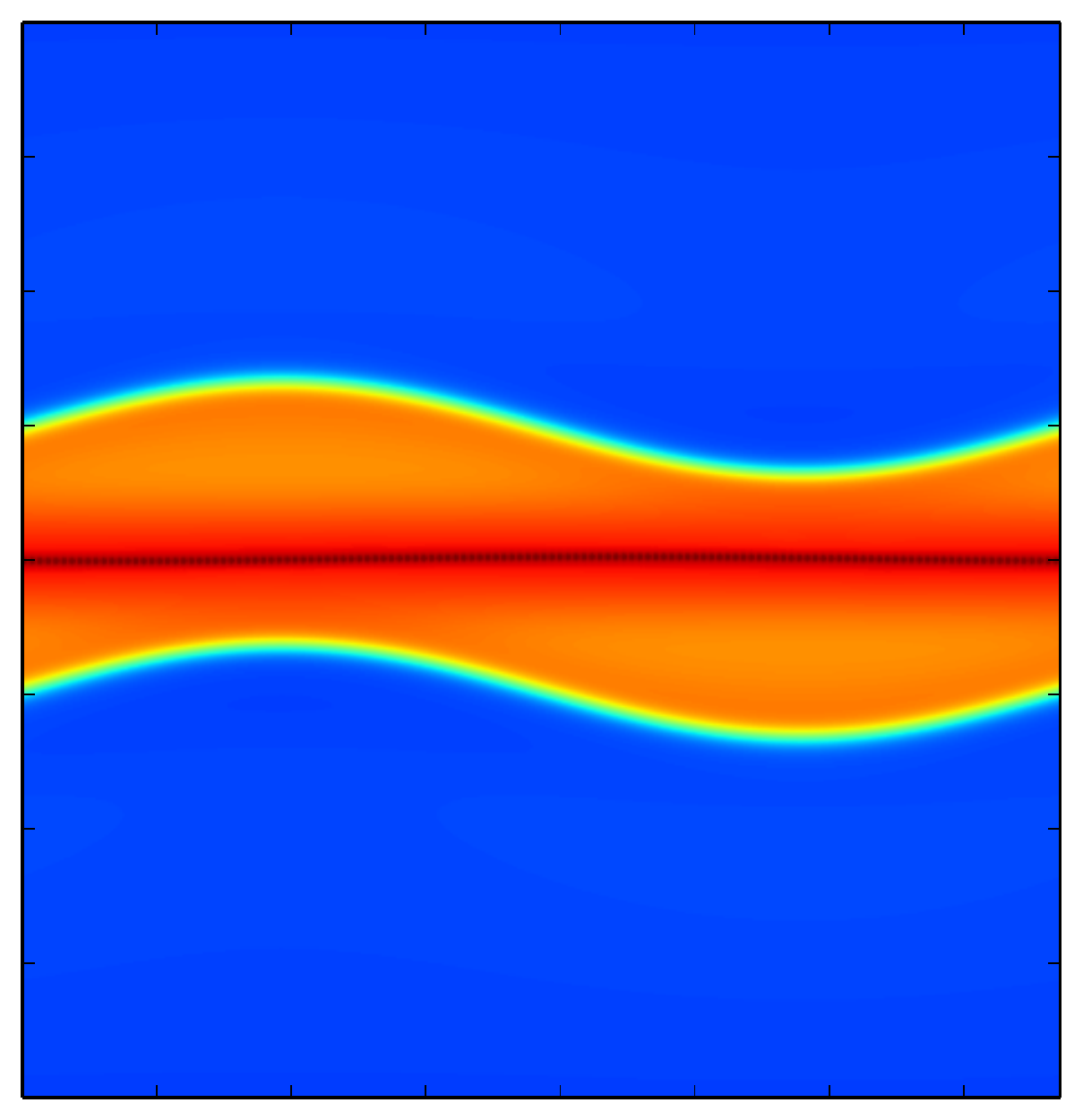}
    \includegraphics[width=0.45\linewidth]{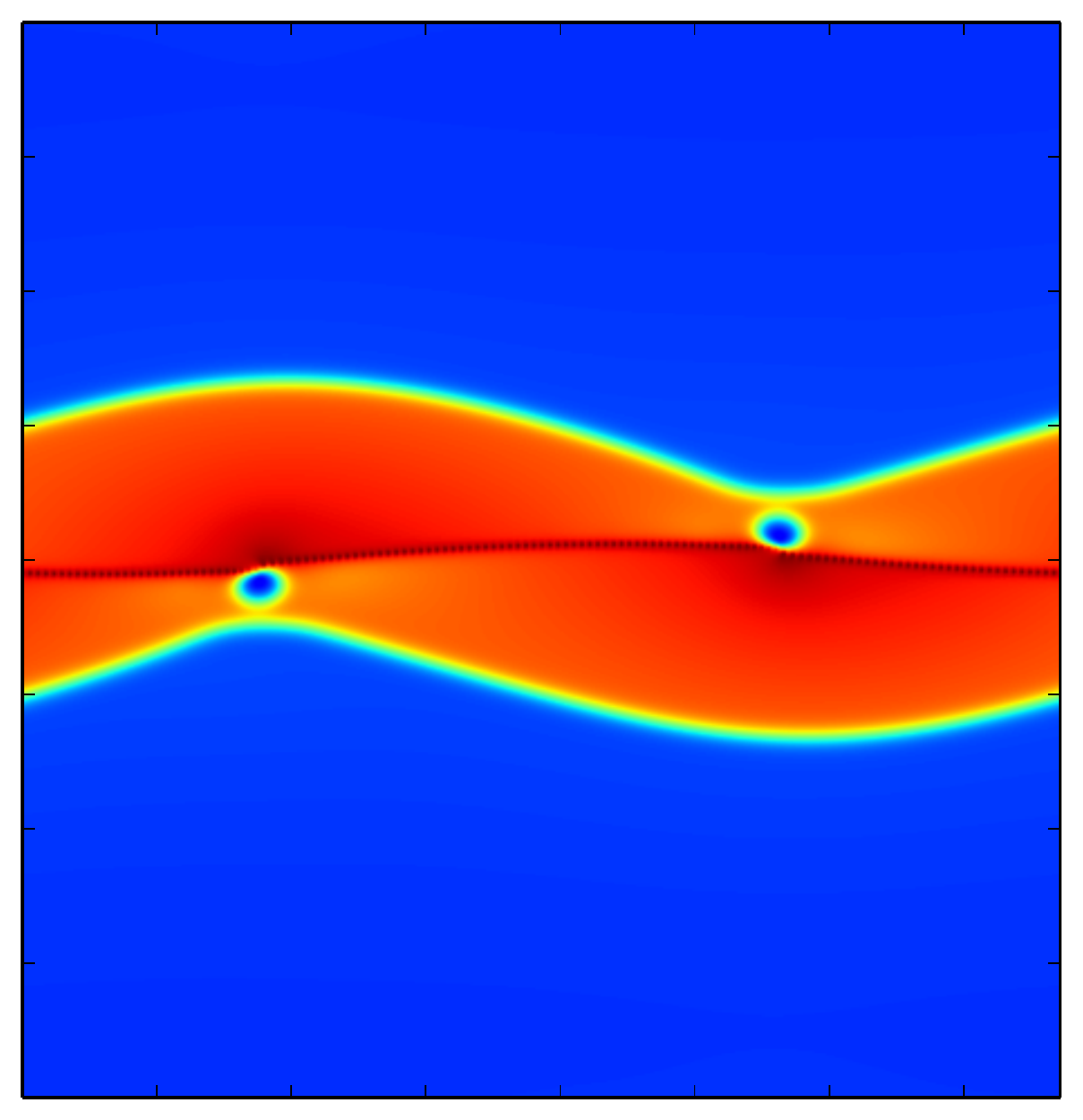}
    \includegraphics[width=0.45\linewidth]{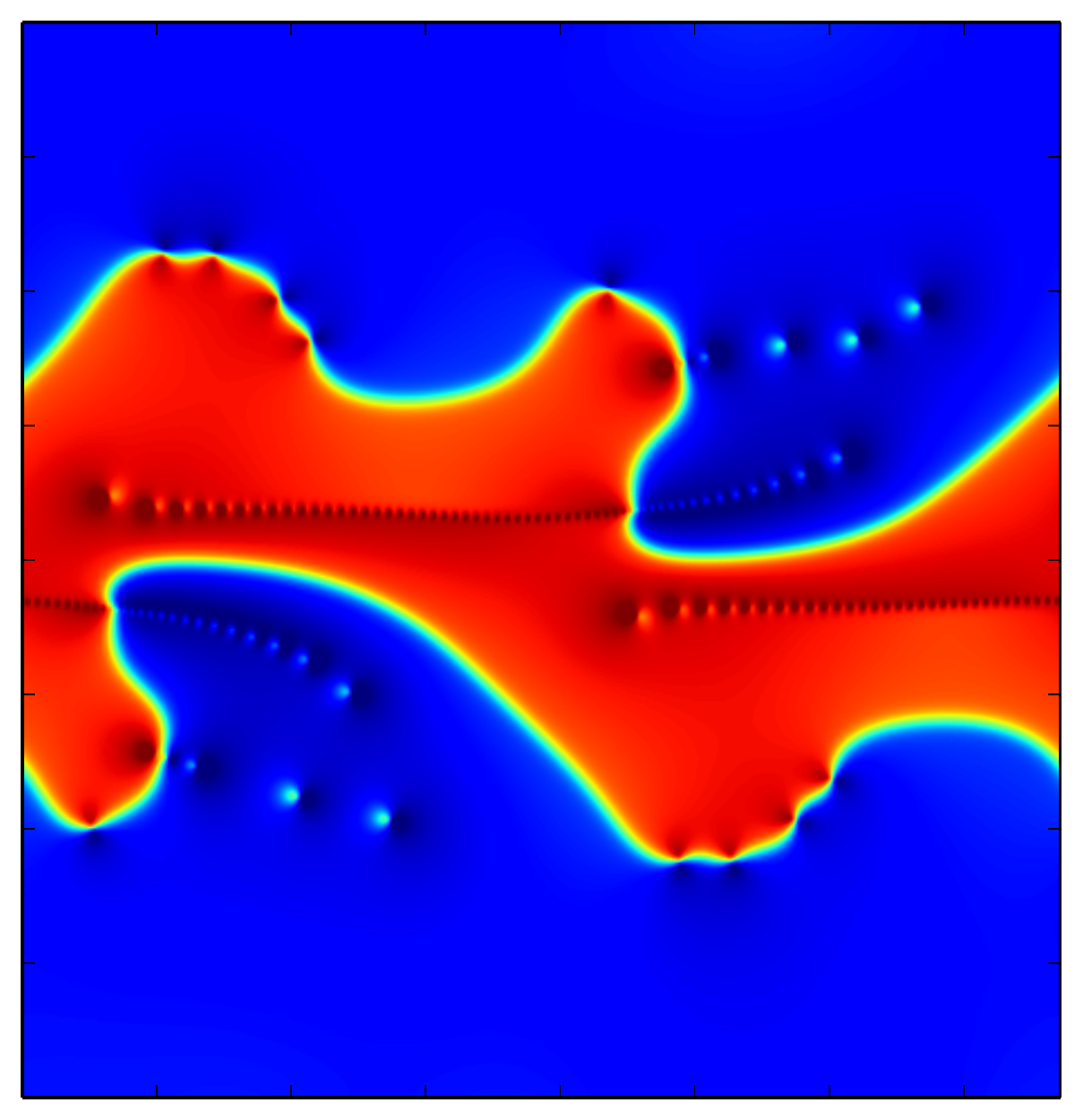}
    \includegraphics[width=0.45\linewidth]{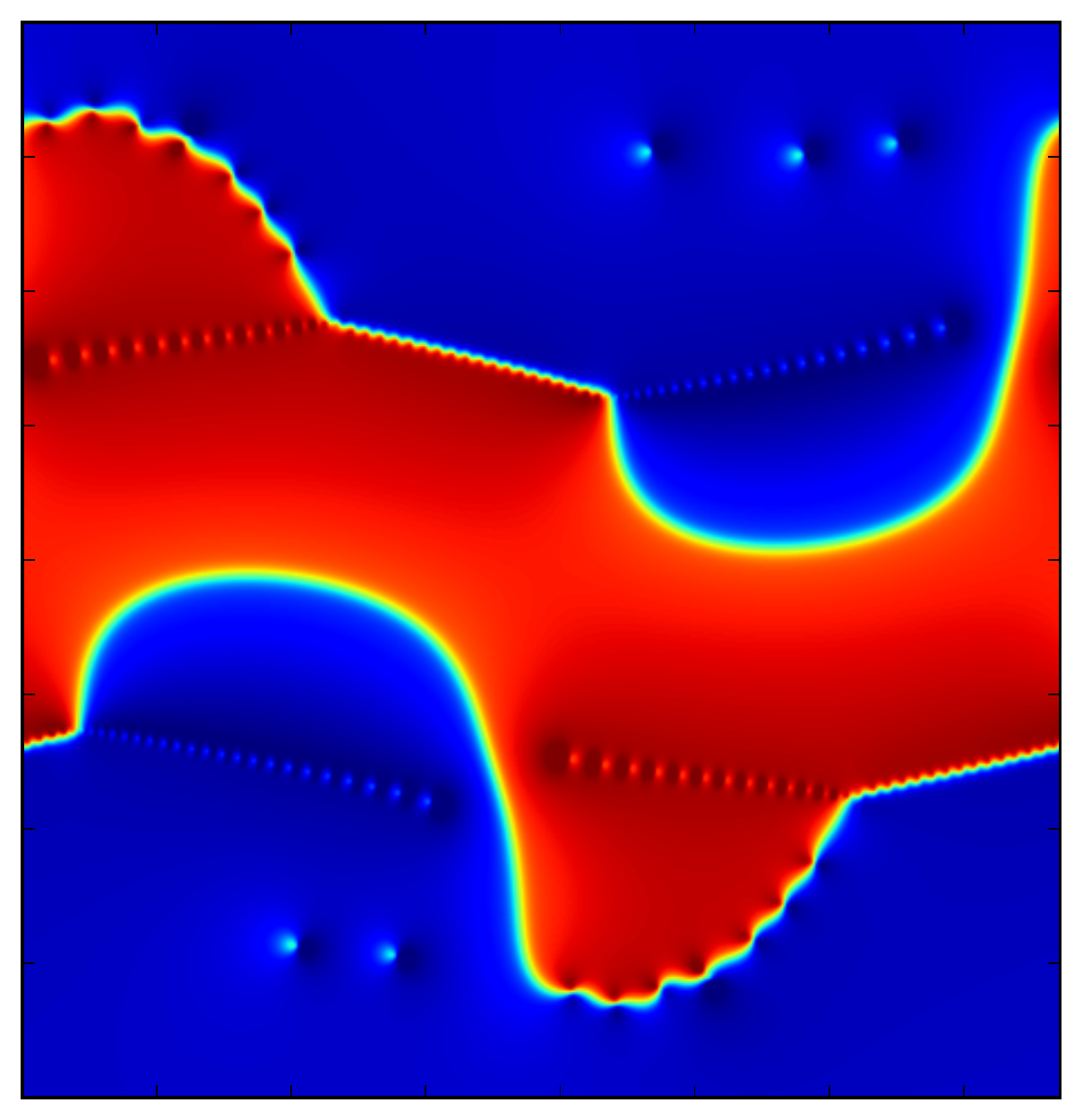}
		
		\begin{picture}(1,0)(0,0)
			\put(-99.,126.) {\mbox{\textcolor{white}{\textbf{(a)}}}}
            \put(5.,126.) {\mbox{\textcolor{white}{\textbf{(b)}}}}
            \put(-99.,19.) {\mbox{\textcolor{white}{\textbf{(c)}}}}
			\put(5.,19.) {\mbox{\textcolor{white}{\textbf{(d)}}}}
		\end{picture}
  \end{center}
   \caption{Snapshots of the concentration field during a simulation performed with
    a GB misorientation angle $\theta=30.4\degree$ centered on a lamellar precipitate of
    eigenstrain $\eps_0=0.086$.
    The system size is $7.8w \times 10.8w$ where $2w$ is the initial width of the precipitate 
    (the vertical length of each frame is smaller than the box dimension).
    The snapshots correspond to dimensionless times (a) $4\times10^3$,
    (b) $7.6\times10^4$, (c) $2.84\times10^5$ and (d) $4.2\times10^6$.
    See online supplementental material\cite{SupplMat} for the movie of this simulation (file movie2.avi).}
   \label{fig:hvbreak}
\end{figure}

\begin{figure}[h]
  \begin{center}
    \includegraphics[width=1.0\linewidth]{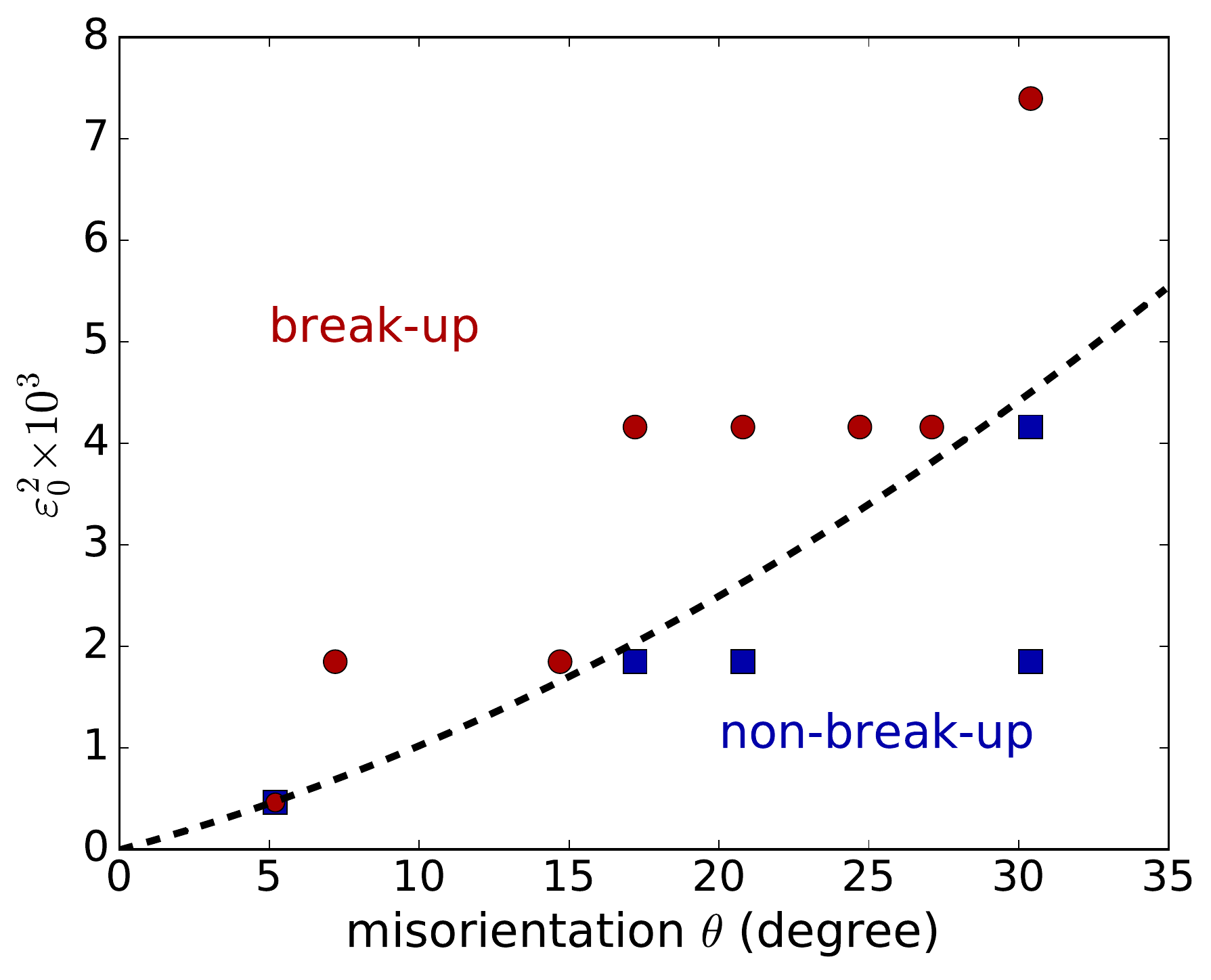}
  \end{center}
  \caption{GB behavior as a function of misorientation angle $\theta$ and misfit strain $\eps_0$. 
  Simulations where the GB breaks up are represented
  with red dots while the ones where the GB is morphologically unstable but remains continuous are shown with blue squares. The dashed
  line is a second order polynomial fit of the boundary between break-up and non-break-up regions of the ($\theta$, $\eps_0^2$) plane.}
  \label{fig:sumbreak}
\end{figure}

In polycrystalline materials, the density and properties of GBs influence significantly the properties
of the bulk material. They are preferred nucleation sites for second phase precipitates \cite{Sutton1995}.
and also facilitate impurities diffusion though a pipe diffusion effect \cite{Hirth1968}. Moreover, GBs are 
natural obstacles to dislocation motion, and fine grain structures often present high yield stresses \cite{Sutton1995}. 
Therefore, controlling the GB density and properties is of first importance to obtain high material properties. 

The instability described in this article affects significantly the GB
and can even lead to the break-up of the GB as shown in \Cref{fig:nonlinear}. 
For this low angle GB ($\theta = 7.2\degree$), $\beta=0.126$ and the perturbation of the GB expressed 
in \Cref{eq:eqH0te} is significant as shown in \Cref{fig:nonlinear}.b. 
In the equilibrium state represented in \Cref{fig:nonlinear}.e., 
the dislocations that were forming the low-angle GB 
decorate the precipitate interface, relaxing the misfit stresses.

Increasing the GB angle does not modify the development nor the growth rate of the instability. However,
for higher misorientation angles, the coupling factor $\beta$ is greater and therefore the amplitude of the GB
perturbation is smaller. So, one can expect the influence of the instability on the GB to be less important.
\Cref{fig:nobreak} shows the late stages of the development of the instability
for a misorientation angle $\theta=30.4\degree$ GB, everything else being identical to the simulation
presented in \Cref{fig:nonlinear}. As expected, the GB is less affected by the instability: its position
is only slightly modified and the precipitate shape evolves until the DBs wet the GB. \Cref{fig:nobreak}.b 
represents the equilibrium state of the system where the GB remains continuous and the precipitate forms
lobes on both sides of the GB. We note here that this destabilization can represent the first stage of development of 
the Widmanst\"{a}tten structure found in steel and Ti-based alloys \cite{DaCostaTeixeira2006,Cheng2010}. It has been shown
experimentally that Widmanst\"{a}tten structures develop in two steps: first, an thin elongated precipitate 
nucleates on the GB and grow laterally; then, the precipitate develops acicular arms growing perpendicularly
to the GB, towards the center of the grain.
The instability presented in this paper and more precisely the morphology shown in \Cref{fig:nobreak}.b 
could trigger the growth of elongated precipitates perpendicular to the GB.

However, increasing the misfit strain can destabilize a high angle GB as well:
\Cref{fig:hvbreak} shows a simulation performed with $\theta=30.4\degree$ and
a misfit eigenstrain of $\eps_0=0.086$. During the development of the instability,
we notice the nucleation of low composition domains close to the GB (blue droplets in \Cref{fig:hvbreak}.b)
promoted by the high compressive stress appearing in the vicinity of the deformed GB. Later in the simulation,
the high angle GB breaks up (\Cref{fig:hvbreak}.c) and the system relaxes
into a configuration presenting two lower angle GBs (\Cref{fig:hvbreak}.d). 
The equilibrium configuration also shows that dislocations decorate the precipitate surface, relaxing
the high misfit stresses.

The appearance of GB break-up then depends on a balance between the misfit stresses and the GB misorientation.
This is summarized in \Cref{fig:sumbreak} where the results of several simulations for various values
of the misorientation angle $\theta$ and eigenstrain $\eps_0$ are presented: the GB break-up occurs for
low angle GBs and high values of $\eps_0^2$. The dashed line separating both regions serves as a guide 
to the eye and is linear for small values of the misorientation angle. It also shows that for a large
enough misfit, the instability break up all GBs.

\section{Interaction between circular precipitates and grain boundaries}
\label{sec:CPG}

\begin{figure}[h]
  \begin{center}
    \includegraphics[width=0.45\linewidth]{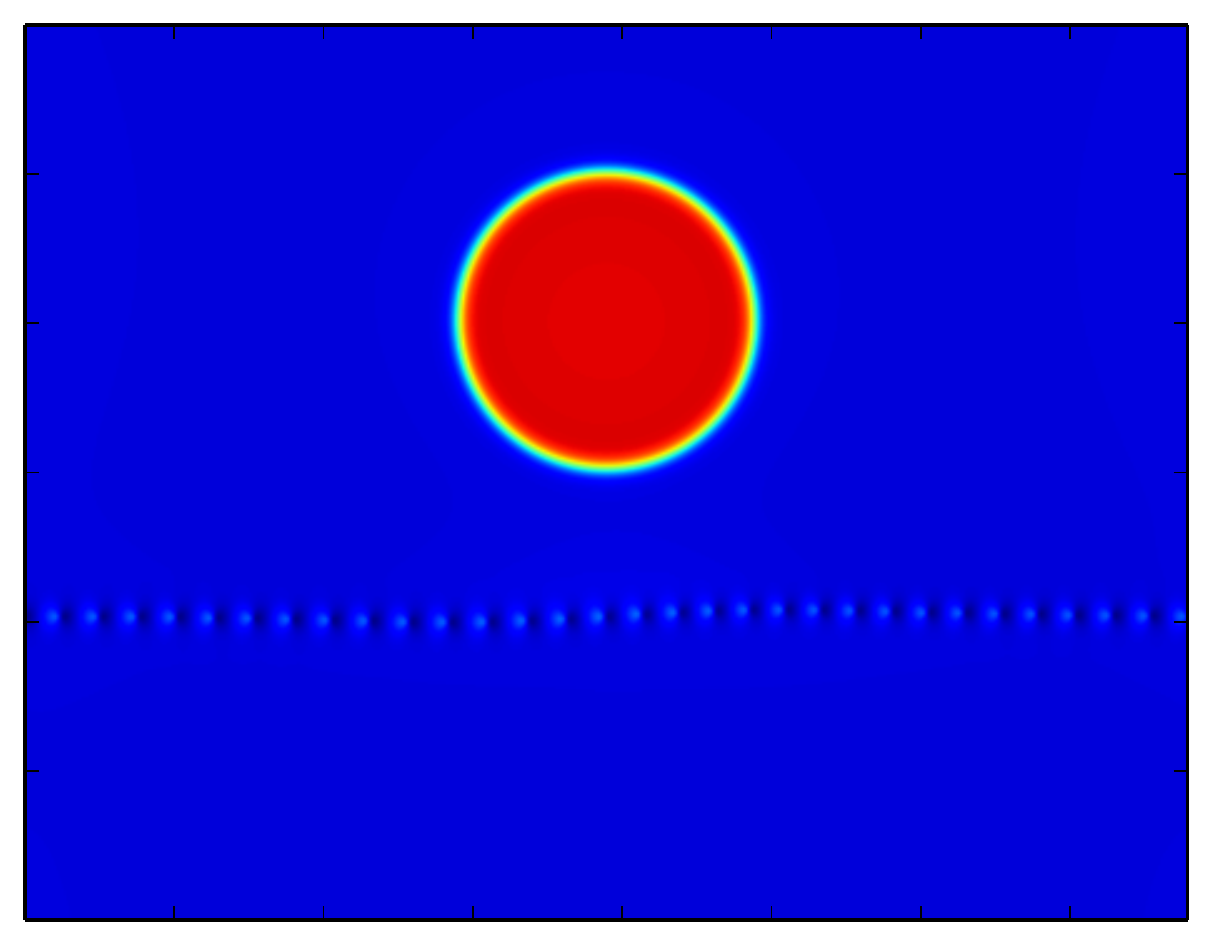}
    \includegraphics[width=0.45\linewidth]{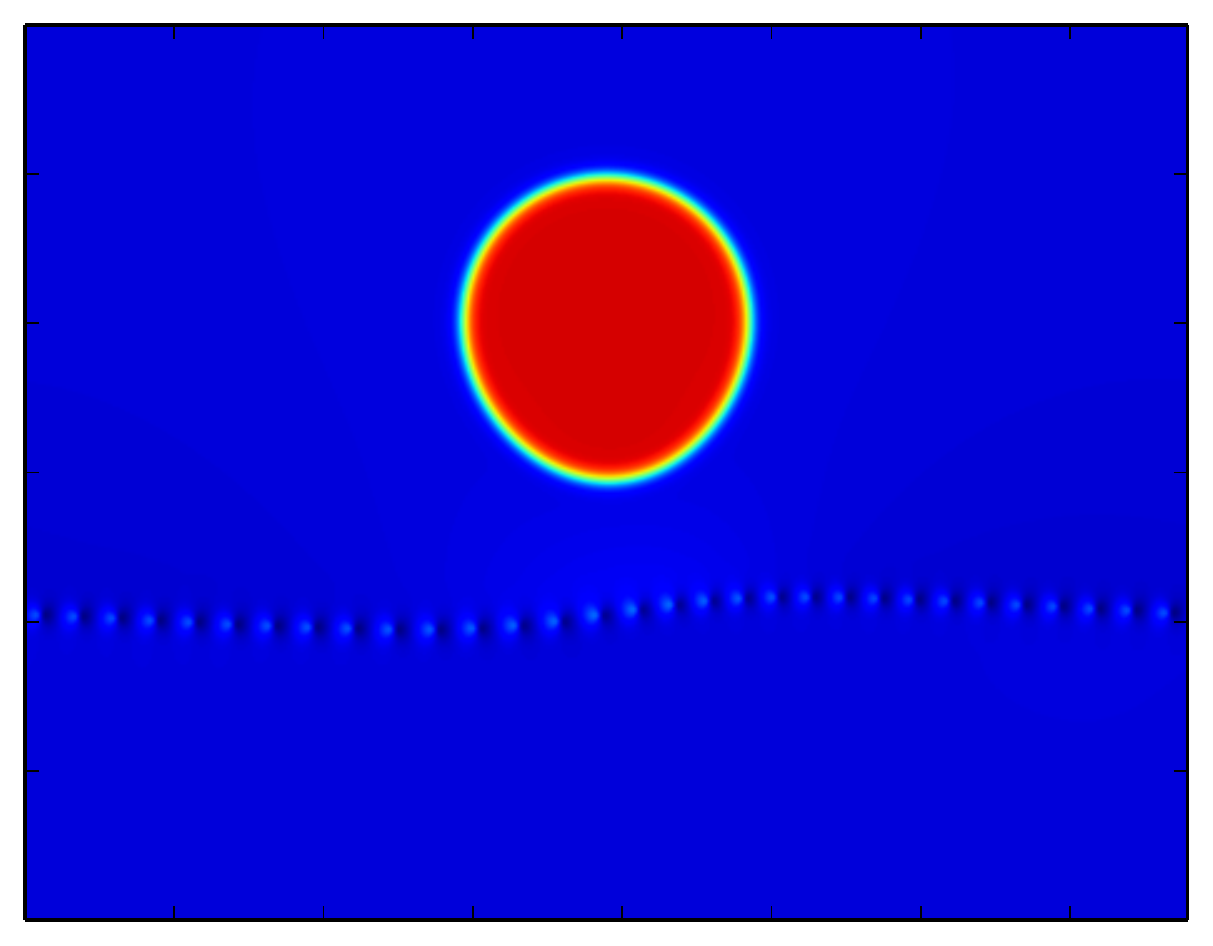}
    \includegraphics[width=0.45\linewidth]{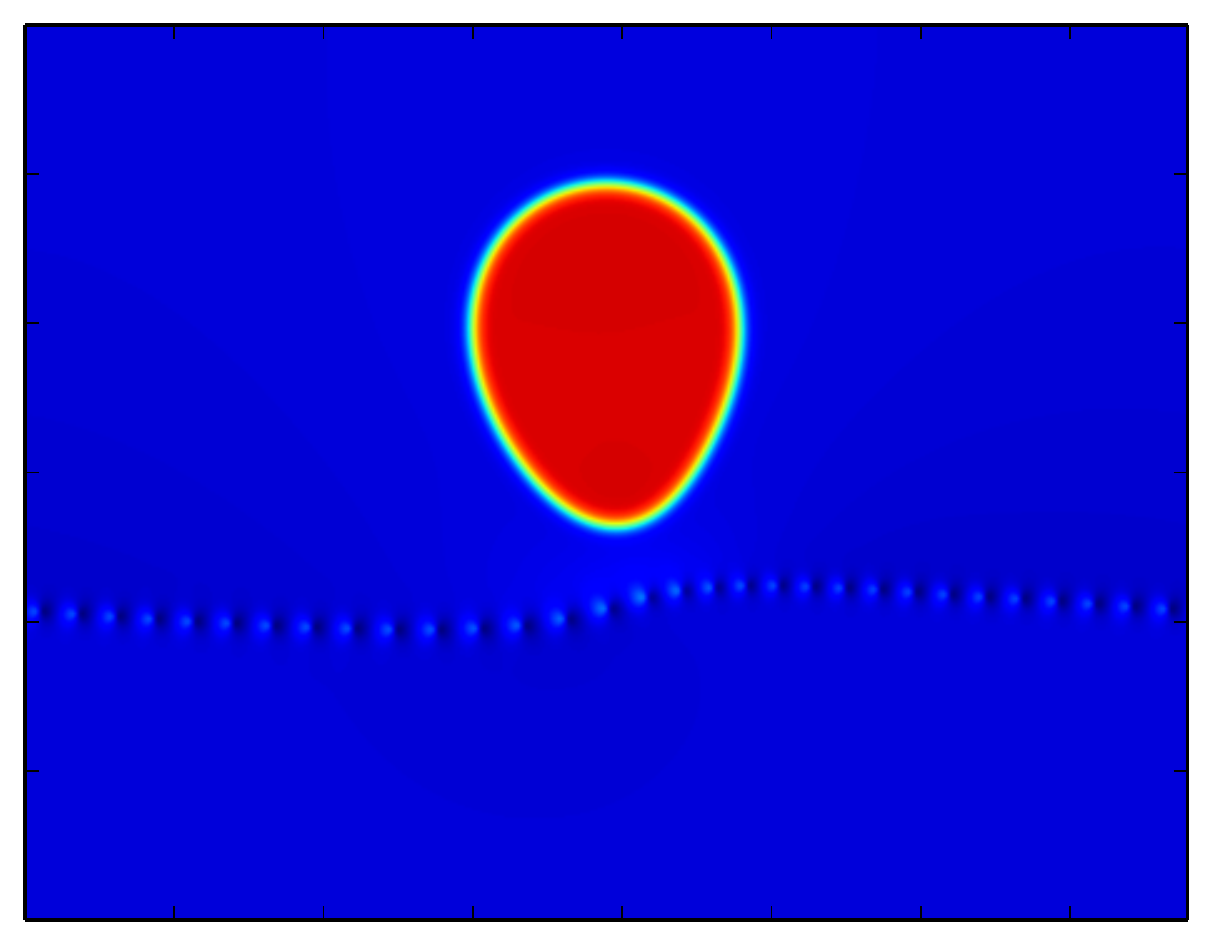}
    \includegraphics[width=0.45\linewidth]{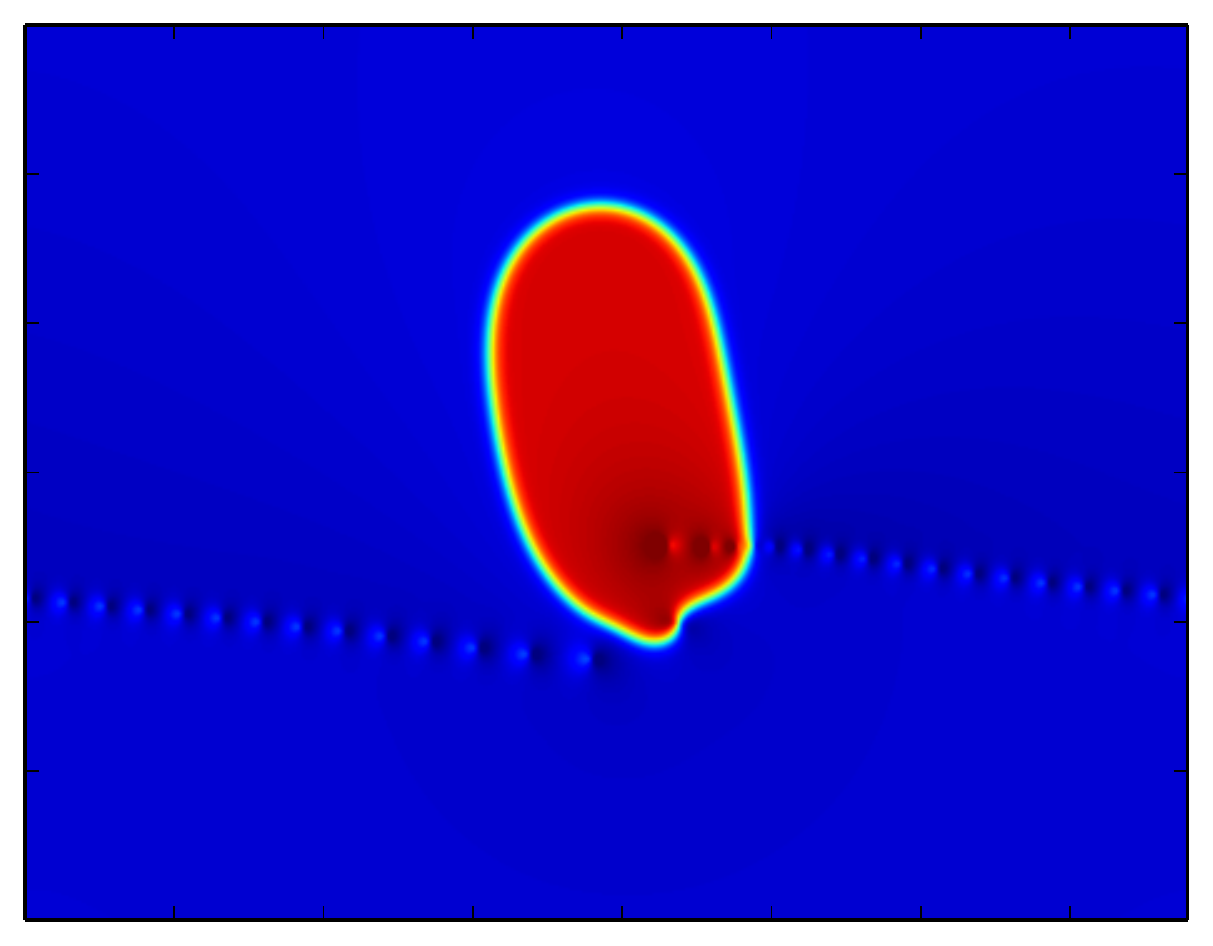}
    \includegraphics[width=0.45\linewidth]{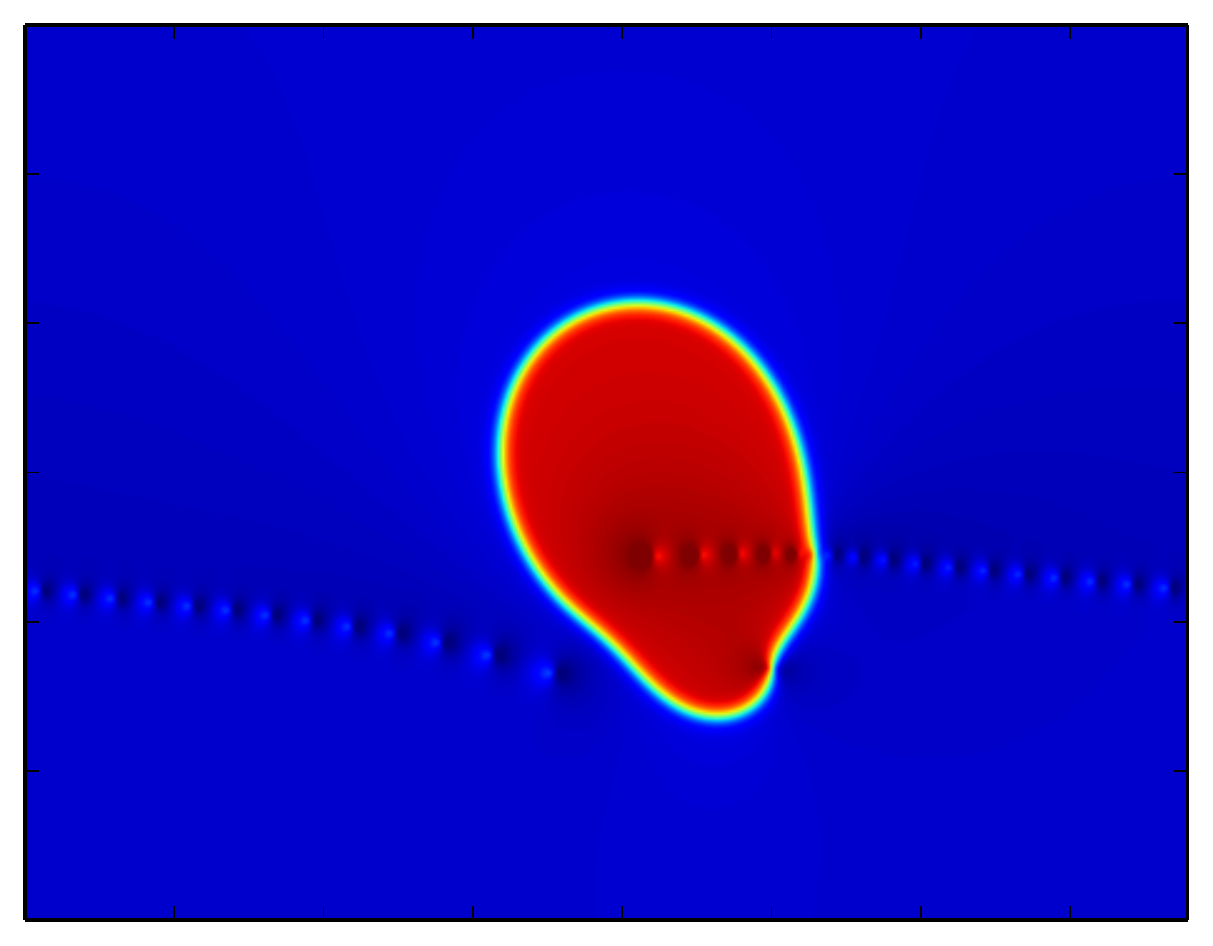}
    \includegraphics[width=0.45\linewidth]{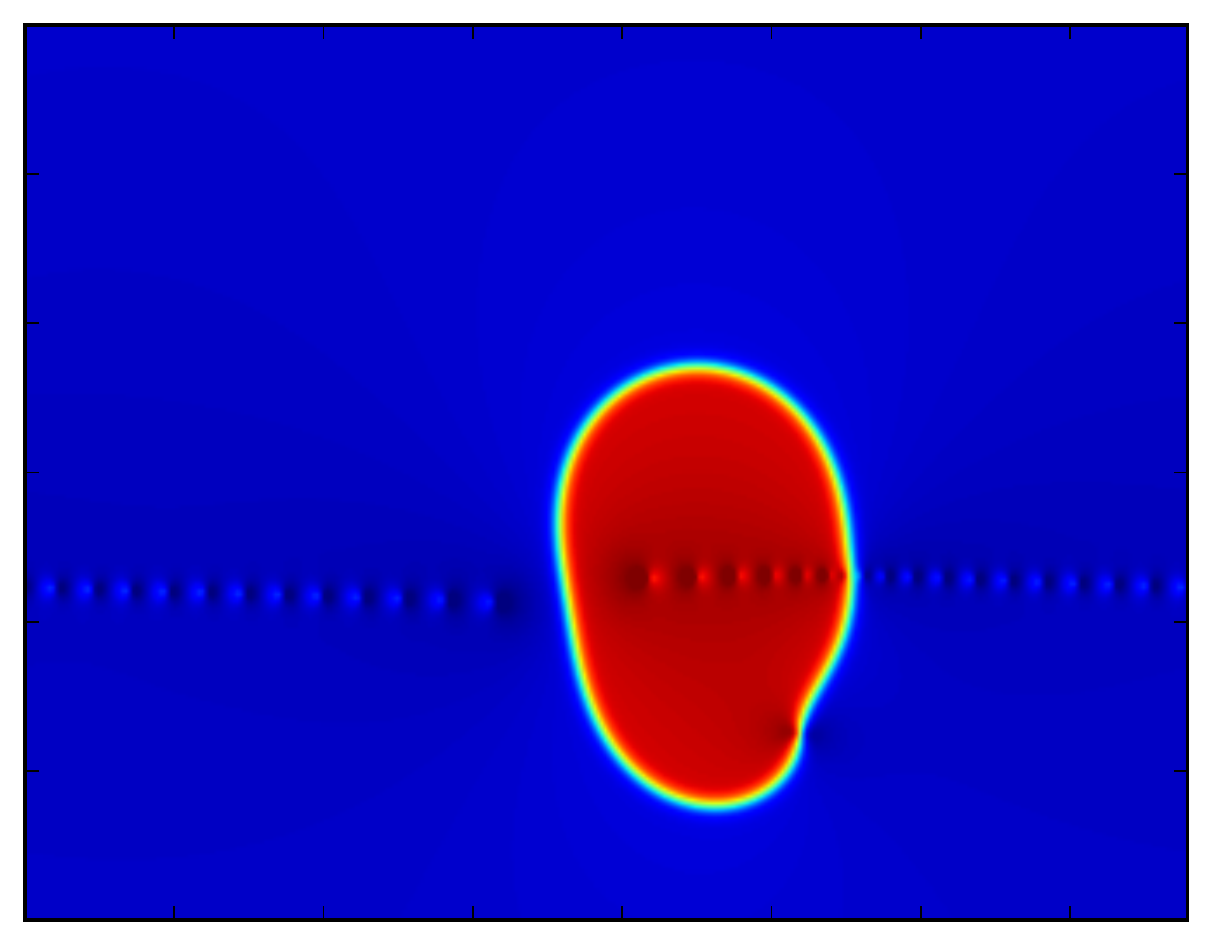}
		
		\begin{picture}(1,0)(0,0)
			\put(-99.,181.) {\mbox{\textcolor{white}{\textbf{(a)}}}}
            \put(5.,181.) {\mbox{\textcolor{white}{\textbf{(b)}}}}
            \put(-99.,100.) {\mbox{\textcolor{white}{\textbf{(c)}}}}
			\put(5.,100.) {\mbox{\textcolor{white}{\textbf{(d)}}}}
            \put(-99.,19.) {\mbox{\textcolor{white}{\textbf{(e)}}}}
            \put(5.,19.) {\mbox{\textcolor{white}{\textbf{(f)}}}}
		\end{picture}
  \end{center}
  \caption{Snapshots of the concentration field during a simulation performed with
    a GB misorientation $\theta=7.2\degree$ and a circular precipitate with a
    misfit eigenstrain $\eps_0=0.043$.
    The system size is $7.8R \times 11.3R$ where $R$ is the radius of the circular 
    precipitate. The snapshots correspond to dimensionless times (a) $1\times10^4$,
    (b) $1.6\times10^5$, (c) $5.1\times10^5$, (d) $1.01\times10^6$, (e) $2.72\times10^6$
    and (f) $6.18\times10^6$. See online supplementental material\cite{SupplMat} for the movie of this simulation (file movie3.avi).}
  \label{fig:circlelow}
\end{figure}

\begin{figure}[h]
  \begin{center}
    \includegraphics[width=0.45\linewidth]{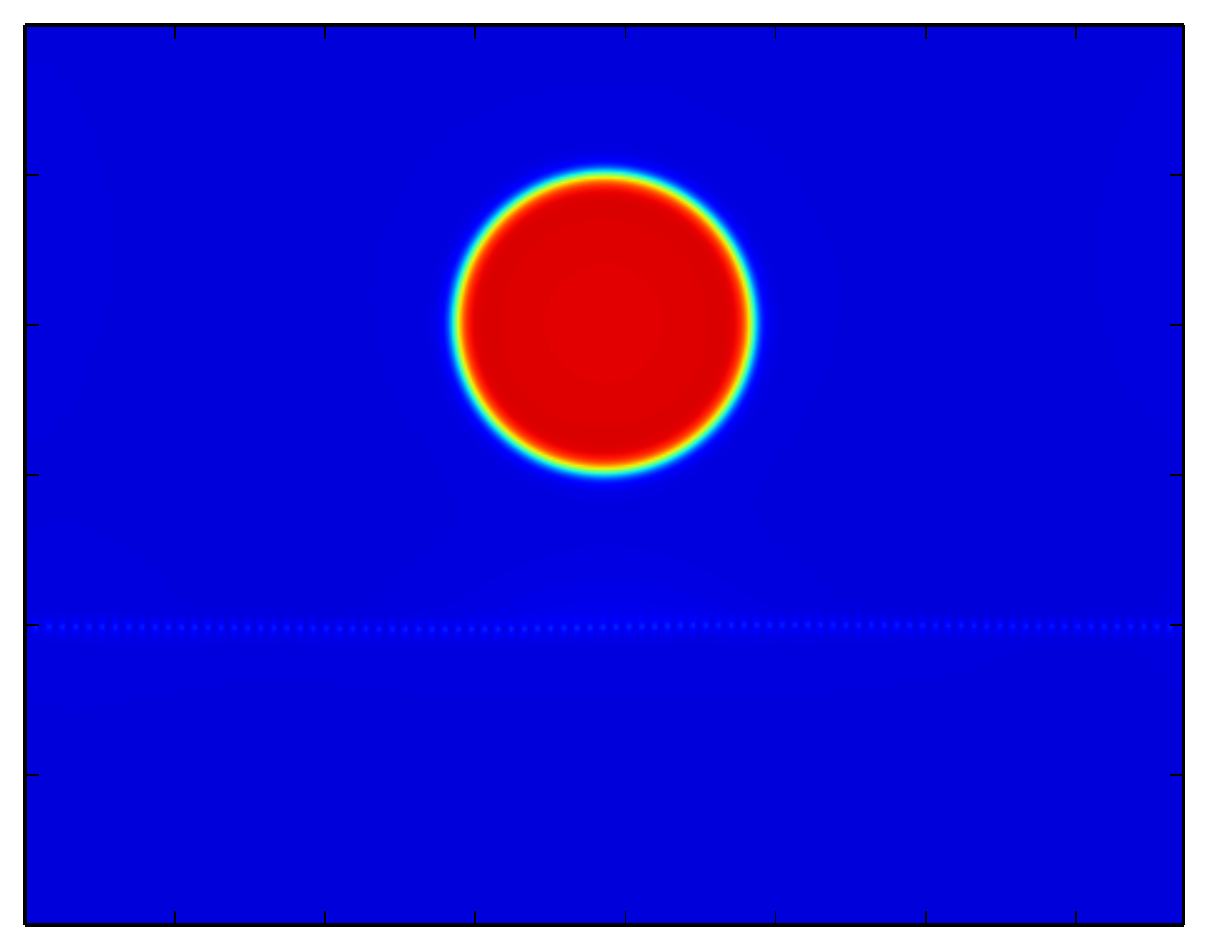}
    \includegraphics[width=0.45\linewidth]{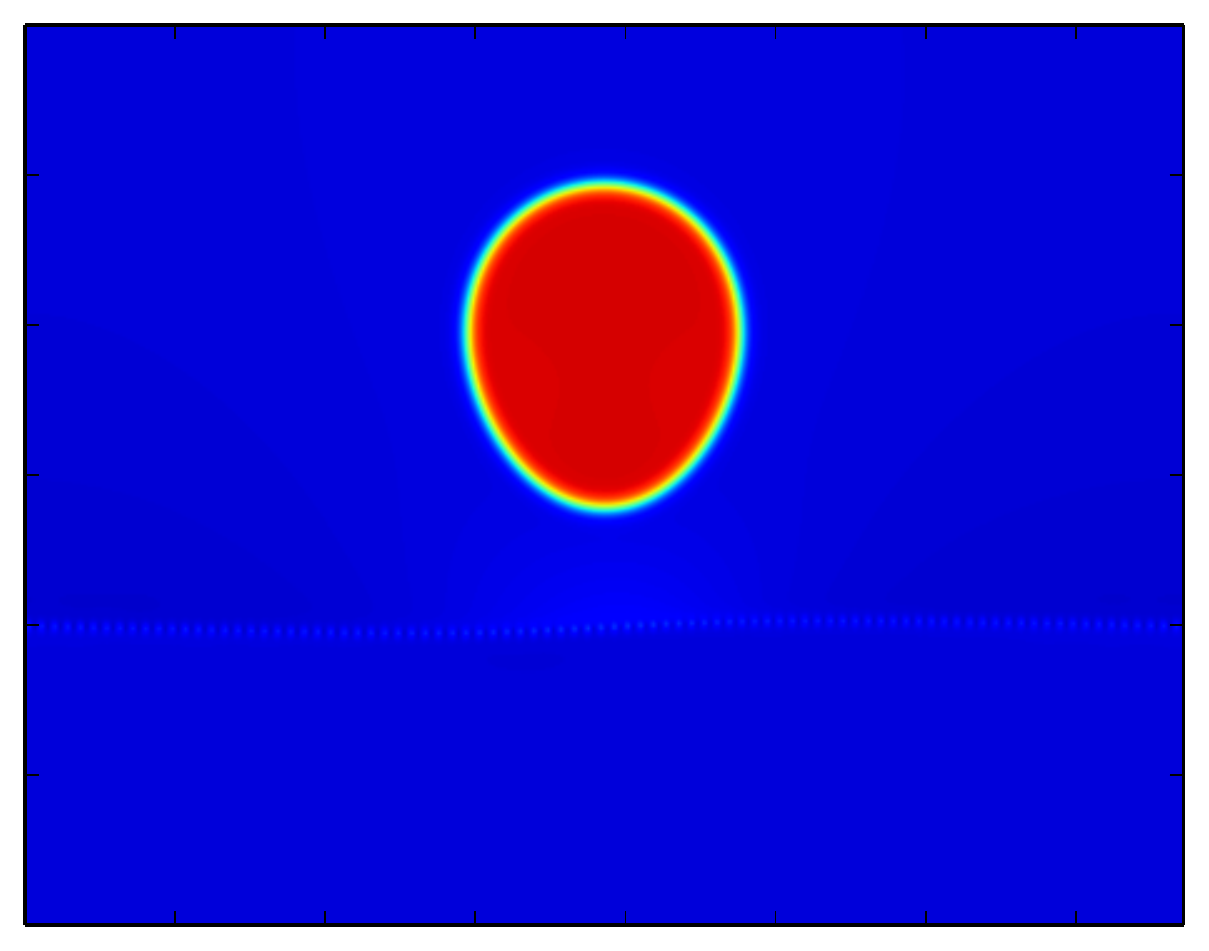}
    \includegraphics[width=0.45\linewidth]{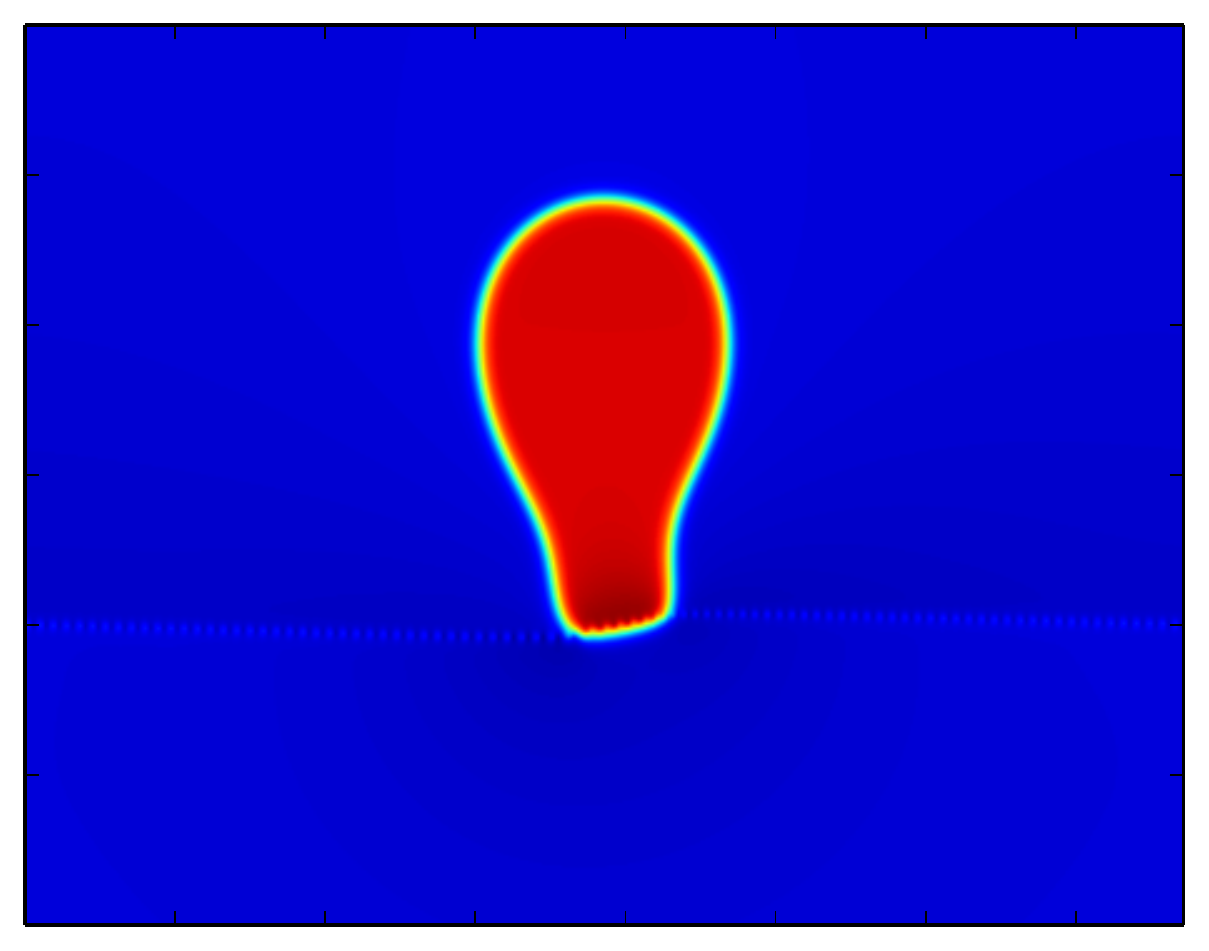}
    \includegraphics[width=0.45\linewidth]{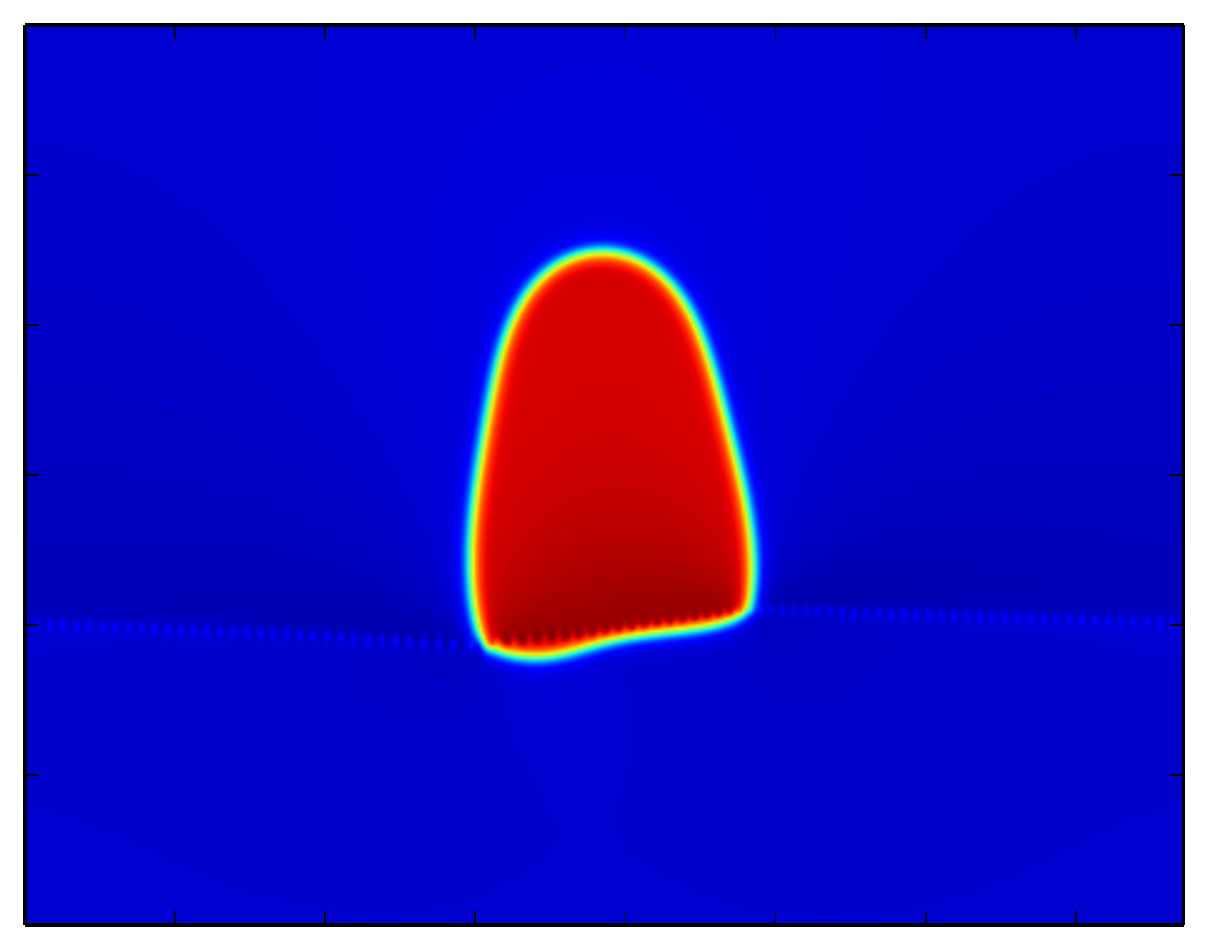}
    \includegraphics[width=0.45\linewidth]{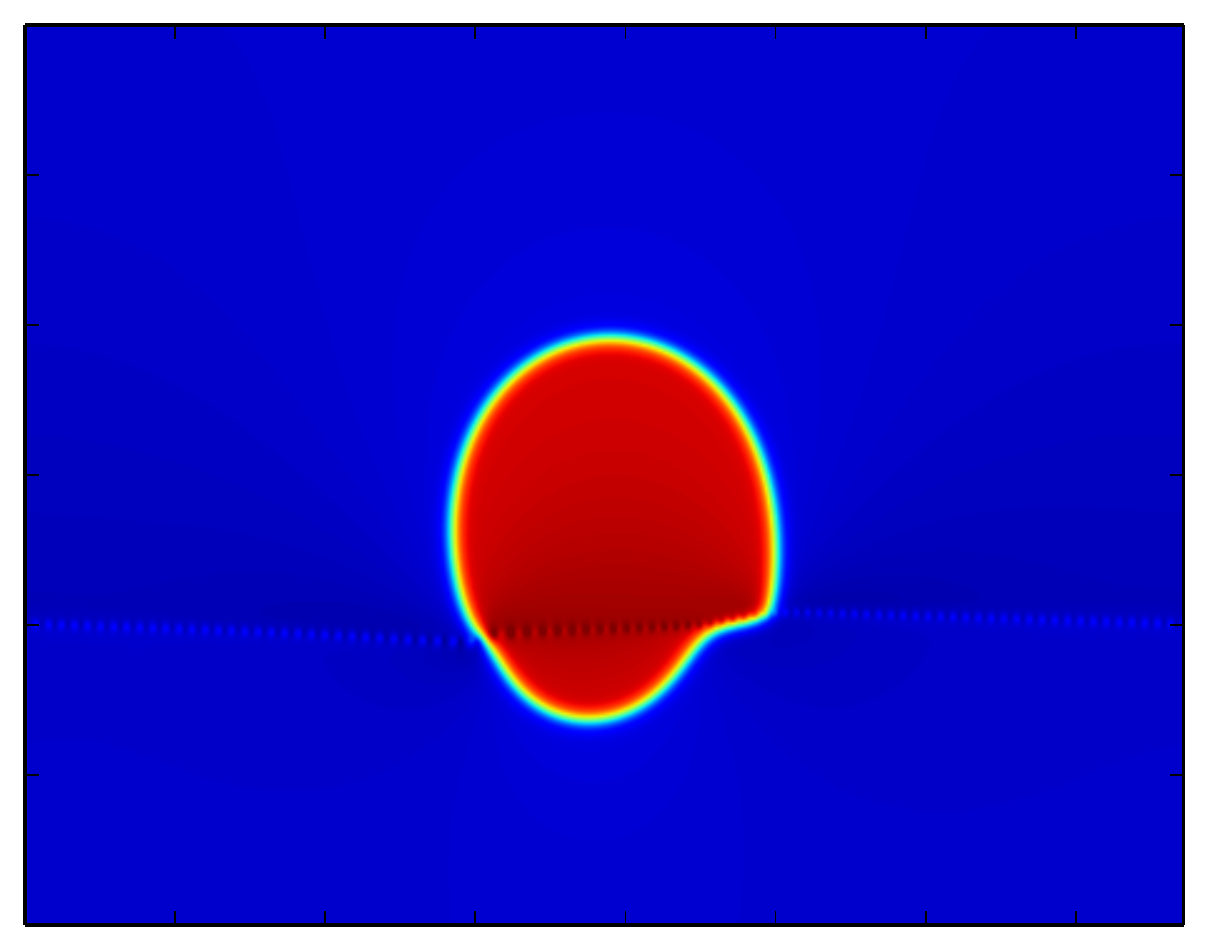}
    \includegraphics[width=0.45\linewidth]{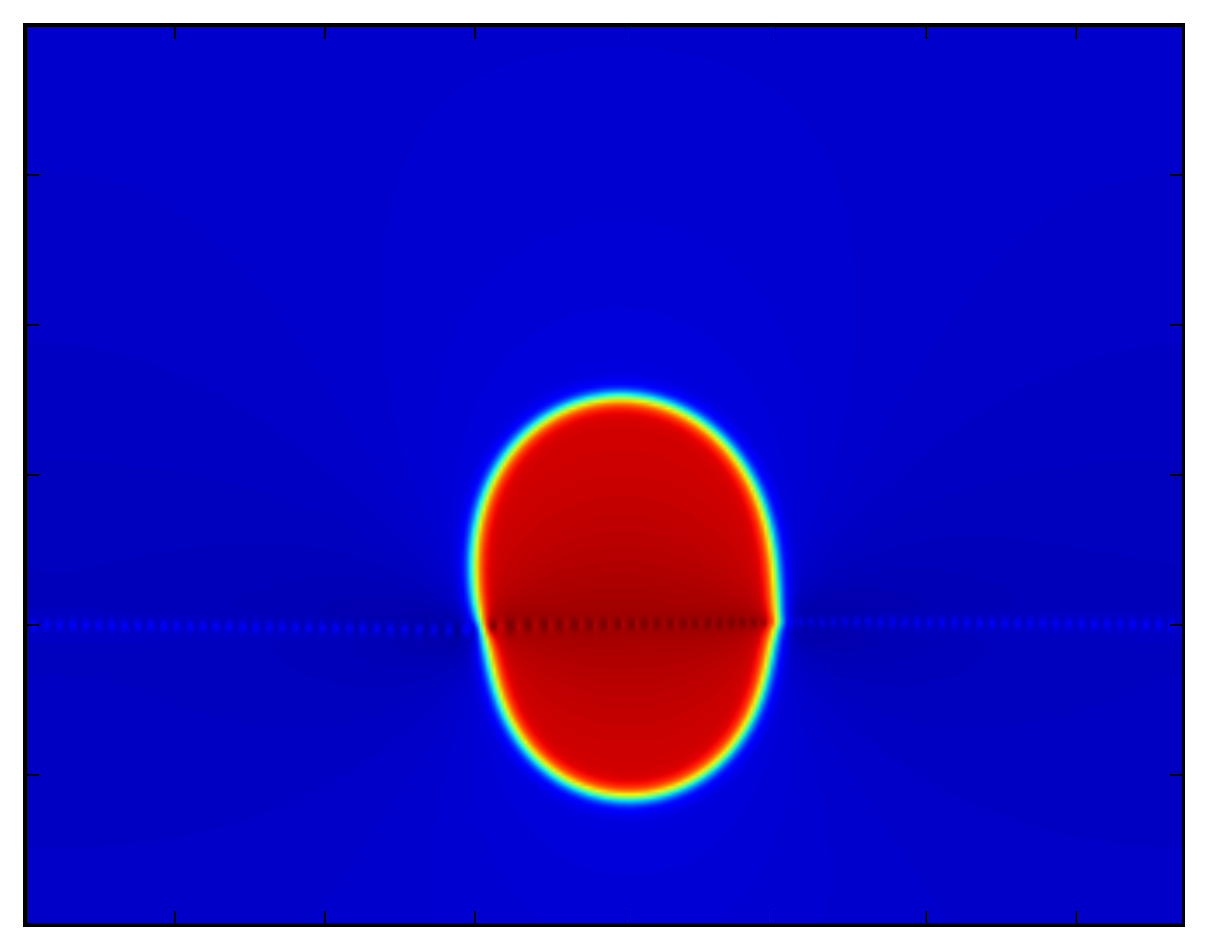}
    \includegraphics[width=1.00\linewidth]{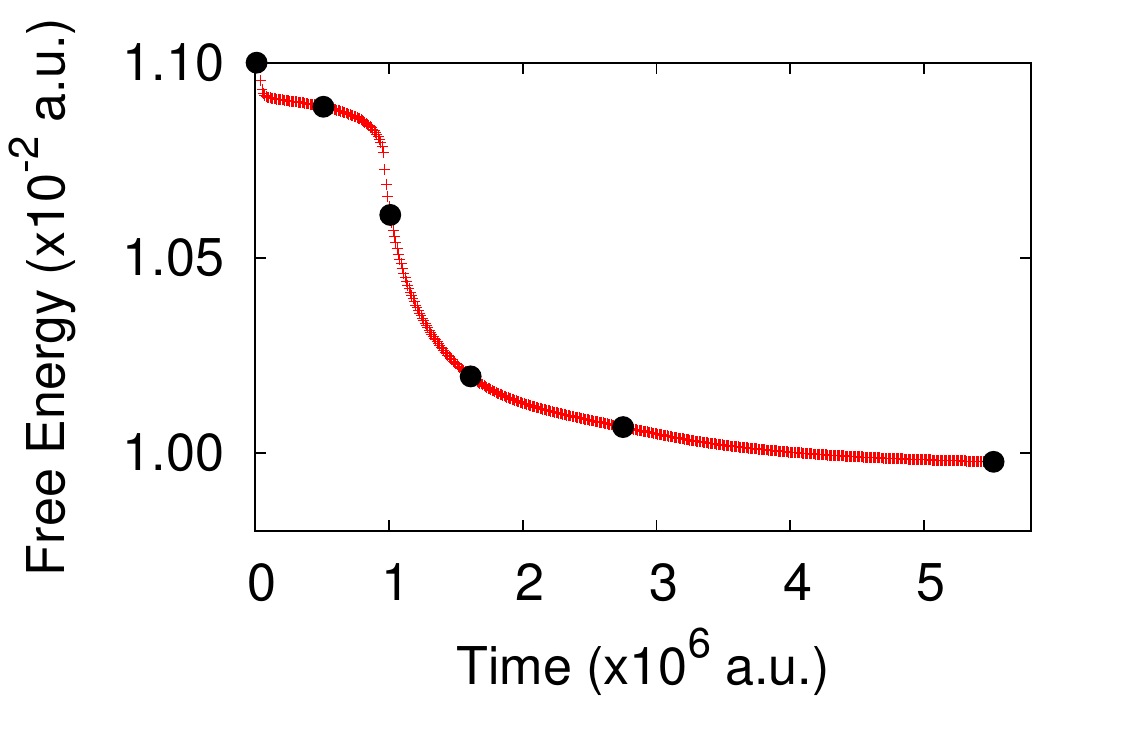}
		
		\begin{picture}(1,0)(0,0)
			\put(-99.,330.) {\mbox{\textcolor{white}{\textbf{(a)}}}}
            \put(5.,330.) {\mbox{\textcolor{white}{\textbf{(b)}}}}
            \put(-99.,249.) {\mbox{\textcolor{white}{\textbf{(c)}}}}
			\put(5.,249.) {\mbox{\textcolor{white}{\textbf{(d)}}}}
            \put(-99.,168.) {\mbox{\textcolor{white}{\textbf{(e)}}}}
            \put(5.,168.) {\mbox{\textcolor{white}{\textbf{(f)}}}}
            \put(-60.,60.) {\mbox{\textcolor{black}{\textbf{(g)}}}}
		\end{picture}
  \end{center}
  \vskip -0.6cm
  \caption{Snapshots of the concentration field during a simulation performed with
    a GB misorientation $\theta=20.8\degree$ and a circular precipitate with a
    misfit eigenstrain $\eps_0=0.043$.
    The system size is $7.7R \times 10.8R$ where $R$ is the radius of the circular 
    precipitate. The snapshots correspond to dimensionless times (a) $1\times10^4$,
    (b) $5.1\times10^5$, (c) $1.01\times10^6$, (d) $1.61\times10^6$, (e) $2.75\times10^6$
    and (f) $5.52\times10^6$. The evolution of the total free energy is represented in panel (g)
    in which black dots locate the snapshots in (a-f). See online supplementental material\cite{SupplMat} for the movie of this simulation (file movie4.avi).}
  \label{fig:circlehigh}
\end{figure}
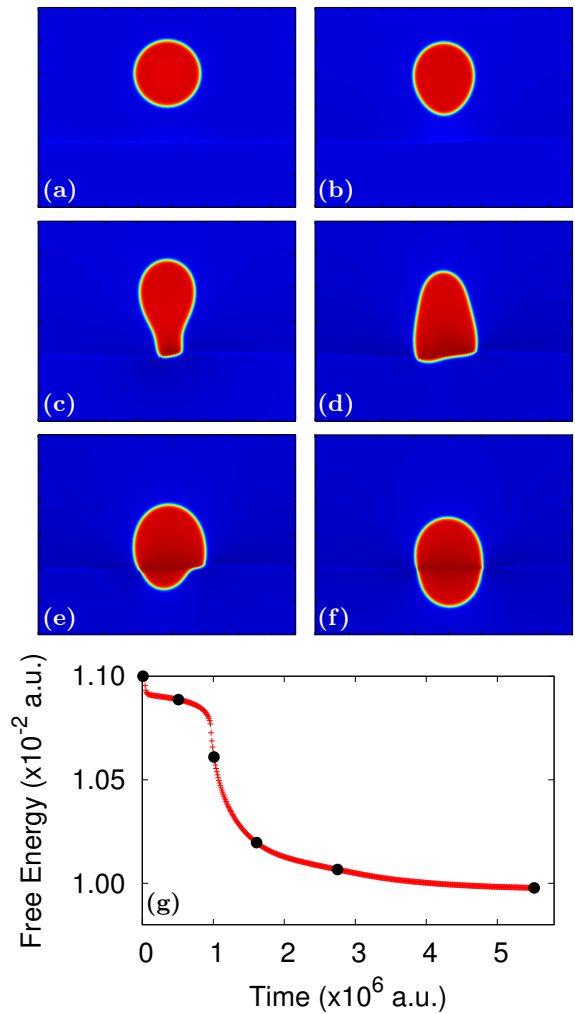

In the previous sections, we considered configurations consisting of a lamellar precipitate 
centered on a GB. Even though this geometry is relevant for heterogeneous nucleation
of precipitates on GBs, circular precipitates also commonly appear 
in the vicinity of a GB as exemplified by Ni-Al superalloys 
\cite{Koul1983,Safari2006,Mitchell2009a}. The $\gamma'$ precipitates in these alloys  
are known to influence the GB morphology by causing
their serration and this mechanism has been shown to improve the creep properties
of the alloy by preventing GB sliding \cite{Koul1983}. The GB serration has been proposed to be due to 
a balance between the elastic energy released by the coherency loss of the precipitate interface 
in contact with the GB and the GB surface tension \cite{Koul1983,Safari2006}, 
two ingredients that are naturally taken into account in the AE model.

We consider a configuration consisting of a circular precipitate of radius $R=32a$ and of misfit $\eps_0=0.043$ 
located at a distance $2R$ of a symmetrical GB. Snapshots of simulations performed with two different
misorientation angles $\theta=7.2\degree$ and $\theta=20.8\degree$ are presented in 
\Cref{fig:circlelow,fig:circlehigh}, respectively. Even though the circular and lamellar precipitate geometries differ significantly, the simulations reveal that the mechanism of the GB instability is similar in both cases. The GB deforms slightly by shear-coupled motion to relax shear stresses produced by the misfitting particle.
In turn, the deformed GB produces an heterogeneous stress field, inducing the migration of the precipitate 
towards the GB (see \Cref{fig:circlelow}.a-c and \Cref{fig:circlehigh}.a-b). This migration
is mediated by the elongation of the precipitate. The additional surface energy caused by this 
elongation is compensated by the relaxation of shear stresses by the GB coupling mechanism. Therefore, the interplay
between the elastic energy and the surface energy lead to the destabilization of the configuration through a 
mechanism similar to the morphological instability of lamellar precipitates described previously. 
Eventually, the misfitting particle enters in contact with the GB. For the low angle GB, 
the misfit stress is large enough to break-up the GB (see \Cref{fig:circlelow}.d), allowing the GB dislocations 
to relax part of the misfit stress. For the higher angle GB, the precipitate interface wets the GB without 
leading to its break-up (see \Cref{fig:circlehigh}.d). In both cases, the equilibrium configuration show a slightly elliptic 
precipitate centered on the somewhat perturbed GB. This configuration relaxes the total energy as 
shown in \Cref{fig:circlehigh}.g.

Despite a large value of the misfit compared to $\gamma'$ precipitates in Ni-Al superalloys, these simulations 
show that elastic interactions between the misfitting particle and the GB
induce a driving force for the migration of the particle, thereby providing a mechanism for how $\gamma'$ precipitates nucleated in the GB vicinity migrate towards the GB. In the case of an isolated misfitting particle, the shear stress induced along the GB by the particle decays as $(R/\ell)^d$ where $R$ and $\ell$ are the particle radius and  its initial perpendicular distance to the GB, respectively, and $d$ is the dimension of space. Hence, in both two and three dimensions, the GB will be deformed on a scale comparable to $\ell$. This deformation will in turn perturb the stress field on a scale $\ell$ , thereby causing the precipitate to migrate at a rate that becomes vanishingly small in the limit $\ell\gg R$. In the case where several particles are present, the interaction between particles and GB is more complex. However, significant migration is generally expected to only occur when a particle is located at a distance from the GB comparable to its size.

\section{Conclusions and outlook}

In summary, we used the AE framework to investigate computationally the interaction between GBs and second phase precipitates in two-phase coherent solids in the presence of misfit strain. We focused on two generic geometries where a GB is centered inside a lamellar precipitate formed by heterogeneous nucleation on the GB, and where the GB is adjacent to a circular precipitate that nucleates inside a grain. We find that, in both geometries, the GB becomes deformed away from its initial planar configuration by a coordinated motion of the GB and the adjacent compositional DB(s) that relaxes the elastic strain energy created by the misfit precipitate. The motion of the GB is driven by shear stresses along the GB (shear-coupled motion) while the motion of the DBs is driven by concentration gradients and controlled by atomic diffusion. 

For the lamellar precipitate geometry, the coordinated motion of the GB and DBs is manifested as a pattern-forming instability with a fastest growing wavelength. This instability bears similarities with the Asaro-Tiller-Grinfeld (ATG) instability \cite{Asaro1972,Grinfeld1986} where destabilization is mediated by the relaxation of the normal stresses at a free surface or a solid-liquid interface. However, the present instability is more complex in that it involves the interaction of two fundamentally different types of interfaces (GB and DBs). Furthermore, it is mediated by the relaxation of a shear stress at the GB. We have characterized analytically this instability by extending our previous linear stability analysis for isotropic elasticity \cite{Geslin2015} to the more complex case of anisotropic elasticity.  The analysis predicts that, if the lamellar precipitate is oriented along an elastically soft direction, elastic anisotropy hinders the instability by reducing the growth rate of the instability and the range of unstable wavelengths. However, anisotropy does not suppress the instability even though the lamellar precipitate would be stable in this configuration in the absence of misfit. Analytical predictions for the growth rate of perturbations and the range of unstable wavelengths are in good overall   quantitative agreement with the results of AE simulations for three-dimensional BCC crystal structures. 

For a circular precipitate adjacent to a planar GB, the coordinated motion of the GB and DB is manifested by an elongation of the precipitate shape and concomitant migration of the precipitate towards the deformed GB. 
The increase of interfacial energy associated with this elongation is compensated by the relaxation of shear stresses by the GB coupling mechanism. Hence, the interplay between elastic and interfacial energy leads to the destabilization of the initial GB-precipitate configuration by a physical mechanism similar to the morphological instability of the GB inside a lamellar precipitate.

Simulations also reveal that, in the lamellar geometry, instability can lead to the break-up of low-angle GBs when the misfit strain exceeds a threshold that depends on the grain boundary misorientation. Stationary equilibrium configurations after break-up can be quite complex and consist of dislocations that reside inside or outside the precipitate and decorate its surface to relax the misfit stress. For the circular precipitate, GB break-up also occurs for low angle GBs even though the final equilibrium configuration is typically an oval shape precipitate centered on an approximately flat GB, at least for the few cases investigated here. For both the lamellar and circular precipitates, dislocation climb is seen to provide an important mechanism to relax the total free-energy in addition to glide. 
 
The present findings should be relevant for interpreting a host of experiments where GBs interact strongly with precipitates, including the aforementioned examples of
Ni-Al superalloys where $\gamma'$ precipitates lead to GB serration \cite{Koul1983,Mitchell2009a} and Widmanst\"{a}tten precipitates in steel and Ti-based alloys, which are observed to grow out in a direction normal to the GB plane \cite{DaCostaTeixeira2006,Cheng2010}.
In the more general setting of spinodal decomposition occurring in a 
polycrystalline material, our results suggest that a large difference of lattice spacing between compositional domains
could influence significantly the grain structure by the break-up of GBs 
or the nucleation of new grains (e.g. \Cref{fig:hvbreak}), thereby affecting
the resulting properties of the bulk material. {\it In situ} experimental observations that characterize the interactions between GBs and precipitates in both controlled bi-crystal geometries and complex networks of GBs remain needed to validate more directly the instability mechanisms highlighted in the present study.  

\acknowledgments{This research was supported by Grant No. DE-FG02-07ER46400 from the U.S. Department of Energy, Office of Basic Energy Sciences.}

\appendix
\begin{widetext}
\section{Amplitude equations for BCC}
\label{sec:AE-BCC}

For BCC ordering, the evolution equation for concentration is the same as \Cref{eq:aemecd}
but with six amplitude variables $A_{1}$, $A_{2}$, $A_{3}$, $A_{4}$, 
$A_{5}$, $A_{6}$. We just list here the six amplitude equations.

  For $A_{1}$:
\begin{align}
  c_w^{-2}\frac{\partial^2 A_{1}}{\partial t^2} + \beta_w\frac{\partial A_{1}}{\partial t} 
  &= \alpha_d^2\left[\Box_{1}^2A_{1} + 2i\eps_0 c\Box_{1} A_{1} + i\eps_0 A_{1}\Box_{1} c +\eps_0\nabla A_{1}\cdot\nabla c -\eps_0^2 c^2A_{1}\right]\\
  & -\frac{1}{12}A_{1} -\frac{1}{45}A_{1}\sum_{j=1}^{6}A_jA_j^{*} +\frac{1}{90}A_{1}|A_{1}|^2 -\frac{1}{45}(A_{4}^*A_{2}A_{5}+A_{4}^*A_{3}A_{6})+\frac{1}{8}(A_{3}A_{5}+A_{6}A_{2})\nonumber
\end{align}
  For $A_{2}$:
\begin{align}
  c_w^{-2}\frac{\partial^2 A_{2}}{\partial t^2} + \beta_w\frac{\partial A_{2}}{\partial t} 
  &= \alpha_d^2\left[\Box_{2}^2A_{2} + 2i\eps_0 c\Box_{2} A_{2} + i\eps_0 A_{2}\Box_{2} c +\eps_0\nabla A_{2}\cdot\nabla c -\eps_0^2 c^2A_{2}\right]\\
  & -\frac{1}{12}A_{2} -\frac{1}{45}A_{2}\sum_{j=1}^{6}A_jA_j^{*} +\frac{1}{90}A_{2}|A_{2}|^2 -\frac{1}{45}(A_{1}A_{4}A_{5}^*+A_{6}^*A_{5}A_{3})+\frac{1}{8}(A_{3}A_{4}+A_{6}^*A_{1})\nonumber
\end{align}
  For $A_{3}$:
\begin{align}
  c_w^{-2}\frac{\partial^2 A_{3}}{\partial t^2} + \beta_w\frac{\partial A_{3}}{\partial t} 
  &= \alpha_d^2\left[\Box_{3}^2A_{3} + 2i\eps_0 c\Box_{3} A_{3} + i\eps_0 A_{3}\Box_{3} c +\eps_0\nabla A_{3}\cdot\nabla c -\eps_0^2 c^2A_{3}\right]\\
  & -\frac{1}{12}A_{3} -\frac{1}{45}A_{3}\sum_{j=1}^{6}A_jA_j^{*} +\frac{1}{90}A_{3}|A_{3}|^2 -\frac{1}{45}(A_{4}^*A_{6}^*A_{1}+A_{6}A_{5}^*A_{2})+\frac{1}{8}(A_{2}A_{4}^*+A_{5}^*A_{1})\nonumber
\end{align}
For $A_{4}$:
\begin{align}
  c_w^{-2}\frac{\partial^2 A_{4}}{\partial t^2} + \beta_w\frac{\partial A_{4}}{\partial t} 
  &= \alpha_d^2\left[\Box_{4}^2A_{4} + 2i\eps_0 c\Box_{4} A_{4} + i\eps_0 A_{4}\Box_{4} c +\eps_0\nabla A_{4}\cdot\nabla c -\eps_0^2 c^2A_{4}\right]\\
  & -\frac{1}{12}A_{4} -\frac{1}{45}A_{4}\sum_{j=1}^{6}A_jA_j^{*} +\frac{1}{90}A_{4}|A_{4}|^2 -\frac{1}{45}(A_{1}^*A_{2}A_{5}+A_{3}^*A_{6}^*A_{1})+\frac{1}{8}(A_{3}^*A_{2}+A_{6}^*A_{5})\nonumber
\end{align}
For $A_{5}$:
\begin{align}
  c_w^{-2}\frac{\partial^2 A_{5}}{\partial t^2} + \beta_w\frac{\partial A_{5}}{\partial t} 
  &= \alpha_d^2\left[\Box_{5}^2A_{5} + 2i\eps_0 c\Box_{5} A_{5} + i\eps_0 A_{5}\Box_{5} c +\eps_0\nabla A_{5}\cdot\nabla c -\eps_0^2 c^2A_{5}\right]\\
  & -\frac{1}{12}A_{5} -\frac{1}{45}A_{5}\sum_{j=1}^{6}A_jA_j^{*} +\frac{1}{90}A_{5}|A_{5}|^2 -\frac{1}{45}(A_{1}A_{4}A_{2}^*+A_{6}A_{5}A_{3}^*)+\frac{1}{8}(A_{3}^*A_{1}+A_{6}A_{4})\nonumber
\end{align}
For $A_{6}$:
\begin{align}
  c_w^{-2}\frac{\partial^2 A_{6}}{\partial t^2} + \beta_w\frac{\partial A_{6}}{\partial t} 
  &= \alpha_d^2\left[\Box_{6}^2A_{6} + 2i\eps_0 c\Box_{6} A_{6} + i\eps_0 A_{6}\Box_{6} c +\eps_0\nabla A_{6}\cdot\nabla c -\eps_0^2 c^2A_{6}\right]\\
  & -\frac{1}{12}A_{6} -\frac{1}{45}A_{6}\sum_{j=1}^{6}A_jA_j^{*} +\frac{1}{90}A_{6}|A_{6}|^2 -\frac{1}{45}(A_{4}^*A_{3}^*A_{1}+A_{5}A_{2}^*A_{3})+\frac{1}{8}(A_{1}A_{2}^*+A_{5}^*A_{4}^*)\nonumber
\end{align}

\section{Solution of the linear system of equations}
\label{sec:RLE}

We list below the solution of the linear system of 13 equations that determines the coefficients $A_i$, $B_i$, $C_i$, $D_i$ and $H_0$.\begin{align}\nonumber
  A_1 =& -\{i h_0 \alpha (1 + \zeta) [(-1 + \zeta) (-M_2 (-(1 + e^{ 2 i k p_1 w}) p_1 + p_2 - e^{2 i k p_1 w} p_2\\\nonumber
    & + M_2 (-1 + p_1 p_2 \zeta + e^{2 i k p_1 w} (1 + p_1 p_2 \zeta))) + M_1 (-2 e^{i k (p_1 + p_2) w} p_2 + M_2^2 ((-1 + e^{2 i k p_1 w}) p_1\\\nonumber
    & + (1 + e^{2 i k p_1 w} - 2 e^{i k (p_1 + p_2) w}) p_2) \zeta - M_2 (1 - p_1 p_2 \zeta + e^{2 i k p_1 w} (1 + p_1 p_2 \zeta) - 2 e^{i k (p_1 + p_2) w} (1 + p_2^2 \zeta)))) C_{12}\\\nonumber
    & - (M_2 - p_2) (p_1 + e^{2 i k p_1 w} p_1 - 2 e^{i k (p_1 + p_2) w} p_1 - p_2 + e^{2 i k p_1 w} p_2 - M_2 (-1 + p_1 p_2 \zeta - 2 e^{i k (p_1 + p_2) w} p_1 p_2 \zeta \\\nonumber
    & + e^{2 i k p_1 w} (1 + p_1 p_2 \zeta)) + M_1 (-1 + p_1 p_2 \zeta + M_2 (-p_1 + p_2) \zeta + e^{i k (p_1 + p_2) w} (2 - 2 M_2 p_2 \zeta)\\\nonumber
    & + e^{2 i k p_1 w} (-1 - p_1 p_2 \zeta + M_2 (p_1 + p_2) \zeta))) C_{ 44}]\}\\
    & /\{2 (M_2 p_1 - M_1 p_2) \zeta (-p_1 + p_2 + M_1 (1 + M_2 (p_1 - p_2) \zeta - p_1 p_2 \zeta) + M_2 (-1 + p_1 p_2 \zeta)) C_{ 44}\}
\end{align}
\begin{align}\nonumber
  B_1 =& -\{i h_0 \alpha (1 + \zeta) [(-1 + \zeta) (-2 e^{i k (p_1 + p_2) w} M_2 p_1 + M_1^2 (1 - 2 e^{i k (p_1 + p_2) w} M_2 p_1 \zeta\\\nonumber
    & + M_2 (p_1 - p_2) \zeta - p_1 p_2 \zeta + e^{2 i k p_2 w} (-1 - p_1 p_2 \zeta + M_2 (p_1 + p_2) \zeta)) + M_1 ((-1 + e^{2 i k p_2 w}) p_1\\\nonumber
    & + (1 + e^{2 i k p_2 w}) p_2 - M_2 (1 - p_1 p_2 \zeta - 2 e^{i k (p_1 + p_2) w} (1 + p_1^2 \zeta) + e^{2 i k p_2 w} (1 + p_1 p_2 \zeta)))) C_{12}\\\nonumber
    & - (M_1 - p_1) (-p_1 + e^{2 i k p_2 w} p_1 + p_2 + e^{2 i k p_2 w} p_2 - 2 e^{i k (p_1 + p_2) w} p_2 - M_2 (1 - 2 e^{i k (p_1 + p_2) w} - p_1 p_2 \zeta\\\nonumber
    & + e^{2 i k p_2 w} (1 + p_1 p_2 \zeta)) + M_1 (1 + M_2 (p_1 - p_2) \zeta - p_1 p_2 \zeta + 2 e^{i k (p_1 + p_2) w} p_1 (-M_2 + p_2) \zeta\\\nonumber
    & + e^{2 i k p_2 w} (-1 - p_1 p_2 \zeta + M_2 (p_1 + p_2) \zeta))) C_{44}]\}\\
  & /\{2 (M_2 p_1 - M_1 p_2) \zeta (-p_1 + p_2 + M_1 (1 + M_2 (p_1 - p_2) \zeta - p_1 p_2 \zeta) + M_2 (-1 + p_1 p_2 \zeta)) C_{44}\}
\end{align}
\begin{align}
  A_2 =& \{i h_0 \alpha (1 + \zeta) (M_2 (-1 + \zeta) C_{12} + (-M_2 + p_2) C_{44})\}/\{2 (M_2 p_1 - M_1 p_2) \zeta C_{44}\}\\
  B_2 =& -\{i h_0 \alpha (1 + \zeta) (M_1 (-1 + \zeta) C_{12} + (-M_1 + p_1) C_{44})\}/\{ 2 (M_2 p_1 - M_1 p_2) \zeta C_{44}\}
\end{align}
\begin{align}\nonumber
  C_2 =& -\{i h_0 \alpha (1 + \zeta) [(-1 + \zeta) (2 e^{i k p_2 w} M_1 (-M_2 + p_2) (-1 + M_2 p_2 \zeta) + e^{i k p_1 w} M_2 (p_1 + p_2 - M_2 (1 + p_1 p_2 \zeta)\\\nonumber
  & + M_1 (-1 - p_1 p_2 \zeta + M_2 (p_1 + p_2) \zeta))) C_{12}+ (M_2 - p_2) (2 e^{i k p_2 w} (M_1 - p_1) (-1 + M_2 p_2 \zeta)\\\nonumber
  & + e^{i k p_1 w} (M_2 - p_1 - p_2 + M_2 p_1 p_2 \zeta + M_1 (1 + p_1 p_2 \zeta - M_2 (p_1 + p_2) \zeta))) C_{44}]\}\\
  & /\{2 (M_2 p_1 - M_1 p_2) \zeta (-p_1 + p_2 + M_1 (1 + M_2 (p_1 - p_2) \zeta - p_1 p_2 \zeta) + M_2 (-1 + p_1 p_2 \zeta)) C_{44})\}
\end{align}
\begin{align}\nonumber
  D_2 =& \{i h_0 \alpha (1 + \zeta) [(-1 + \zeta) (-2 e^{i k p_1 w} M_2 (-M_1 + p_1) (-1 + M_1 p_1 \zeta) - e^{i k p_2 w} M_1 (p_1 + p_2 - M_2 (1 + p_1 p_2 \zeta)\\\nonumber
  &+ M_1 (-1 - p_1 p_2 \zeta + M_2 (p_1 + p_2) \zeta)))  C_{12}- (M_1 - p_1) (2 e^{i k p_1 w} (M_2 - p_2) (-1 + M_1 p_1 \zeta)\\\nonumber
  & + e^{i k p_2 w} (M_2 - p_1 - p_2 + M_2 p_1 p_2 \zeta + M_1 (1 + p_1 p_2 \zeta - M_2 (p_1 + p_2) \zeta))) C_{44}]\}\\
  & /\{2 (M_2 p_1 - M_1 p_2) \zeta (-p_1 + p_2 + M_1 (1 + M_2 (p_1 - p_2) \zeta - p_1 p_2 \zeta) + M_2 (-1 + p_1 p_2 \zeta)) C_{44}\}
\end{align}
\begin{align}
  A_3 =& -\{i h_0 \alpha (1 + \zeta) (M_2 (-1 + \zeta) C_{12}+ (-M_2 + p_2) C_{44})\}/\{2 (M_2 p_1 - M_1 p_2) \zeta C_{44}\}\\
  B_3 =& \{i h_0 \alpha (1 + \zeta) (M_1 (-1 + \zeta) C_{12}+ (-M_1 + p_1) C_{44})\}/\{2 (M_2 p_1 - M_1 p_2) \zeta C_{44}\}
\end{align}
\begin{align}\nonumber
  C_3 =& \{i h_0 \alpha (1 + \zeta) [(-1 + \zeta) (2 e^{i k p_2 w} M_1 (-M_2 + p_2) (-1 + M_2 p_2 \zeta) + e^{i k p_1 w} M_2 (p_1 + p_2 - M_2 (1 + p_1 p_2 \zeta)\\\nonumber
  & + M_1 (-1 - p_1 p_2 \zeta + M_2 (p_1 + p_2) \zeta)))  C_{12}+ (M_2 - p_2) (2 e^{i k p_2 w} (M_1 - p_1) (-1 + M_2 p_2 \zeta)\\\nonumber
  & + e^{i k p_1 w} (M_2 - p_1 - p_2 + M_2 p_1 p_2 \zeta + M_1 (1 + p_1 p_2 \zeta - M_2 (p_1 + p_2) \zeta))) C_{44}]\}\\
  & /\{2 (M_2 p_1 - M_1 p_2) \zeta (-p_1 + p_2 + M_1 (1 + M_2 (p_1 - p_2) \zeta - p_1 p_2 \zeta) + M_2 (-1 + p_1 p_2 \zeta)) C_{44}\}
\end{align}
\begin{align}\nonumber
  D_3 =& \{i h_0 \alpha (1 + \zeta) [-(-1 + \zeta) (-2 e^{i k p_1 w} M_2 (-M_1 + p_1) (-1 + M_1 p_1 \zeta) - e^{i k p_2 w} M_1 (p_1 + p_2 - M_2 (1 + p_1 p_2 \zeta)\\\nonumber
  & + M_1 (-1 - p_1 p_2 \zeta + M_2 (p_1 + p_2) \zeta)))  C_{12}+ (M_1 - p_1) (2 e^{i k p_1 w} (M_2 - p_2) (-1 + M_1 p_1 \zeta)\\\nonumber
  & + e^{i k p_2 w} (M_2 - p_1 - p_2 + M_2 p_1 p_2 \zeta + M_1 (1 + p_1 p_2 \zeta - M_2 (p_1 + p_2) \zeta))) C_{44}]\}\\
  & /\{2 (M_2 p_1 - M_1 p_2) \zeta (-p_1 + p_2 + M_1 (1 + M_2 (p_1 - p_2) \zeta - p_1 p_2 \zeta) + M_2 (-1 + p_1 p_2 \zeta)) C_{44}\}
\end{align}
\begin{align}\nonumber
  A_4 =& \{i h_0 \alpha (1 + \zeta) [(-1 + \zeta) (-M_2 (-(1 + e^{ 2 i k p_1 w}) p_1 + p_2 - e^{2 i k p_1 w} p_2 + M_2 (-1 + p_1 p_2 \zeta + e^{2 i k p_1 w} (1 + p_1 p_2 \zeta)))\\\nonumber
  &+ M_1 (-2 e^{i k (p_1 + p_2) w} p_2 + M_2^2 ((-1 + e^{2 i k p_1 w}) p_1 + (1 + e^{2 i k p_1 w} - 2 e^{i k (p_1 + p_2) w}) p_2) \zeta\\\nonumber
  &- M_2 (1 - p_1 p_2 \zeta + e^{2 i k p_1 w} (1 + p_1 p_2 \zeta) - 2 e^{i k (p_1 + p_2) w} (1 + p_2^2 \zeta)))) C_{12}- (M_2 - p_2) (p_1 + e^{2 i k p_1 w} p_1\\\nonumber
  &- 2 e^{i k (p_1 + p_2) w} p_1 - p_2 + e^{2 i k p_1 w} p_2 - M_2 (-1 + p_1 p_2 \zeta - 2 e^{i k (p_1 + p_2) w} p_1 p_2 \zeta + e^{2 i k p_1 w} (1 + p_1 p_2 \zeta))\\\nonumber
& + M_1 (-1 + p_1 p_2 \zeta + M_2 (-p_1 + p_2) \zeta + e^{i k (p_1 + p_2) w} (2 - 2 M_2 p_2 \zeta) + e^{2 i k p_1 w} (-1 - p_1 p_2 \zeta + M_2 (p_1 + p_2) \zeta))) C_{44}]\}\\
  &/\{2 (M_2 p_1 - M_1 p_2) \zeta (-p_1 + p_2 + M_1 (1 + M_2 (p_1 - p_2) \zeta - p_1 p_2 \zeta) + M_2 (-1 + p_1 p_2 \zeta)) C_{44}\}
\end{align}
\begin{align}\nonumber
  B_4 =& \{i h_0 \alpha (1 + \zeta) [(-1 + \zeta) (-2 e^{i k (p_1 + p_2) w} M_2 p_1 + M_1^2 (1 - 2 e^{i k (p_1 + p_2) w} M_2 p_1 \zeta + M_2 (p_1 - p_2) \zeta - p_1 p_2 \zeta\\\nonumber
  &+ e^{2 i k p_2 w} (-1 - p_1 p_2 \zeta + M_2 (p_1 + p_2) \zeta)) + M_1 ((-1 + e^{2 i k p_2 w}) p_1 + (1 + e^{2 i k p_2 w}) p_2\\\nonumber
  &- M_2 (1 - p_1 p_2 \zeta - 2 e^{i k (p_1 + p_2) w} (1 + p_1^2 \zeta) + e^{2 i k p_2 w} (1 + p_1 p_2 \zeta)))) C_{12}- (M_1 - p_1) (-p_1 + e^{2 i k p_2 w} p_1 + p_2\\\nonumber 
  & + e^{2 i k p_2 w} p_2 - 2 e^{i k (p_1 + p_2) w} p_2 - M_2 (1 - 2 e^{i k (p_1 + p_2) w} - p_1 p_2 \zeta + e^{2 i k p_2 w} (1 + p_1 p_2 \zeta))\\\nonumber
& + M_1 (1 + M_2 (p_1 - p_2) \zeta - p_1 p_2 \zeta + 2 e^{i k (p_1 + p_2) w} p_1 (-M_2 + p_2) \zeta + e^{2 i k p_2 w} (-1 - p_1 p_2 \zeta + M_2 (p_1 + p_2) \zeta))) C_{44}]\}\\
  & /\{2 (M_2 p_1 - M_1 p_2) \zeta (-p_1 + p_2 + M_1 (1 + M_2 (p_1 - p_2) \zeta - p_1 p_2 \zeta) + M_2 (-1 + p_1 p_2 \zeta)) C_{44}\}\\\nonumber
  H_0 =& \{2 i h_0 (M_1 p_1 - M_2 p_2) \alpha (1 + \zeta) [(e^{i k p_1 w} M_2 (-M_1 + p_1) - e^{i k p_2 w} M_1 (-M_2 + p_2)) (-1 + \zeta) C_{12}\\\nonumber
  &+ (e^{i k p_1 w} - e^{i k p_2 w}) (-M_1 + p_1) (-M_2 + p_2) C_{44}]\}\\
  & /\{(M_2 p_1 - M_1 p_2) \beta (-p_1 + p_2 + M_1 (1 + M_2 (p_1 - p_2) \zeta - p_1 p_2 \zeta) + M_2 (-1 + p_1 p_2 \zeta)) C_{44}\}
\end{align}
\end{widetext}


%

\end{document}